\documentclass{aastex62}

\usepackage{lineno}
\usepackage{amsmath}
\received{Month ??, 2021}
\revised{, 2021}
\accepted{Month ??, 2021}%\today}
\submitjournal{ApJ}

\shorttitle{The NIR-MIR SEDs of QSOs}
\shortauthors{Mart\'inez-Paredes et al.}
\begin{document}

\title{Modeling the unresolved NIR-MIR SEDs of local ($z<0.1$) QSOs}

\correspondingauthor{M. Mart\'inez-Paredes}
\email{mariellauriga@kasi.re.kr, mariellauriga@gmail.com}

\author[0000-0002-0786-7307]{M. Mart\'inez-Paredes}
\affil{Korea Astronomy and Space Science Institute 776, Daedeokdae-ro, Yuseong-gu, Daejeon, Republic of Korea (34055)}

\author{O. Gonz\'alez-Mart\'in}
\affil{Instituto de Radioastronom\'ia y Astrof\'isica UNAM 
Apartado Postal 3-72 (Xangari), 58089 Morelia, Michoac\'an, Mexico} 

\author{K. HyeongHan}
\affil{Korea Astronomy and Space Science Institute 776, Daedeokdae-ro, Yuseong-gu, Daejeon, Republic of Korea (34055)}
\affil{Yonsei University, 50 Yonsei-ro Seodaemun-gu, Seoul, Republic of Korea (03722)}

\author{S. Geier}
\affiliation{Instituto de Astrof\'isica de Canarias (IAC),E-38205 La Laguna, Tenerife, Spain}
\affiliation{Gran Telescopio Canarias (GRANTECAN), Cuesta de San José s/n, 38712 Breña Baja, La Palma, Spain}
	
\author{I. Garc\'ia-Bernete}
\affiliation{Instituto de Física de Cantabria (CSIC-UC), Avenida de los Castros, E-39005 Santander, Spain; Visiting Fellow, Centre for Extragalactic Astronomy, Durham University, South Road, Durham DH1 3LE, UK}

\author{C. Ramos Almeida}
\affiliation{Instituto de Astrof\'isica de Canarias (IAC),E-38205 La Laguna, Tenerife, Spain}
\affiliation{Departamento de Astrof\'isica, Universidad de La Laguna (ULL), E-38206 La Laguna, Tenerife, Spain}

%Alphabetic order
\author{A. Alonso-Herrero}
\affiliation{Centro de Astrobiolog\'ia, CSIC-INTA, ESAC Campus, E-28692 Villanueva de la Cañada, Madrid, Spain}

\author{I. Aretxaga}
\affiliation{Instituto Nacional de Astrof\'isica, \'Optica y Electr\'onica (INAOE), Luis Enrrique Erro 1, Sta. Ma. Tonantzintla, Puebla, Mexico}

\author{M. Kim}
\affil{Korea Astronomy and Space Science Institute 776, Daedeokdae-ro, Yuseong-gu, Daejeon, Republic of Korea (34055)}

\author{B. W. Sohn}
\affil{Korea Astronomy and Space Science Institute 776, Daedeokdae-ro, Yuseong-gu, Daejeon, Republic of Korea (34055)}

\author{J. Masegosa}
\affil{Instituto de Astrofisica de Andalucia, E 18008 Granada, Spain}

%\collaboration{(AAS Journals Data Scientists collaboration)}
%% Note that the \and command from previous versions of AASTeX is now
%% depreciated in this version as it is no longer necessary. AASTeX 
%% automatically takes care of all commas and "and"s between authors names.

%% AASTeX 6.2 has the new \collaboration and \nocollaboration commands to
%% provide the collaboration status of a group of authors. These commands 
%% can be used either before or after the list of corresponding authors. The
%% argument for \collaboration is the collaboration identifier. Authors are
%% encouraged to surround collaboration identifiers with ()s. The 
%% \nocollaboration command takes no argument and exists to indicate that
%% the nearby authors are not part of surrounding collaborations.

%% Mark off the abstract in the ``abstract'' environment. 
\begin{abstract}

To study the nuclear ($\lesssim1\,$kpc) dust of nearby ($z<0.1$) type 1 Quasi Stellar Objects (QSOs) we obtained new near-infrared (NIR) high angular resolution ($\sim0.3$ arcsec) photometry in the H and Ks bands, for 13 QSOs with  available mid-infrared (MIR) high angular resolution spectroscopy ($\sim7.5-13.5\,\mu$m). We find that in most QSOs the NIR emission is unresolved. We subtract the contribution from the accretion disk, which decreases from NIR ($\sim35\%$) to MIR ($\sim2.4\%$). We also estimate these percentages assuming a bluer accretion disk and find that the contibution in the MIR is nearly seven time larger. We find that the majority of objects ($64\%$, 9/13) are better fitted by the Disk+Wind H17 model \citep[][]{Hoenig17}, while others can be fitted by the Smooth F06  \citep[$14\%$, 2/13,][]{Fritz06}, Clumpy N08 \citep[$7\%$, 1/13,][]{Nenkova08a,Nenkova08b}, Clumpy H10 \citep[$7\%$, 1/13,][]{Hoenig10b}, and Two-Phase media S16 \citep[$7\%$, 1/13,][]{Stalev16} models. However, if we assume the bluer accretion disk, the models fit only 2/13 objects. We measured two NIR to MIR spectral indexes, $\alpha_{NIR-MIR(1.6,8.7\,\mu\text{m})}$ and $\alpha_{NIR-MIR(2.2,8.7\,\mu\text{m})}$, and two MIR spectral indexes, $\alpha_{MIR(7.8, 9.8\,\mu\text{m})}$ and $\alpha_{MIR(9.8, 11.7\,\mu\text{m})}$, from models and observations. From observations, we find that the NIR to MIR spectral indexes are $\sim-1.1$ and the MIR spectral indexes are $\sim-0.3$. Comparing the synthetic and observed values, we find that none of the models simultaneously match the measured NIR to MIR and $7.8-9.8\,\mu$m slopes. However, we note that measuring the $\alpha_{MIR(7.8, 9.8\,\mu\text{m})}$ on the starburst-subtracted {\it Spitzer}/IRS  spectrum, gives values of the slopes ($\sim-2$) that are similar to the synthetic values obtained from the models.

\end{abstract}

%% Keywords should appear after the \end{abstract} command. 
%% See the online documentation for the full list of available subject
%% keywords and the rules for their use.

%% From the front matter, we move on to the body of the paper.
%% Sections are demarcated by \section and \subsection, respectively.
%% Observe the use of the LaTeX \label
%% command after the \subsection to give a symbolic KEY to the
%% subsection for cross-referencing in a \ref command.
%% You can use LaTeX's \ref and \label commands to keep track of
%% cross-references to sections, equations, tables, and figures.
%% That way, if you change the order of any elements, LaTeX will
%% automatically renumber them.
%%
%% We recommend that authors also use the natbib \citep
%% and \citet commands to identify citations.  The citations are
%% tied to the reference list via symbolic KEYs. The KEY corresponds
%% to the KEY in the \bibitem in the reference list below. 
\section{Introduction} \label{sec:introduction}
The nuclear dust that surrounds the central engine (supermassive black hole, accretion disk, and broad-line region) of the active galactic nuclei (AGNs) is widely known as the dusty torus of AGNs \citep[e.g.,][]{Krolik_Begelman88,Antonucci93,Urry_Padovanni95,Robson95,Peterson97}. This structure plays a fundamental role in the description of an AGN since it represents the connection between the central engine and the kiloparsec scales \citep[the narrow line region and the host galaxy][]{Ramos_Almeida17}. Some observations of Seyfert galaxies have suggested that a large fraction of the obscuring material is distributed into a toroidal structure. However, the exact geometry of this structure is still unknown \citep[see e.g.,][]{Hoenig19, Combes19}. Some authors have claimed the detection of a dusty structure from mid-infrared (MIR) and sub-millimeter interferometric data   \citep[e.g.,][]{Jaffe04, Garcia_Burillo16, Imanishi18, Alonso-Herrero19}. They have found sizes that range from few parsecs to few tens of pc \citep[e.g.,][]{Jaffe04, Packham05b, Radomski08, Garcia_Burillo16, Imanishi18}, and sometimes with more than a single component \citep[e.g.,][]{Lopez_Gonzaga14,Tristram14, Garcia_Burillo19}. Unfortunately, these studies are limited to a small number of nearby AGNs.
 
The obscuring material in the dusty torus absorbs a significant part of the optical-UV radiation emitted by the accretion disk and re-emits it at infrared (IR) wavelengths between $\sim1-1000\,\mu$m, \citep[e.g.,][]{Krolik_Begelman88,Sanders89,Antonucci93,Elvis94, Peterson97, Neugebauer79}, peaking around $\sim5-35\,\mu$m \citep[e.g.,][]{Stein74, Rees69, Burbidge_Stein70, Rieke_Low72, Sanders89}.
Therefore, fitting the IR  spectral energy distribution (SED) of AGNs is a good technique to derive the physical properties of this dusty structure \citep[e.g.,][]{Gonzalez-Martin19b}. During the last decades, there have been great efforts to understand the assemble 
of the nuclear dust responsible for the observed IR SED of AGNs. For example, \citet{Fritz06} proposed a homogeneous distribution of the nuclear dust into a toroidal structure around the central engine (Smooth F06 model hereafter). This model reproduces well the observed IR SEDs of type 1 and 2 Seyfert galaxies and  predicts a deep $10\,\mu$m silicate absorption feature, but  predicts a narrower IR bump \citep[e.g.,][]{Dullemond_vanBemme05}. \citet{Nenkova02, Nenkova08a, Nenkova08b} proposed a dusty torus model in which the dust is distributed in clumps in a toroidal geometry around the central engine (Clumpy N08 model hereafter). 
This model predicts a more attenuated $10\,\mu$m silicate emission feature and broader IR bump. However, this model fails in reproducing the apparent ``excess" of near-IR (NIR) emission presented in some type 1 AGNs \citep[e.g][]{Mor09}, which has been attributed to the presence of hot dust in the inner part of the torus \citep[e.g.,][]{Braatz93,Cameron93,Bock00, Hoenig13}. 
Trying to resolve these issues \citet{Stalev16} proposed a two-phase medium torus model, composed of high-density clumps immersed into a low-density medium (Two-phase media S16 model hereafter). According to this model, the low-density dust medium significantly contributes to the NIR emission in type 1 AGNs. \citet{Feltre12} compared the smooth and clumpy models and found that both models can produce similar MIR continuum shapes for different parameters of the models. Additionally, they found that  there are very different NIR spectral slopes between the models, probably due to the different dust composition and central radiation assumed.

Due to the difficulties to consistently reproducing the IR SEDs of AGNs with the torus models, and the growing evidence for a component of  dust located in the polar region of the AGNs in addition to the equatorial component  \citep[e.g.,][]{Hoenig12, Hoenig13, Tristram14, Lopez_Gonzaga16, Leftley18}, \cite{Hoenig17} 
proposed a different geometry composed by a compact and geometrically thin clumpy dusty disk in the equatorial region plus an extended and elongated polar structure of dust, which is assumed to be co-spatial with the outermost layers of the outflowing gas (i.e. distributed in a hollow-cone geometry). Hereafter we refer to this model as the Disk+Wind H17 model. They claim that the more compact distribution of dust in the thin disk allows this model to better explain the $3.5-5\,\mu$m bump seen in several type 1 AGNs. 

Several groups have used both low angular resolution ($\sim3-5\,$arcsec) {\it Spitzer}/IRS spectra ($\sim5-35\,\mu$m) and high angular resolution ($\sim0.3\,$arcsec) data 
(spectra and photometry) in the infrared range to model the putative dusty torus in AGNs. For example, some authors \citep[e.g.,][]{Ramos_Almeida09, Ramos_Almeida11, Alonso_Herrero11, Ichikawa15, Fuller16, Garcia_Bernete19} used high angular resolution data to constrain the parameters of the Clumpy N08 model in a sample of Seyfert galaxies. Similar studies in QSOs are limited to low angular resolution data \citep[e.g.,][]{Mor09,Nikutta09,Mateos16}, partly due to the lack of high angular resolution data at IR wavelength. For example, using the 
{\it Spitzer}/IRS spectrum of 26 QSOs and the Clumpy N08 model \cite{Mor09} found that to reproduce the data it is necessary to add a hot dust component ($T\sim1400$ K). Later, using
the same model and the {\it Spitzer}/IRS spectrum of 25 QSOs at $z\sim2$, \cite{Deo11} showed that by adding a hot dust component to
the spectral fitting, it is possible to simultaneously fit the NIR SED and $10\,\mu$m silicate feature. \citet{Alonso_Herrero11} suggested that this hot dust component might not be necessary when using
high angular resolution IR data. Nevertheless, they found that some type 1 Seyfert still show an excess of emission at NIR despite the high angular resolution data, which they attributed to a hot dust
component \citep[see also][]{Ichikawa15, Garcia_Bernete19}. 
To model the low angular resolution {\it Spitzer} and {\it AKARI} ($2.5-5\,\mu$m) spectra of 85 luminous QSOs ($L_{bol}>10^{45.5}$erg s$^{-1}$, with a redshift $z$ between 0.17 and 6.42) \cite{Hernan_Caballero16} assumed a single accretion disk template plus two black-bodies for the dust emission. They successfully reproduced the $0.1-10\,\mu$m SED. However, they noted that the best-fitting model systematically underpredicts the observed flux around $1\,\mu$m. After comparing the NIR excess and the NIR to optical luminosity ratio, they concluded that this excess of emission originates in hot dust near the sublimation radius. 

Assuming the Clumpy N08 dusty torus model, \citet{Martinez_Paredes17} used the starburst-subtracted {\it Spitzer}/IRS spectrum  ($\sim8-15\,\mu$m) plus the point spread function (PSF) photometric point at H-band ($1.6\,\mu$m) obtained from NICMOS data on the {\it Hubble Space Telescope}  (available for 9 out of 20 objects), to investigate the physical and geometrical properties of the dusty torus in a sample of 20 nearby ($z<0.1$) QSOs.
They found that by including the spectral range between ($\sim5-8\,\mu$m), the SEDs modelling results in a bad
fit of the $10\,\mu$m silicate feature, producing parameters inconsistent with their optical classification as type 1 AGNs. Therefore, they excluded this part of the spectrum from the fitting. 
A similar result and approach was reported by \citet{Nikutta09} 
for a couple of QSOs. From the good fits \citet{Martinez_Paredes17} found that 3/9 objects show an 
excess of emission at H band, probably because the unresolved emission could be still contaminated by emission
from the host galaxy \citep{Veilleux06, Veilleux09b}, and/or because in
these cases 
the addition of a hot dust component could be necessary. 
Additionally, \citet{Martinez_Paredes17} found that for 50 percent of
QSOs the viewing angles range from $\sim50-88$ degrees. They claimed that the addition of a second photometric point in the NIR would improve the determination of the viewing angle as suggested by \cite{Ramos_Almeida14}. They found that to well constrain the six free parameters of the Clumpy N08 model for 
Seyfert galaxies, it is necessary to build an unresolved IR SED 
composed of at least two photometric points in the NIR, a photometric point in the MIR, plus the high angular resolution spectrum in the N-band ($\sim7.5-13.5\mu$m).

On the other hand, recently \cite{Martinez_Paredes20} investigated which dusty models better reproduce the AGN-dominated {\it Spitzer}/IRS spectrum ($\sim5-35\,\mu$m) of local ($z<0.1$) type 1 AGNs (including some QSOs). They explored the Smooth F06 model, Clumpy N08 model, Clumpy torus model of \cite{Hoenig10a} (hereafter Clumpy H10 model), Two-phase media S16 model, and the Disk+Wind H17 model. They found that in most cases and with all models, it is always necessary to add the stellar component to fit the bluer spectral range ($\sim5-8\,\mu$m), but that none of the models can well reproduce the spectral shape between $\sim5-30\,\mu$m. However, they noted that the Disk+Wind H17 model  reproduces well the spectrum with flatter residuals, especially at high bolometric luminosities. 

Therefore, considering the difficulties in simultaneously modelling the 
NIR-MIR SEDs of QSOs,
we obtain new NIR high angular resolution data ($\sim0.3$ arcsec) from the NIR cameras on the 10.4m 
Gran Telescopio CANARIAS (GTC)
to build for the first time a set of well-sampled NIR to MIR SEDs of QSOs 
that allows us to investigate which model, among the Clumpy N08 and H10, Smooth F06, Two-phase media S16, and Disk+Wind H17 
better reproduces the data and constrains the parameters.

Throughout this paper, we assume $\Lambda$CDM cosmology with $H_0=70$ km~s$^{-1}$Mpc$^{-1}$, $\Omega_m=0.3$, and $\Omega_{\Lambda}=0.7$. 

\begin{table*}
\begin{minipage}{1.\textwidth}
\caption{{\bf Basic properties of the sample.} Column 1 lists the name of the object, columns 2-3, 4 and 5 list the coordinates, redshift, and angular scale taken from NED. Column 6 list the corrected hard (2-10 keV) x-ray luminosity.} \label{tab:basic_info}
\centering
%\resizebox{18cm}{!}{
\begin{tabular}{l|ccccc}
				\hline
Name  &  \multicolumn{2}{c}{Coord. J2000} & $z$  & Angular scale& log L$_{X}^{a}$\\
     & AR &DEC & & kpc/$''$& (erg s$^{-1}$)\\
%     &  & & &  & \\     
\hline
Mrk~509     & 20:44:09.7&$-10:43:25$ & 0.034& 0.677& 44.68\\
PG~0050+124 & 00:53:34.9 & $+12:41:36$ & 0.034& 0.677& 43.85\\
PG~2130+099 & 21:32:27.8& $+10:08:19$& 0.0630& 1.213& 43.51\\
PG~1229+204 &12:32:03.6 & $+20:09:29$& 0.063&1.213 & 43.49\\
PG~0844+349 &08:47:42.4 & $+34:45:04$ & 0.064& 1.231& 43.74\\
PG~0003+199 & 00:06:19.5 & $+20:12:10$ & 0.026& 0.523& 43.28\\
PG~0804+761 & 08:10:58.6 & $+76:02:43$ & 0.100& 1.844& 44.46\\
PG~1440+356 & 14:42:07.4& $+35:26:23$& 0.08& 1.510& 43.76 \\
PG~1426+015 &14:29:06.6 & $+01:17:06$ &0.09 &1.679 &44.11 \\
PG~1411+442 & 14:13:48.3& $+44:00:14$& 0.09& 1.679& 43.40\\
PG~1211+143 & 12:14:17.7& $+14:03:13$& 0.081&1.527 & 43.70 \\
PG~1501+106 & 15:04:01.2 & $+10:26:16$ &0.04  & 0.791& 43.89\\
MR~2251-178 & 22:54:05.9& $-17:34:55$ & 0.064& 1.231& 44.46\\
\hline\\
 & \multicolumn{5}{c}{Additional objects observed in the NIR}\\
\hline
PG~2214+139 & 22:17:12.2 &$+14:14:21$ & 0.066& 1.266&43.82 \\
PG~0923+129 & 09:26:03.3& $+12:44:04$& 0.029& 0.581& 43.41\\
PG~0007+106 & 00:10:31.0& $+10:58:30$& 0.089& 1.662& 44.15\\
\hline
\end{tabular}
%}
\\
Note.- $^{a}$The X-ray luminosity is from \cite{Zhou2010}, except for PG~0007+1006 and PG~0923+129, for which the value is taken from \cite{Piconcelli05} and \cite{Shu10}, respectively.
\end{minipage}
\end{table*}

\section{The sample} \label{sec:sample}
We selected the 13/20 local type 1 QSOs in the sample of
\cite{Martinez_Paredes17} that have high angular resolution spectroscopy in the N band ($\sim7.5-13.5\,\mu$m). The 20 QSOs were selected  and observed with the MIR camera CanariCam (CC) on GTC. They set the following criteria: 1) a redshift $z < 0.1$; 2) a nuclear flux at N-band (aperture $\lesssim5$ arcsec) $f_N > 0.02$ Jy; and 3) $\rm L_X\,(2-10\,keV)$ $>$ $10^{43}$ erg sec$^{-1}$. The redshift combined with the high angular resolution offered by the combo CC/GTC ($\sim0.3''$) allow them to observe the inner nuclear region within 1 kpc, and the MIR flux allow them to detect these objects with CC in the N and Si2 ($8.7\,\mu$m) bands. The hard X-ray flux was used as an intrinsic indicator of the AGN activity. They imaged in the Si2-band 19/20 QSOs, and obtained the spectrum in the N band for 11/13. The data were published by \citet{Martinez_Paredes17} and \citet{Martinez_Paredes19}. The remaining two objects
had high angular resolution spectra at N band from VISIR on the Very Large Telescope \citep[VLT,][see Table~\ref{tab:basic_info}]{Hoenig10a, Burtscher13}. Note that we include in the sample three additional objects that we observe in the NIR wavelengths, and report in this paper. 

%We use the cameras CIRCE and EMIR on the 10.4m GTC to image at NIR 10 of the 13 QSOs that have high angular resolution spectroscopy, five at H band ($1.6\,\mu$m) and 12 at 
%K band ($2.1\,\mu$m). The other three have ancillary data from the literature, See Table~\ref{tab:ancillary_data}. Our final sample is composed by the 13 QSOs that have both 
%MIR high angular resolution spectrum at N band, at least a photometric point at MIR, and two photometric points at NIR, with the exception of the PG~1501+106 for which there is not NIR information. Note that we also image at NIR three objects that do not have high angular resolution 
%spectrum at MIR. 
\section{The data}
\subsection{New high angular resolution NIR data and reduction}
Using the Canarias InfraRed Camera Experiment \citep[CIRCE,][]{Garner16, Eikenberry18} on the GTC between October 2016 and February 2017 through the open Spain-Mexico collaborative time, we imaged four objects in the H-band ($\lambda_{c}=1.63\,\mu$m) and seven objects in Ks-band ($\lambda_{c}=2.16\,\mu$m), respectively (PI: M. Mart\'{i}nez-Paredes, ID: GTC2-16BIACMEX). CIRCE is a NIR camera that operates in the spectral range between 1 and 2.5 $\mu$m, which has a total field of view of $3.4'\times3.4'$ with plate scale of $0.1''$/pixel.
In order to complete our observations, we used the NIR Espectr\'ografo Multiobjeto Infra-Rojo \citep[EMIR,][]{Balcells00} on the GTC between September 2018 and May 2019 through the open Mexico time (PI: M. Mart\'{i}nez-Paredes, ID: GTC2-18BMEX). EMIR has a field of view of $6.67'\times6.67'$ and a plate scale of $0.2''$/pixel. We observed five objects in the Ks-band and one in the H-band. The observations were obtained under good weather conditions. The full width at half maximum (FWHM) of the standard star ranges from $0.50-1.0$ arcsec for the CIRCE observations and from $0.68-0.84$ arcsec for the EMIR observations, while the airmass ranges were $\sim1-1.6$ and $\sim1-1.5$, respectively. In Table~\ref{tab:obs_inf} we list 
the basic information of the observations.

For all the targets, we imaged a photometric standard star immediately before or after the science observation. The standard star was chosen to be as close as possible in celestial coordinates to the science object. We used the standard star for the flux calibration and to determine the PSF.

We reduced the CIRCE images using a custom pipeline written in Interactive Data Language (IDL). All images got dark-subtracted to account for electronic offsets. For sky-subtraction, a median combination of the sky exposures was used, which was then scaled to the background level of the object exposures. In order to align the images, the offsets were then determined by measuring the centroid of the object. The aligned sky-subtracted images were then mean-combined. Bad pixels were treated by discarding them in the combination.
For the EMIR images, we requested the reduced data to the observatory. In general, the reduction includes the following procedures; sky subtraction, cosmic ray cleaning, stacking of the 
individual images and rejection of bad images \citep[see e.g.,][]{Ramos_Almeida19}.

\begin{table*}
	\begin{minipage}{1.\textwidth}
		\caption{{\bf NIR observations.} Column 1 lists the name of the object. Columns 2, 3, 4, 5, and 6 list the filter used, the date of the observation, the on-source time, the name of the spectro-photometric standard star, and the time offset between the standard and the QSO. Columns 7, 8, and 9 list the full width at half maximum (FWHM) of the standard star, the average airmass during the observation, and the telescope/instrument used. \label{tab:obs_inf}}
\centering
\resizebox{18cm}{!}{
\begin{tabular}{l|cccccccc}
				\hline
Name &Filter & Date & $t_{on-source}$ & STD &  $t_{offset}$ &  FWHM$_{STD}$ & Airmass  & Telescope/Instrument \\
   &       & AAAA.MM.DD &	(sec) &   & (min)  & (arcsec)&  &    \\     
\hline
Mrk~509     & H   &  2016.10.31   &  45  &  S813-D  & 25  &  0.54  & 1.40   &  GTC/CIRCE  \\
Mrk~509     & Ks   & 2016.10.31     &  65.47  & S813-D   &  14 &  0.56  &  1.36  & GTC/CIRCE   \\
PG~0050+124   & Ks   &  2017.01.09   &  13.5  & FS-102   & 10  & 0.50    & 1.15    & GTC/CIRCE    \\
PG~2130+099   & Ks   &  2016.10.31   &  45  & P576-F   & 4  & 0.52      & 1.08    & GTC/CIRCE    \\
PG~1229+204 & Ks & 2019.05.17 & 2 & FS-131 & 13 & 0.84     & 1.02 & GTC/EMIR\\
PG~0844+349   & Ks   & 2017.02.06   &  45  & P0345-R    & 19  & 0.68   & 1.56   & GTC/CIRCE    \\
PG~0003+199   & H   &  2016.10.23   &  15  &  FS-102  & 21  &  0.65  & 1.11    &  GTC/CIRCE  \\
PG~0003+199   & Ks   &  2016.10.23   &  9  &  FS-102  & 29  & 0.64  &  1.23  &  GTC/CIRCE  \\
PG~0804+761   & H   &  2017.02.06   &  90  & P0345-R   & 31  & 0.84    &  1.56  & GTC/CIRCE   \\
PG~0804+761&Ks & 2019.02.12& 2 & P091-D &23&0.74                       &1.48& GTC/EMIR\\
PG~1211+143 & Ks & 2019.0517 & 2 & S860-D & 5 & 0.74                   &1.04 & GTC/EMIR\\
MR~2251-178&H &2018.09.27 & 2 & S667-D & 4 & 0.68                  &1.44 & GTC/EMIR\\
MR~2251-178 & Ks & 2018.09.27 &1.5 & FS-32 & 3 & 0.68                 &1.5& GTC/EMIR\\
\hline
PG~2214+139   & Ks   &  2016.10.31   &  45  & FS-31    & 12   & 0.62   & 1.04    & GTC/CIRCE   \\
PG~0923+129   & H   & 2017.02.06    & 45   & P0345-R    & 44   & 0.84   & 1.04    & GTC/CIRCE    \\
PG~0923+129 &Ks & 2017.02.09 &5&FS-17&7&1.0&1.05&GTC/CIRCE\\
%PG~0923+129 & Ks & 2019.02.12 & 2 & FS-126 & 5 & 0.30                &1.22 & GTC/EMIR\\
PG~0007+106$^{*}$   & Ks   &  2016.10.23   &  15  & FS-102   & 8   & 0.64    & 1.19    & GTC/CIRCE    \\
%PG~0923+129 & Ks &2017.02.09 & 5 & FS-17 & 7& 0.54 &1.05 & GTC/CIRCE \\
				\hline
			\end{tabular}}\\
			Notes.-$^{*}$Radio loud quasar.
		\end{minipage}
	\end{table*}

\subsection{GTC/CanariCam imaging in the Si2 filter (8.7 micron) and N-band spectroscopy. }

We use the MIR high angular resolution unresolved photometry in the Si2-band ($\lambda_{c}=8.7\,\mu$m) measured within an aperture radius of $1''$.  We also use the high angular resolution spectra in the N-band ($\sim7.5-13.5\,\mu$m, slit width $=0.52''$) from \cite{Martinez_Paredes17}. These data were obtained with GTC/CC for 19/20 and 11/20 QSOs, respectively. These data were obtained as part of the CC guaranteed time (P.I: C. Packham), ESO-GTC time (P.I: A. Alonso-Herrero), and open Mexico time (P.I: I. Aretxaga and M. Mart\'inez-Paredes).

\begin{table*}
\begin{minipage}{1.\textwidth}
\caption{{\bf Observational IR information.} Column 1 lists the name of the object, columns 2 and 3 list the source of the ground-based high angular resolution spectroscopy in the N band and the {\it Spitzer}/IRS spectrum, respectively. Columns 4 and 5 list the source of the spatial ({\it HST}/NICMOS) and ground-based high angular resolution data at H and Ks bands. Columns 6 and 7 list the source of the ground-based high angular resolution MIR photometry.} \label{tab:ancillary_data}
%\centering
\resizebox{18cm}{!}{
\begin{tabular}{l|cccccc}
				\hline
Name  &  \multicolumn{2}{c}{MIR Spectrum} & \multicolumn{4}{c}{Photometry}\\
     &  ground-based & {\it Spitzer}& \multicolumn{2}{c}{NIR}& \multicolumn{2}{c}{MIR}\\
     &  ($7.5-13.5\,\mu$m)  & ($5-35\,\mu$m) & H (1.63$\mu$m) &Ks(2.16$\mu$m) &Si-2 (8.7$\mu$m) & Others\\     
\hline
Mrk~509     & VLT/VISIR & Yes & GTC/{\bf CIRCE}& GTC/{\bf CIRCE}& VLT/VISIR & VLT/VISIR: Si2, Si5, NEII, PAH2, SIV \\
PG~0050+124 & VLT/VISIR & Yes &{\it HST}/NICMOS &GTC/{\bf CIRCE} & GTC/CC&VLT/VISIR: PAH2$\_2$; Subaru/COMICS: N11p7\\
PG~2130+099 & GTC/CC & Yes &{\it HST}/NICMOS &GTC/{\bf CIRCE} &GTC/CC & VLT/VISIR: NEII, PAH2, SIV; Subaru/COMICS: N11p7\\
PG~1229+204 & GTC/CC & Yes & {\it HST}/NICMOS & GTC/{\bf EMIR}&GTC/CC & ...\\
PG~0844+349 & GTC/CC & Yes &{\it HST}/NICMOS &GTC/{\bf CIRCE} & GTC/CC& Subaru/COMICS: N11p7 \\
PG~0003+199     & GTC/CC & Yes$^{a}$ &GTC/{\bf CIRCE} &GTC/{\bf CIRCE} & GTC/CC& ... \\
PG~0804+761     & GTC/CC & Yes &GTC/{\bf CIRCE} & GTC/{\bf EMIR}& GTC/CC& ... \\
PG~1440+356 & GTC/CC & Yes &{\it HST}/NICMOS &Gemini/QUIRC$^{*}$ & GTC/CC& ... \\
PG~1426+015 & GTC/CC & Yes &{\it HST}/NICMOS & Gemini/QUIRC$^{*}$&GTC/CC & ... \\
PG~1411+442 & GTC/CC & Yes & {\it HST}/NICMOS & Gemini/QUIRC$^{*}$&GTC/CC & ... \\
PG~1211+143 & GTC/CC & Yes & Gemini/QUIRC$^{*}$ &GTC/{\bf EMIR} &GTC/CC & ... \\
PG~1501+106     & GTC/CC & Yes &... & ...& GTC/CC& ... \\
MR~2251-178     & GTC/CC & Yes$^{b}$ &GTC/{\bf EMIR} &GTC/{\bf EMIR} &GTC/CC & VLT/VISIR: NEII$\_$2, PAH2$\_$2, SIV \\
\hline\\
                &\multicolumn{4}{c}{Additional objects observed in the NIR} \\
                \hline
PG~2214+139 & No & Yes &{\it HST}/NICMOS & GTC/{\bf CIRCE}& Bad quality& ... \\
PG~0923+129     & No & Yes & GTC/{\bf CIRCE}& GTC/{\bf CIRCE}&GTC/CC & ...\\
PG~0007+106     & No & Yes &{\it HST}/NICMOS &GTC/{\bf CIRCE} &GTC/CC & ...\\
				\hline
			\end{tabular}}\\
		Note.- The new data are highlighted in boldface. $^{*}$Corresponding fluxes are considered as upper limits. 
		$^{a}$For this object the {\it Spitzer}/IRS spectrum ranges from $\sim5-14\,\mu$m. $^{b}$For this object we use the high spectral resolution 
		{\it Spitzer}/IRS spectrum from $\sim10-35\,\mu$m.
		
	\end{minipage}
	\end{table*}

\subsection{Ancillary MIR and NIR high angular resolution data}\label{ancillary}

Five objects have high angular resolution photometry in the MIR obtained with the camera VISIR on the Very Large Telescope (VLT) \cite{Hoenig10a} and/or COMICS on Subaru \citep[see][and references therein]{Asmus16}. We also used the high angular resolution spectrum in the N-band obtained with VLT/VISIR by \citet{Hoenig10a} and \citet{Burtscher13} for 2 out of the 13 objects in our sample, see Table~\ref{tab:ancillary_data}. 

In a previous work, \citet[][]{Martinez_Paredes17} found that only nine objects have
unresolved PSF photometry available 
in the literature from the {\it HST/NICMOS} \citep[][]{Veilleux06, Veilleux09b} in the H band (F160W, $\lambda=1.6\,\mu$m). From the ground-based observations obtained by \cite{Guyon06} between November 
2000 and April 2002
with the Quick Infrared Camera \citep[QUIRC,][]{Hodapp96} on the Gemini telescope, we found that one 
object has unresolved PSF photometry in the H band, and three  
in the Ks-band (see Table~\ref{tab:ancillary_data}).

%\OGM{Como se comparan los datos estos con los del paper de 2017? Son los mismos? Deberías referirte al paper si los usaste alli. }

\subsection{{\it Spitzer}/IRS spectra}
Low resolution (R $\sim$ 60-127) {\it Spitzer}/IRS spectra were retrieved from the Cornell Atlas database of CASSIS v6 \citep[][]{Lebouteiller11} for 13 objects, using optimal extraction, i.e, a point source extraction. 
%\OGM{A veces tienes 13 objetos y otras 20.} 
CASSIS provides the low resolution spectra observed in different modes such as SL1$\sim7.4-14.5\,\mu$m, SL2$\sim5.2-7.7\,\mu$m, LL1$\sim19.9-39.9\,\mu m$ and LL2$\sim13.9-21.3\,\mu$m with different slit widths ($3.6-11.1$ arcsec), see Table~\ref{tab:ancillary_data}. We stitched and scaled the different spectra by setting SL2 as a reference flux to calibrate the effect of different slit widths and the background contamination due to the different slit orientation \citep[see e.g.,][]{Martinez_Paredes17}. Note that because of the lack of the low resolution IRS spectra between $\sim5-35\,\mu$m, for one object we used the IRS spectrum from $\sim5-14\,\mu$m, and for another the high spectral resolution spectrum from $\sim10-35\,\mu$m. 

\begin{figure*}
\includegraphics[width=1\columnwidth]{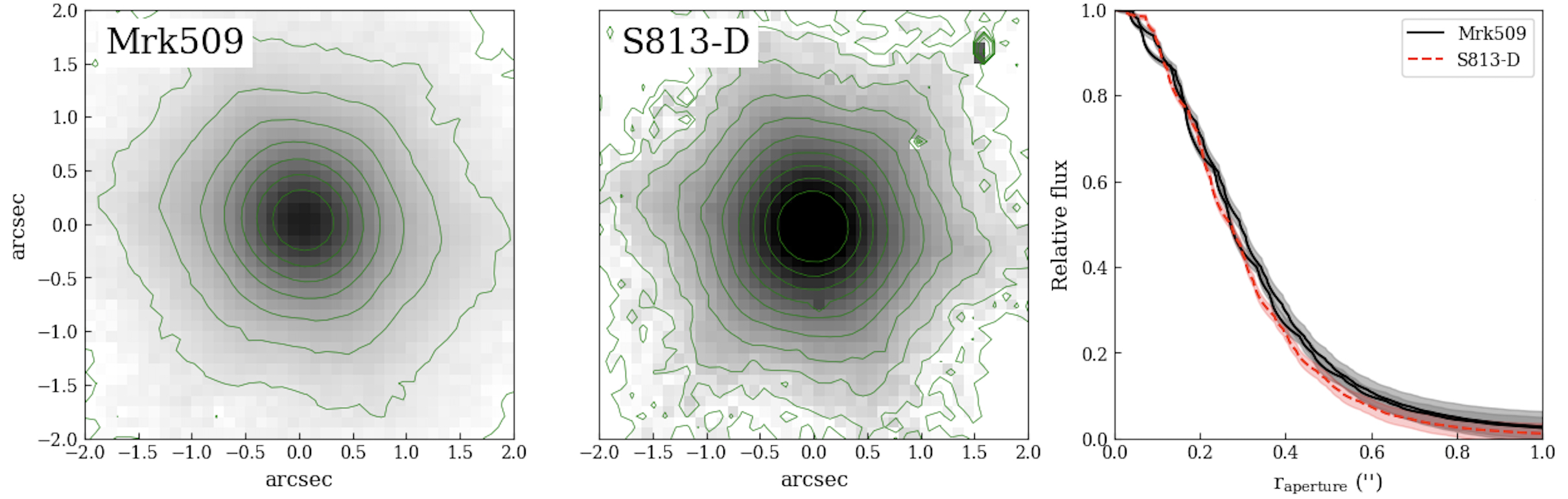}\\
\includegraphics[width=1\columnwidth]{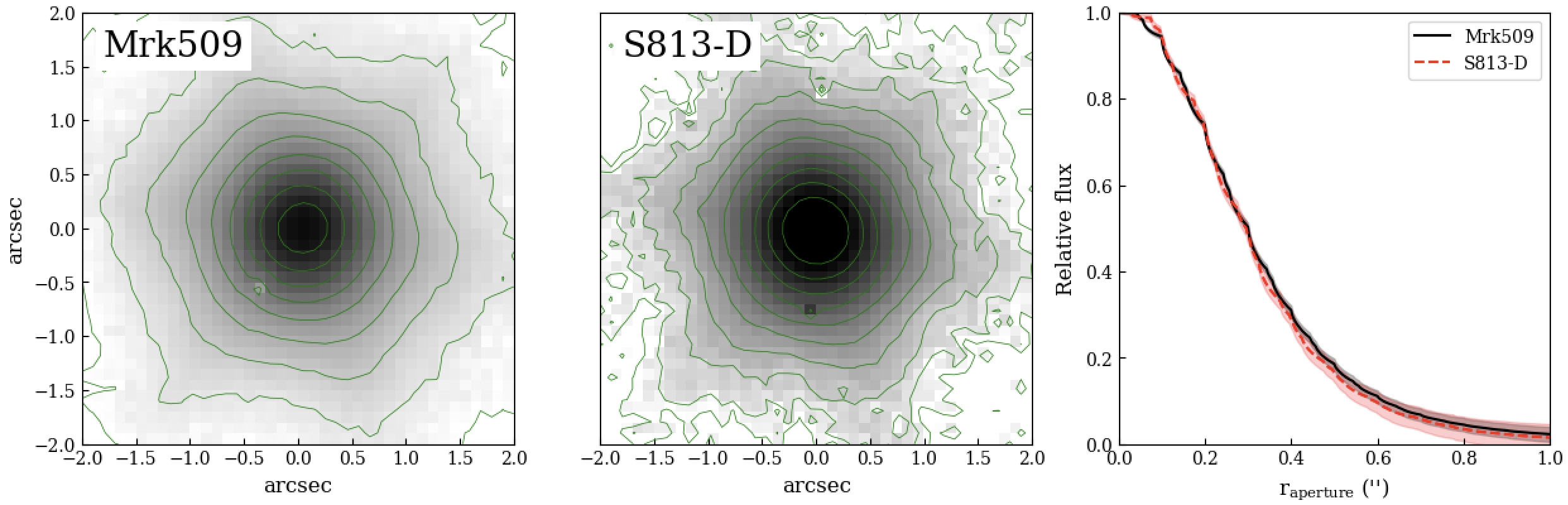}
\caption{{\bf Images and radial profiles of Mrk~509 and S813-Dat H (Upper panel) and Ks (Bottom-panel) bands}. From left to are the 
science image, the standard star, and their radial profiles. The radial profile of the QSO and its uncertainty are plotted as a black solid-line and the grey shaded region. The radial profile of the standard star and its uncertainty are plotted as a red dashed-line and pink shaded region. In all images the lowest contour is 3$\sigma$ over the background. The next contours are traced in 2$\sigma$ steps.}
\label{radial_profile_Mrk509}
\end{figure*}

\begin{figure*}
    \centering
    \includegraphics[width=1\columnwidth]{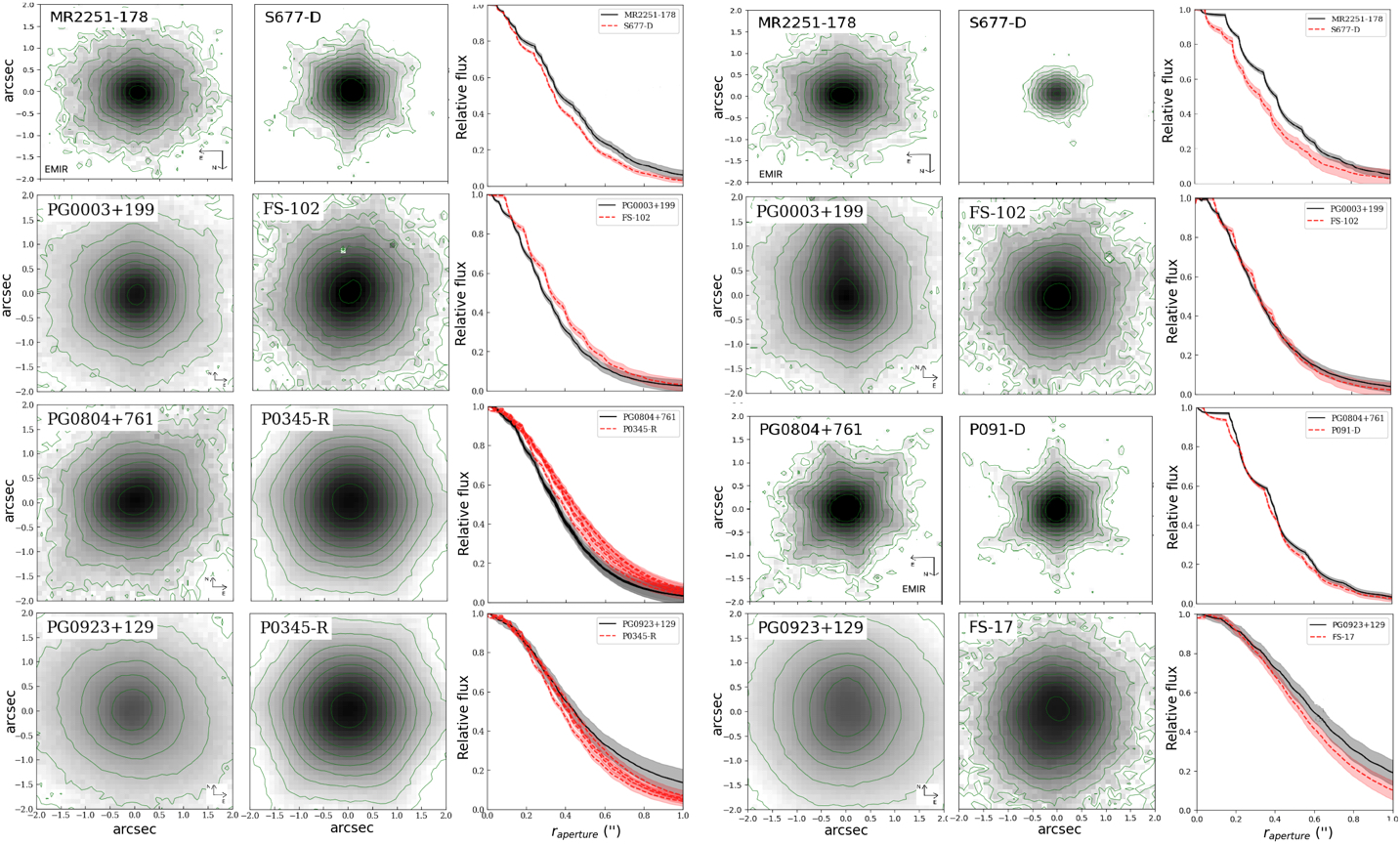}\\
    \caption{{\bf Images and radial profiles of QSOs and the standard at H (left panel) and Ks (right panel) bands}. From left to right there are the 
science image, the standard star, and their radial profiles. The radial profile of the QSO and its uncertainty are plotted as a black solid-line and the grey shaded region. The radial profile of the standard star and its uncertainty are plotted as a red dashed-line and pink shaded region. In all images the lowest contour is $3\sigma$ over the background. The next contours are traced in $2\sigma$ steps.}
    \label{H_Ks_imgages}
\end{figure*}

\begin{figure*}
    \centering
    \includegraphics[width=1\columnwidth]{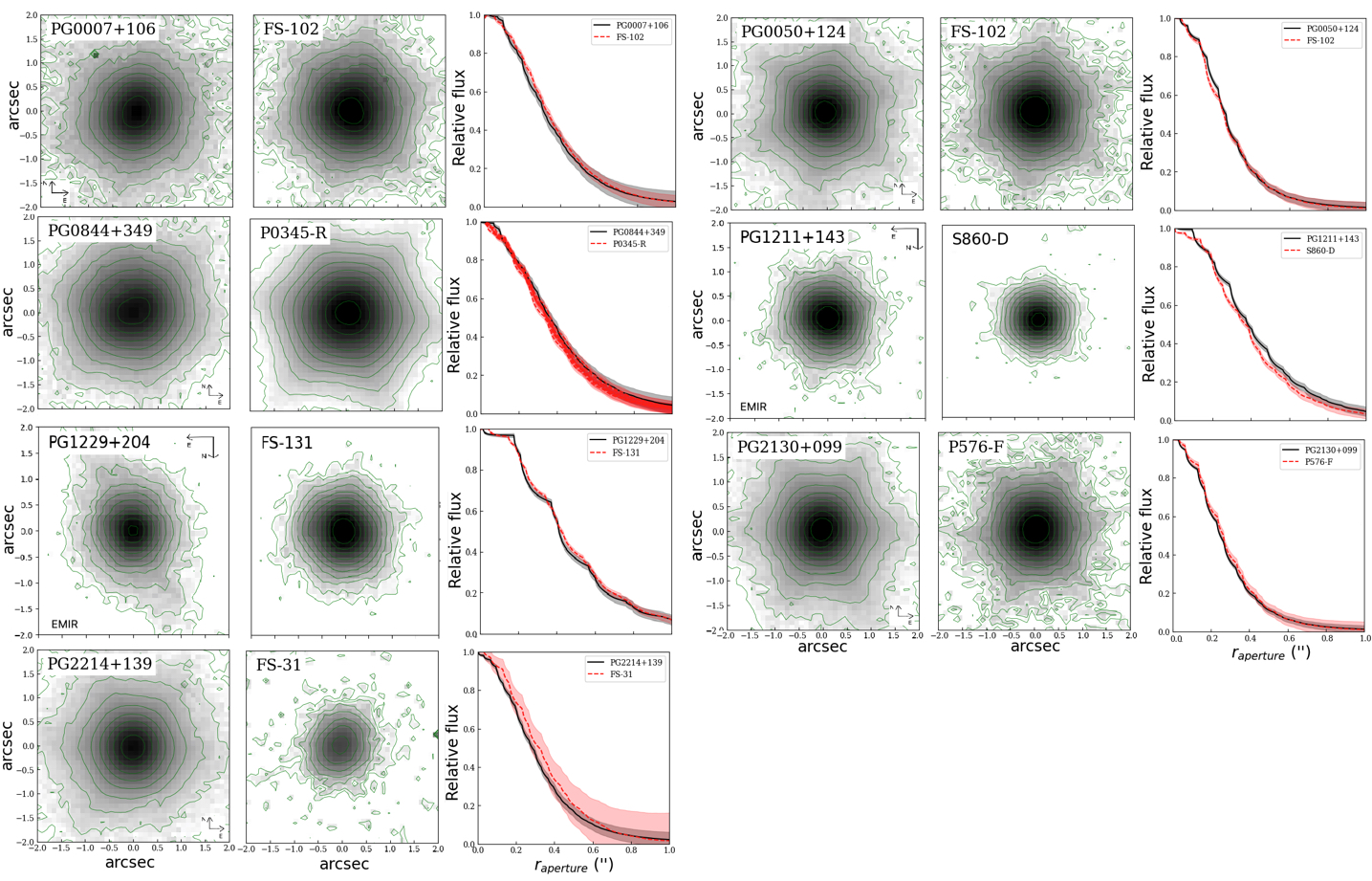}\\
    \caption{{\bf Images and radial profiles of QSOs and the standard in the Ks band}. From left to right there are the 
science image, the standard, and their radial profiles. The radial profile of the QSO and its uncertainty are plotted as a black solid-line and the grey shaded region, respectively. The radial profile of the standard and its uncertainty are plotted as a red dashed-line and pink shaded region. In all images the lowest contour is 3$\sigma$ over the background. The next contours are traced in 2$\sigma$ steps.}
    \label{Ks_images}
\end{figure*}

\begin{table*}
	\begin{minipage}{1.\textwidth}
		\caption{{\bf New NIR unresolved photometry measured within an aperture radius of 1 arcsec}. Column 1 lists the name of the objects. Columns 2 and 3 list the flux and the source to standard FWHM ratio in the H band. Columns 4 and 5 list the flux and the source to standard FWHM ratio in the Ks band. In parenthesis we indicate the flux previously measured by \cite{Guyon06} using QUIRC/Gemini. In the case of Mrk~509 the flux in parenthesis was reported by \citet{Fischer06} using ISAAC/VLT.\label{tab:NIR_flux}}
\centering
\begin{tabular}{l|cccc}
				\hline
Name & $f_{\nu}$ (mJy) & & $f_{\nu}$ (mJy) &  \\
  &  H ($1.63\,\mu$m) & $\frac{FWHM_{target}}{FWHM_{STD}}$& Ks ($2.16\,\mu$m) & $\frac{FWHM_{target}}{FWHM_{STD}}$\\  
%    &  & & & \\     
\hline
\hline
Mrk~509   &  $15.6\pm2.2$ ($<14.1$)&1.0 & $21.1\pm3.0$ ($<22.3$)& 1.0 \\
PG~0050+124 & ... &... & $26.6\pm3.8$ ($<39.6$)  &1.1 \\
PG~2130+099 & ... &... & $13.0\pm1.9$ ($<25.5$) & 0.9\\
PG~1229+204 & ...   &... &$4.3\pm0.6$ ($<7.0$)& 1.0\\
PG~0844+349 & ... &... & $10.3\pm1.5$ & 1.0\\
PG~0003+199 & $13.5\pm1.9$ &0.9 & $20.6\pm2.9$ &1.0 \\
PG~0804+761 & $6.8\pm1.0$ ($<13.2$) &0.8 & $9.9\pm1.4$ ($<26.2$)& 1.1\\
PG~1211+143 & ...   &... &$8.0\pm1.1$  & 1.0\\
MR~2251-178 & $7.0\pm1.0$ &1.1 &$7.9\pm1.1$  &1.2 \\
\hline
PG~2214+139$^{*}$ & ... &... & $9.6\pm1.4$  &1.04 \\
PG~0923+129$^{*}$ & $6.4\pm0.9$ &1.0 & $6.8\pm1.0$ &1.1 \\
%                  &             &  & $5.4\pm0.8^{b}$ &1.3$^{c}$ \\
PG~0007+106$^{*}$ & ...  &... & $4.3\pm0.6$ ($<19.3$)& 0.9\\
				\hline
			\end{tabular}\\
Note.-$^{*}$The object is not modeled because we do not have MIR high angular resolution spectrum. 
%$^{b}$Flux measured from EMIR image. 
%The PSF flux is $3.2\pm0.5$ mJy.
	\end{minipage}
	\end{table*}

\section{Analysis} \label{sec:analysis}

\subsection{NIR} \label{sec:nir_phot}
%{\it Agregar un parrafo describiendo un poco la emision NIR de los QSOs..ver Veilleux et al. y Guyon et al.}

We performed aperture photometry on the QSOs and its corresponding standard star to look for possible extended emissions and measure the unresolved flux in the H and Ks bands. We use the photometry task tools {\sc photutils} \citep{Bradley19} from the analysis package {\sc Astropy}. 
We measure the relative flux through the radius aperture of the standard star and QSOs and build their radial profiles. Figure\,\ref{radial_profile_Mrk509} shows, as an example,
the images in the H- and Ks-band of Mrk\,509 and its standard star, and their corresponding radial profiles in black and red dashed lines, respectively. Additionally, 
in Figures~\ref{H_Ks_imgages} and Figure~\ref{Ks_images} we show the images and radial profiles of the remaining objects observed in both H and Ks bands and of the objects observed only in the Ks-band, respectively. The 
uncertainties in the radial profiles are estimated according to the 
following expression, $\sqrt{\sigma_{back}^2N_{pix}+\sigma_{back}^2N_{pix}^2/N_{pix-ring}}$ where $\sigma_{back}$ is the standard deviation of the background level, $N_{pix}$ is the number of pixels within an aperture, and $N_{pix-ring}$ is the number of pixels within a 80-pixel width annulus \citep[see e.g.,][]{Reach05}. 
Additionally, we add in quadrature the uncertainties due to the  PSF-variability ($\sim14\%$), which is estimated from the standard stars observed within a single night for multiple times with time intervals of few minutes, and the flux calibration ($\sim6\%$).

Considering that in most cases the radial profile of the QSOs and standard stars are compatible within the uncertainties we assumed that the emission of these QSOs is unresolved and measured the flux within an aperture radius of one arcsecond ($<1$ kpc, see Table~\ref{tab:NIR_flux}). 
The exception is for MR~2251-178, for which we find that $\sim20\,\%$ of the emission is due to the extended component in both bands.
Note that the aperture corrected fluxes measured within an aperture radius of $\sim2\times FWHM$ gives similar results within the uncertainties. A similar analysis was done by \citet[][]{Martinez_Paredes17} to estimate the unresolved emission of these QSOs at Si2 band inside a radius aperture of $1''$, which correspond to a physical scales $<1$ kpc. 

In Table~\ref{tab:NIR_flux} we also list 
the flux previously reported in the literature \citep[e.g.,][]{Guyon06,Fischer06} from ground-based 
observations for some objects. 
In general, we find that our fluxes are lower, probably since their studies being focused on the morphology of the host galaxy more than in the measurement of the PSF, which is the goal in our analysis. In the case of Mrk~509, our measurements in the H and Ks bands are similar to those previously measured by \citet{Fischer06} within an aperture radius of 1.5 arcsec.

\subsection{Disentangling the accretion disk and torus}

To decontaminate the IR data from the possible contribution 
of the accretion disk, especially at NIR wavelengths, we explored two broken power laws as the SED of the accretion disk $\lambda F_{\lambda}\propto\lambda^{\alpha}$.
The first one is a classic accretion disk SED, which is in agreement with theoretical predictions and observational evidences \citep[e.g.,][]{Hubeny01, Davis11, Slone12, Capellupo15, Stalev16} ,

\begin{equation}
\lambda  F_{\lambda}\propto \left\{
\begin{array}{ll}
	\lambda^{1.2} &\,  \, \, 0.001\le\lambda\le0.01\,\mu\text{m} \\
    \lambda^{0} & \,  \, \, 0.01<\lambda\le0.1\,\mu\text{m} \\
    \lambda^{-0.5} & \,  \, \,  0.1<\lambda\le5\,\mu\text{m} \\
	\lambda^{-3} & \,  \, \,  5<\lambda\le50\,\mu\text{m} \\
\end{array} 
\right.
\label{equ1}
\end{equation}

The second one is a bluer broken power law derived from QSOs observations  \citep[e.g.,][]{Zheng97, Manske98, Vanden01, Scott04, Kishimoto08},

\begin{equation}
\lambda  F_{\lambda}\propto \left\{
\begin{array}{ll}
	\lambda^{1} &\,  \, \, \lambda<0.03\,\mu\text{m} \\
	\lambda^{0} & \,  \, \, 0.03<\lambda\le0.3\,\mu\text{m} \\
	\lambda^{-4/3} & \,  \, \,  0.3<\lambda\le3\,\mu\text{m} \\
	\lambda^{-3} & \,  \, \,  \lambda>3\,\mu\text{m} \\
\end{array} 
\right.
\label{equ2}
\end{equation}

We compile the dereddened UV nuclear photometry from the Galaxy Evolution Explorer space telescope (GALEX) at Far-ultraviolet (FUV, 0.153 $\mu$m) and Near-ultraviolet (NUV, 0.231 $\mu$m) bands \citep[][]{Bianchi11, Monroe16, Bianchi17}. Two objects do not have GALEX fluxes 
in the literature. For one of them (PG\,0844+349) we use the dereddened PSF photometry in the \emph{u} band (0.352 $\mu$m) from the SDSS DR12 SkyServer, and for the another one (PG\,0804+761) the
continuum emission at 110 and 130 $\mu$m reported by \cite{Stevans14} using the Cosmic Origins Spectrograph on the {\it HST}. We use these data to 
normalise the broken power-law components and subtract it
from both the NIR and MIR photometry, and from both the ground-based high angular resolution and {\it Spitzer}/IRS spectra.

\begin{figure*}
	\centering
	\begin{tabular}{cc}
		\includegraphics[width=0.5\columnwidth]{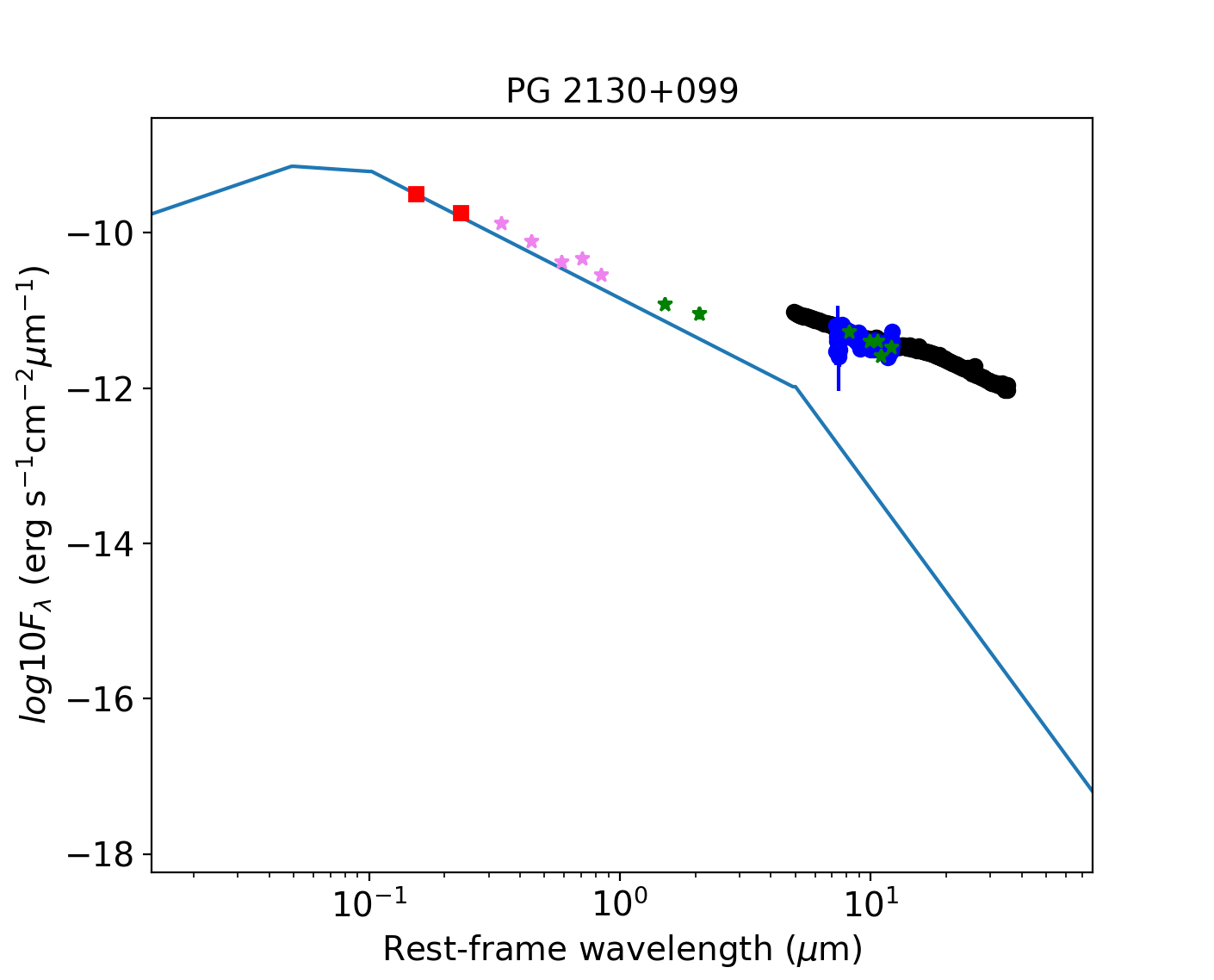}&\includegraphics[width=0.54\columnwidth]{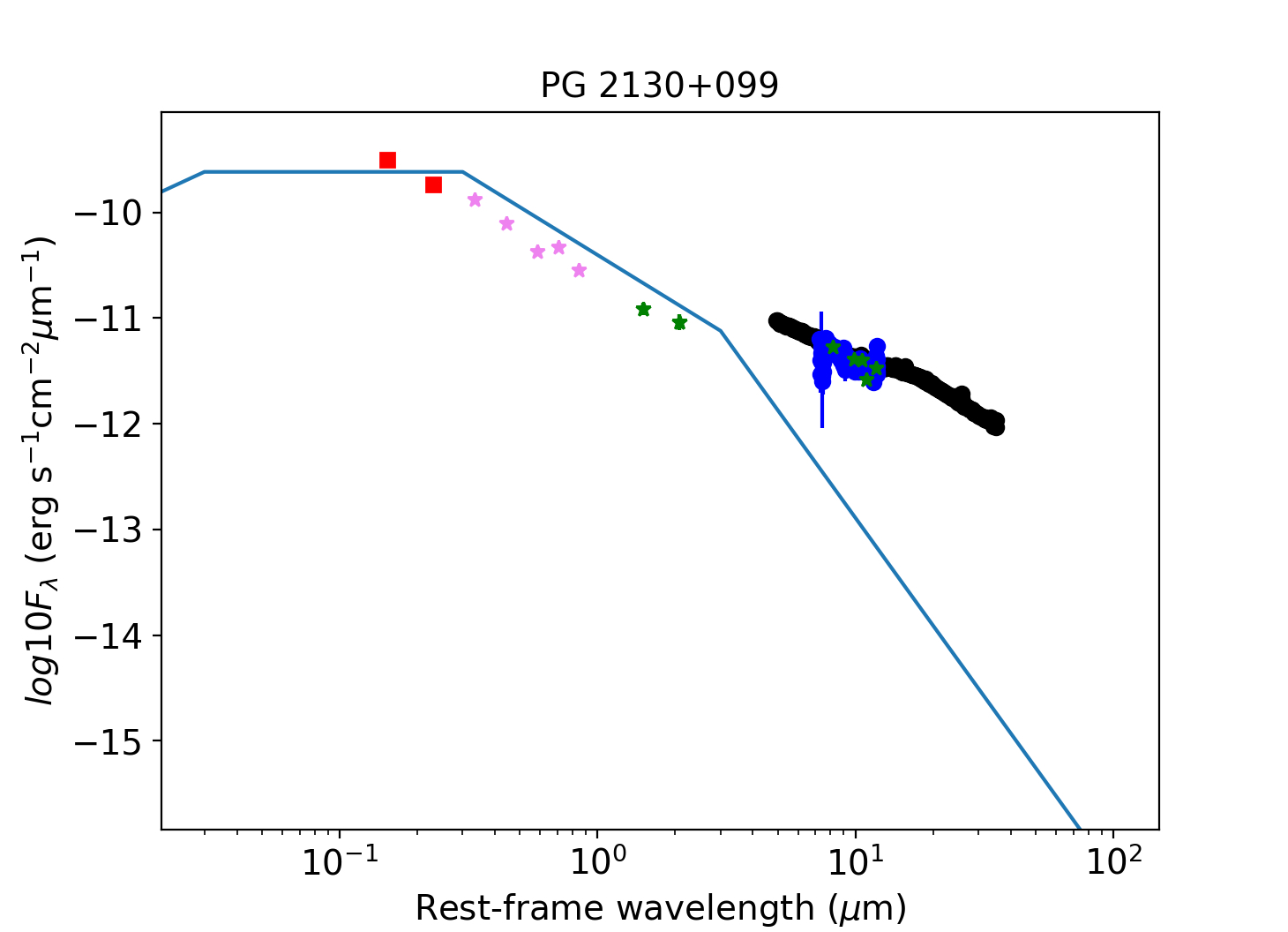}
	\end{tabular}
	\caption{{\bf Left panel}: The solid blue line is the normalized broken power law from \citet{Stalev16}, which is scaled to the photometric UV PSF fluxes at FUV (0.153 $\mu$m) and NUV (0.231 $\mu$m) bands from GALEX (red points). Pink symbols are the optical PSF fluxes from the SDSS bands (u, g, r, i, and z), which are included as an indicator of nuclear optical emission. The green photometric points are the NIR fluxes from HST/NICMOS and GTC/CIRCE in the H and Ks bands, and the MIR fluxes from GTC/CC, VLT/VISIR, and Subaru/COMICS. The black and blue points are the {\it Spitzer}/IRS  and GTC/CC spectra, respectively. {\bf Right panel}: as left panel but for the bluer accretion disk.} 
	\label{IIZw136_power_law}
\end{figure*}

\begin{figure*}
	\centering
	\begin{tabular}{cc}
		\includegraphics[width=0.5\columnwidth]{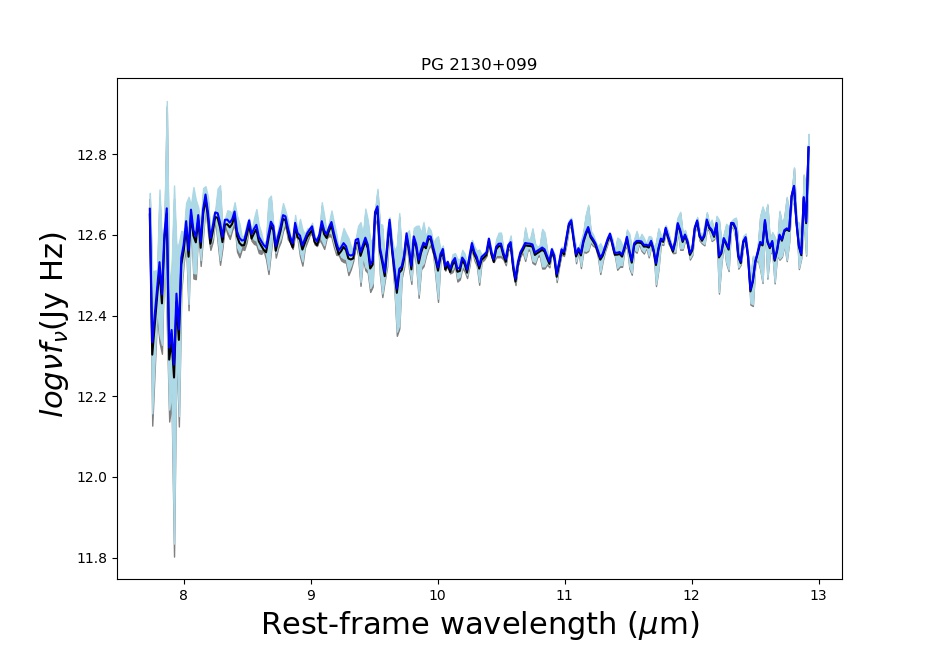}&\includegraphics[width=0.5\columnwidth]{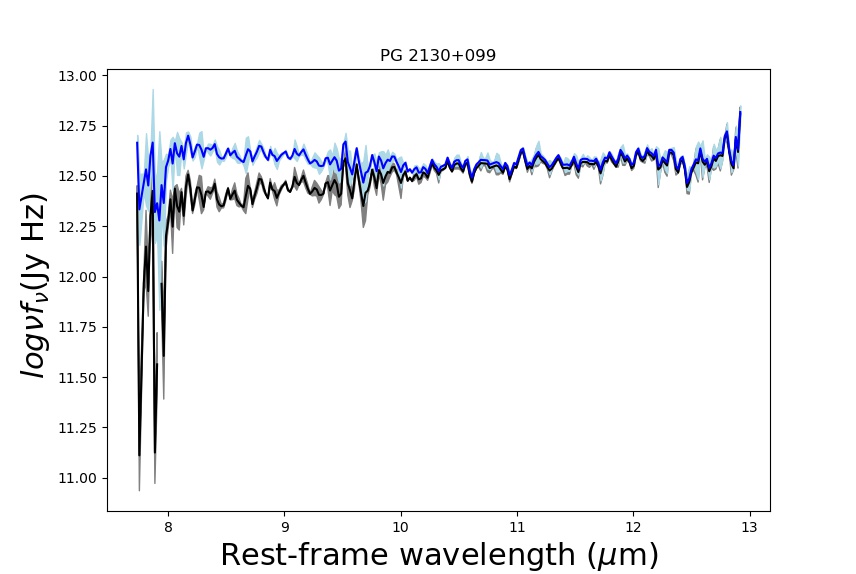}\\
		\includegraphics[width=0.5\columnwidth]{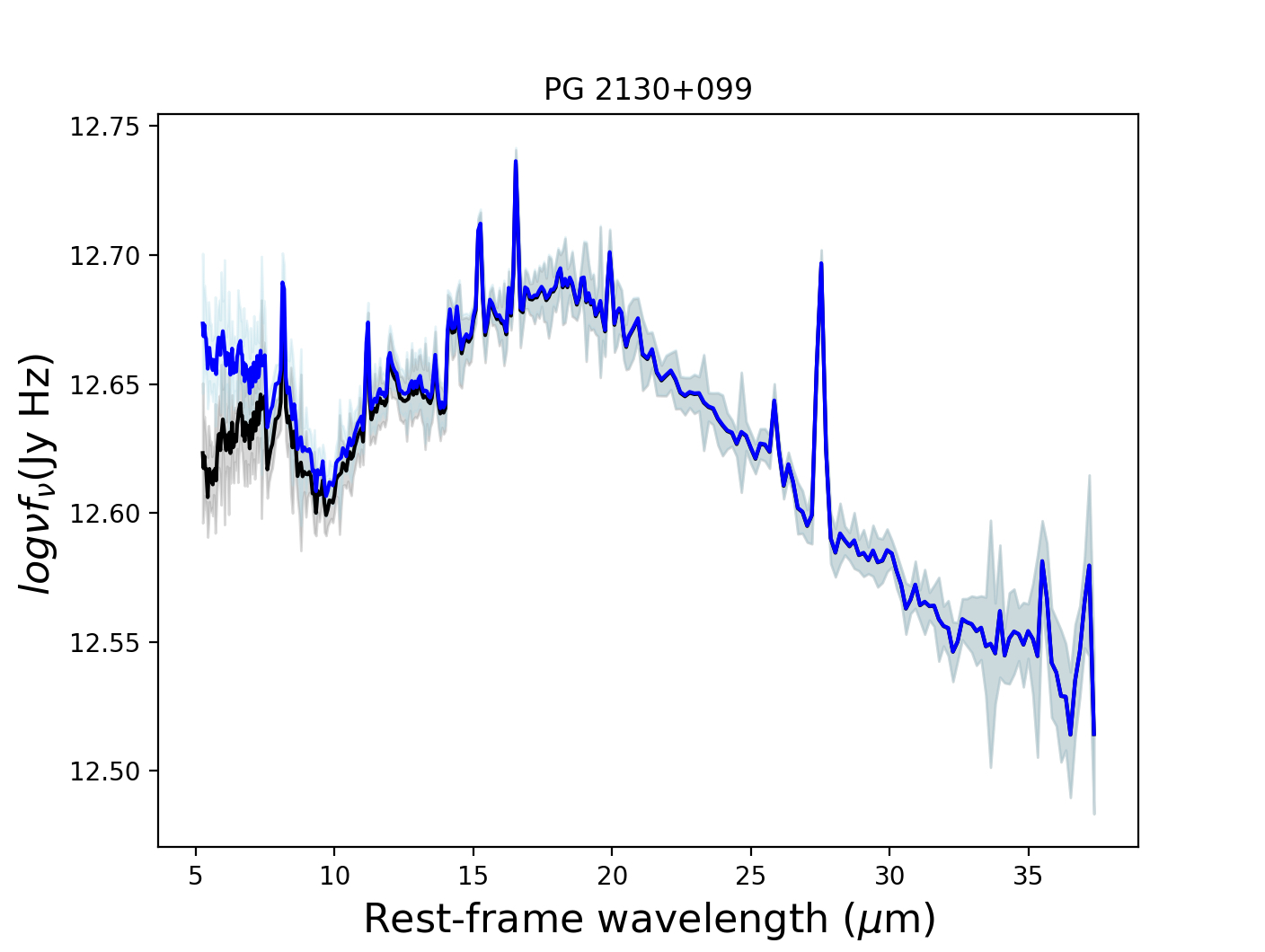}&\includegraphics[width=0.5\columnwidth]{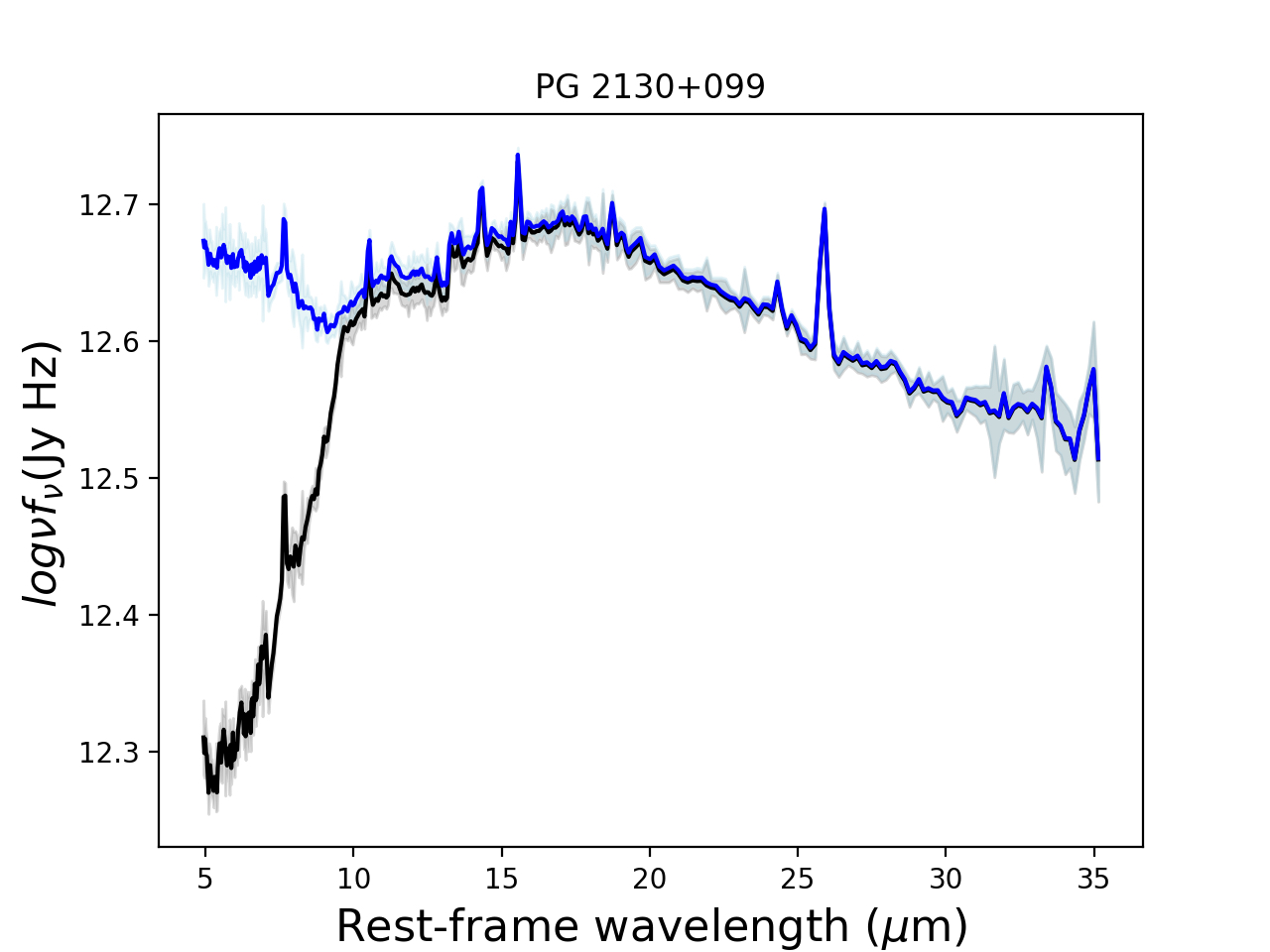}
	\end{tabular}
	\caption{{\bf Upper panels}: in blue and black is the high angular resolution spectrum before and after we removed the contribution from the classic (left) and bluer (right) accretion disk, respectively. {\bf Bottom panels}: the same but for the {\it Spitzer}/IRS spectra.} 
	\label{IIZw136_spectra}
\end{figure*}

As an example, in Figure\,\ref{IIZw136_power_law}, we show the normalized classic and bluer broken power-law models for PG~2130+099, see  Eqs. \eqref{equ1} and \eqref{equ2}. The UV photometric points are plotted in red. Additionally, we plot the dereddened PSF  fluxes from the SDSS bands (u, g, r, i, and z) obtained from the DR12 SkyServer\footnote{http://skyserver.sdss.org/dr12} as an indicator of the PSF emission at optical wavelength. We also plot in green points the NIR and MIR unresolved fluxes, the CC/GTC high angular resolution spectrum (in blue), and the {\it Spitzer}/IRS spectrum (in black) before subtracting the accretion disk.  
As can be seen from the figure, the  
bluer accretion disk overestimates the emission of the optical and NIR photometric points. The same happens for the rest of QSOs, except for PG~0050+124, PG~1440+356, and PG~1411+442. Although for the two latter objects, only the Ks band is above the  extrapolated accretion disk SED.  In the case of the classic accretion disk, there are only three objects (PG~1229+204, PG~0844+349, and PG~0804+761) for which the extrapolation of the SED drops slightly above the NIR photometric points. 

For those cases in which the accretion disk is above the photometric points, we do not subtract its emission and use the photometry as upper limits.

We aslo compare the spectrum before and after subtracting the accretion disk contribution. As an example, we plot the 
high angular resolution and {\it Spitzer}/IRS spectra of PG~2130+099 in Figure\,\ref{IIZw136_spectra}. We find that 
if we assume the classic accretion disk there are no differences between the high angular resolution spectra before and after subtraction. However, in the {\it Spitzer}/IRS spectra we find that in most objects (Mrk\,509, PG~2130+099, PG\,0003+199,PG\,1440+356, PG\,1426+015, PG\,1411+442, PG\,1211+143, and PG\,1501+106) there is an average contribution from the accretion disk of $\sim2.4\%$ between $\sim5-7.5\mu$m. On the other hand, when we assume the bluer accretion disk, then the spectral shape of both the high angular resolution and {\it Spitzer}/IRS spectra change after the disk component subtraction. We find that the bluer accretion disk contributes $\sim17.5\%$ and $\sim5.5\%$ to the {\it Spitzer} and high angular resolution spectra between $\sim5-10\,\mu$m, respectively. For PG~0804+761 the extrapolation of the bluer accretion disk overstimate  the emission of both spectra until $\sim10\,\mu$m.

On average, we find that the classic accretion disk contributes  $\sim35\%$ and $\sim30\%$ (12 objects), in the H and Ks bands, respectively. These are lower than the contributions we find assuming the bluer accretion disk,  $\sim41\%$ (two objects) and $\sim54\%$ (four objects), in the same bands.
\citet{Hernan_Caballero16} found that the contribution of the accretion disk at 1, 2, and 3 $\mu$m (rest-frame wavelength) are 63, 17, and 8 $\%$ for a sample of QSOs with a redshift $z$ between 0.17 and 6.42. On the other hand, \citet{Garcia_Bernete19} found that the contribution from the accretion disk is $\sim46$, $\sim23$, and $\sim11$ $\%$ in the J, H, and K bands for a sample of type 1 Seyferts.

\section{Modeling the NIR to MIR SEDs}
\subsection{The Dusty Torus Models} \label{sec:models}

\subsubsection{\citet{Fritz06} - Smooth F06 model}

The \citet{Fritz06} model has a flared disk with open polar cone regions. In this model the dust homogeneously and continuously fills the torus volume. The following parameters describe it. First, the inner and outer radius ($\rm Y=R_{out}/R_{in}$) ratio. The inner radius is defined by the dust sublimation temperature (1500~K) under the intense radiation by the central AGN. It is defined as $R_{in}\eqsim1.3\sqrt{L_{46}^{AGN}}T_{1500}^{-2.8}$ pc, for a typical graphite grain with a radius of 0.05\,$\mu$m \citep[][]{Barvainis87}. Other parameters are the angular width of the torus $\Theta$, the viewing angle $i$, the index of the polar and radial gas distribution $\gamma$, and $\beta$, respectively, and, the optical depth at $9.7\mu\mbox{m}$.
The dust is mainly composed of graphite and silicate grains in nearly equal percentages. The central point-like source is represented by a broken power-law that illuminates the dust.

\subsubsection{\citet{Nenkova08a, Nenkova08b} - Clumpy N08 model}
The \citet{Nenkova08a, Nenkova08b} model assumes that the dust within a torus-like geometry is distributed in the form of identical spherical clumps. They enshroud the central engine emitting the energy following a broken power-law SED and are assumed to have the standard Galactic ISM (i.e. 47~\% graphite and 53~\% silicate). The following parameters describe the model, the viewing angle ($i$), the number of clouds along the line of sight ($N_{0}$), the angular width of the torus ($\sigma$), the radial extension ($\rm Y=R_{out}/R_{in}$), the index of the radial distribution {\it q}, and the optical depth of the clouds ($\tau_{V}$). The inner radius in this model is defined as $R_{in}\eqsim0.4\sqrt{L_{45}^{AGN}}T_{1500}^{-2.6}$ pc. They argued that the differences in the power of the sublimation temperature reflect the more detailed radiative transfer calculations performed in this model.

\subsubsection{\citet{Hoenig10b}- Clumpy H10 model}
The \citet{Hoenig10b} model inherites the strategy in \citet{Nenkova08a}, however, used a 3D Monte Carlo radiative transfer simulation to treat the dust distribution in a probabilistic manner. It assumes the central point source described by a broken power-law and three different dust compositions: the standard ISM, large ISM grains ($0.1\sim1~\mu$m), and intermediate grains ($0.05\sim0.25~\mu$m) with 70~\% graphite and 30~\% silicate. The torus parameters are the power-law index of the radial distribution ({\it a}), the power-law index of the cloud size distribution ({\it b}), the viewing angle ($i$), the number of clouds along the equatorial plane ($N_{0}$), the outer radius ($R_{out}$), and the cloud size at the inner-most torus radius. In this model the inner radius is defined as $R_{in}\eqsim0.36\sqrt{L_{45}^{AGN}}$ pc assuming ISM dust, while for ISM large grain this is $R_{in}\eqsim0.5\sqrt{L_{46}^{AGN}}$ pc \citep[see][]{Hoenig10b}.

\subsubsection{\citet{Stalev16} - Two-phase media S16 model}
They model the dusty torus with dusty clumps enshrouded by a smooth low-density  nebulosity called a two-phase medium. The torus is heated by a central point source with anisotropic emission \citep{Netzer87}, which is the strongest for the polar direction and the weakest along the equatorial plane. A power-law dictates the standard ISM dust distribution along the radial ($p$) and polar ($q$) direction. Additionally, the model depends on the viewing angle ($i$), the half-opening angle ($\sigma$), the optical depth at $9.7~\mu m$ ($\tau$), and the radial extension ($\rm Y=R_{out}/R_{in}$). The inner radius is defined as in the smooth model of \citet{Fritz06}. 

\subsubsection{\citet{Hoenig17} - Disk+Wind H17 model}
This model is composed of a clumpy polar wind and a compact disk. The dust clouds are assumed to lie along a radial
distance from the central black hole $r$, following a power-law distribution described by the power-law index $a$ (for the disk), and the power-law index $a_{w}$ (for the wind). 
The disk is also described by the dimensionless scale height $h$ and the number of clouds along the equatorial line $N_{0}$. While the wind by the half-opening angle $\theta$ and the angular width of the hollow cone $\sigma$.
Other parameters are the wind-to-disk ratio of the number of clouds $f_{wd}$ and the viewing angle $i$. Different sets of dust composition have been implemented, assuming the standard ISM and the standard ISM plus larger grains dust. 

\begin{figure*}
\centering
\includegraphics[width=1\columnwidth]{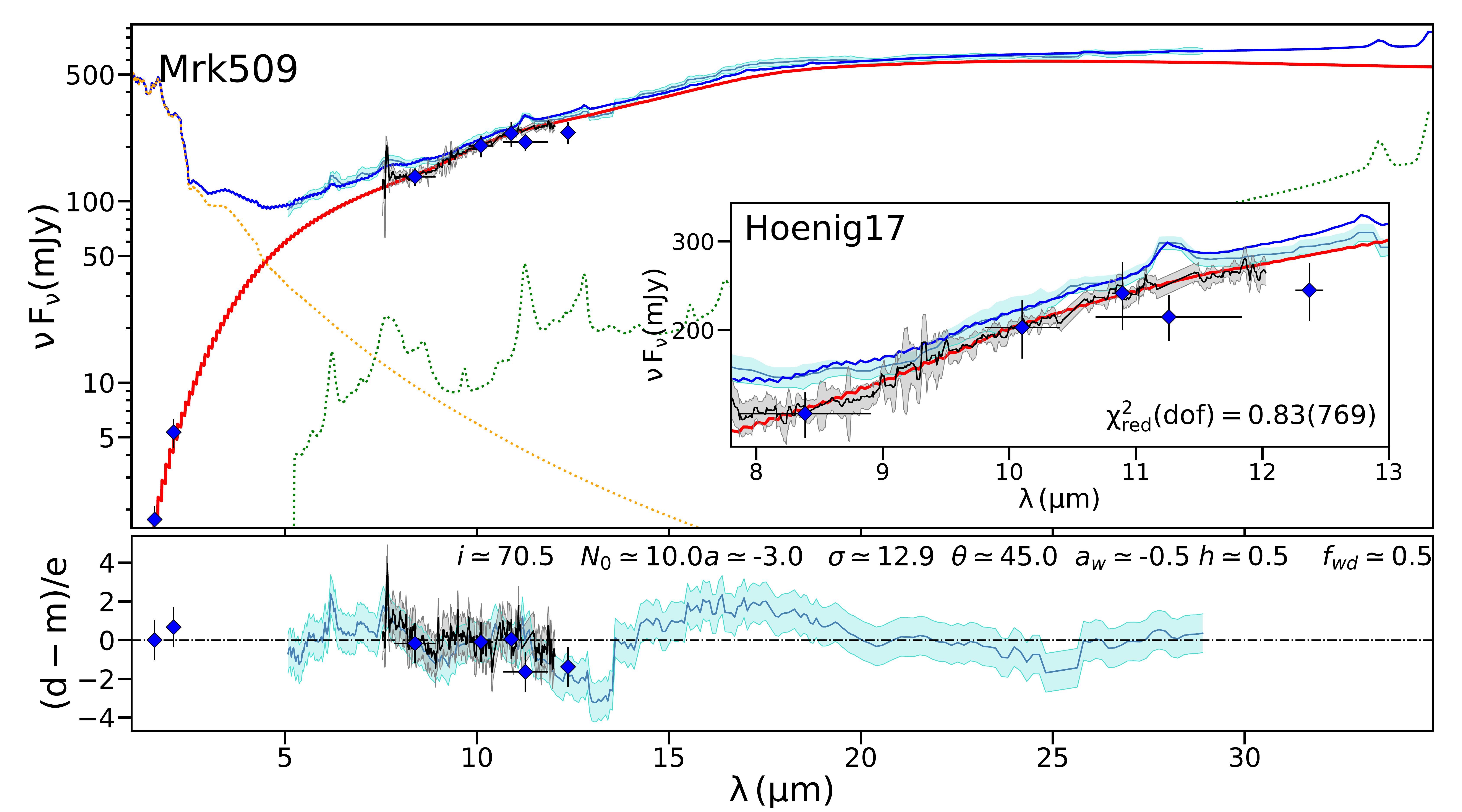}
\caption{{\bf The SED of Mrk~509 fitted by the Disk+Wind H17 model}. The high angular resolution photometric points are plotted as blue points with their $1\sigma$ error bars, while the high angular resolution spectrum is plotted with a black solid line, the grey shaded region represent the errors. The {\it Spitzer}/IRS spectrum is represented with a dark cyan solid line and its error with a cyan shaded region. The red solid line is the best model resulting from 
fitting the high angular resolution data. The green and yellow dotted lines are the starburst and stellar components, respectively. The blue solid line represents the sum of the stellar, starburst and torus components that best fit the {\it Spitzer}/IRS spectrum.\label{models_Mrk509}}
\end{figure*}

\begin{figure*}
\centering
\includegraphics[width=0.9\columnwidth]{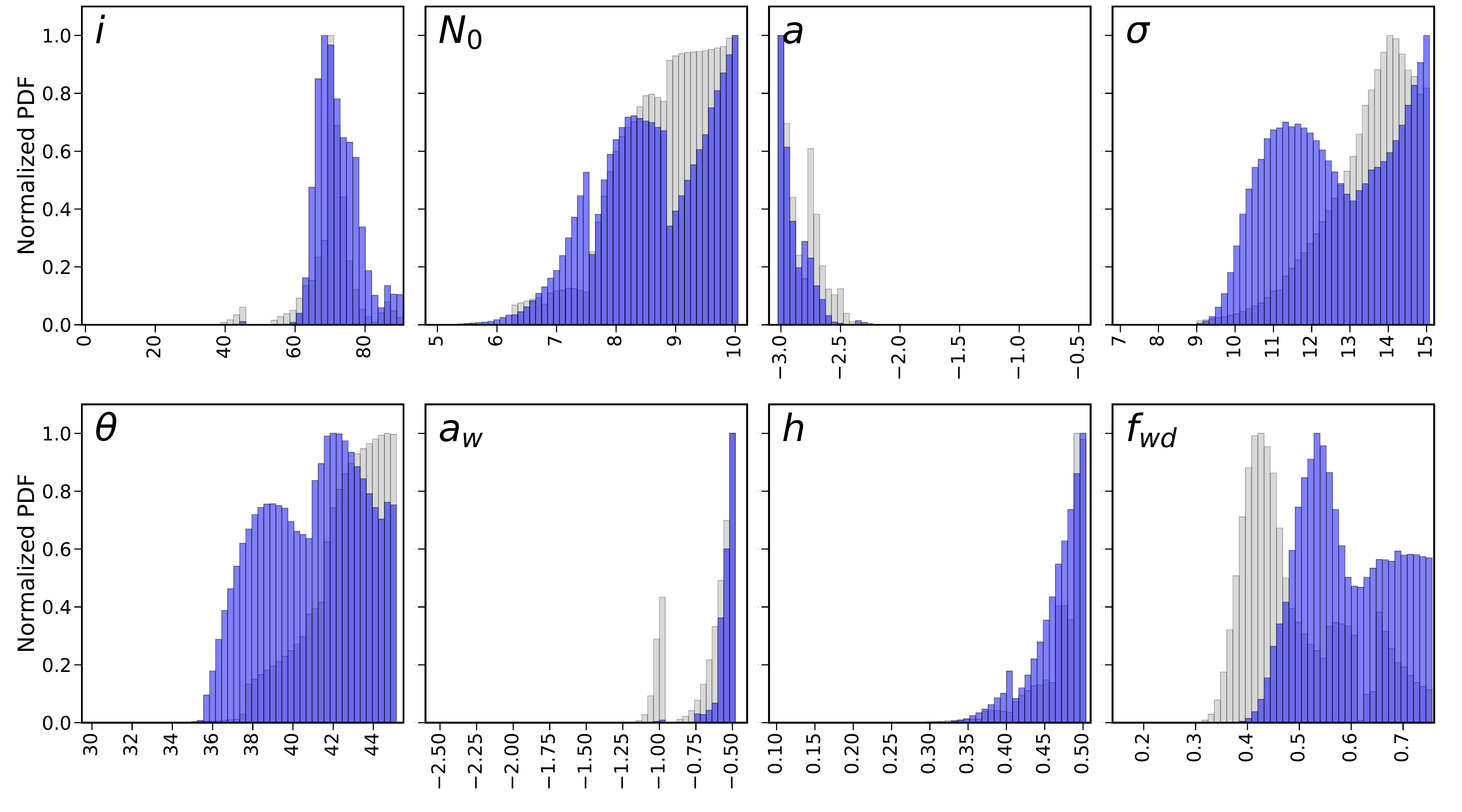}\\
\begin{tabular}{cc}
\includegraphics[width=0.3\columnwidth]{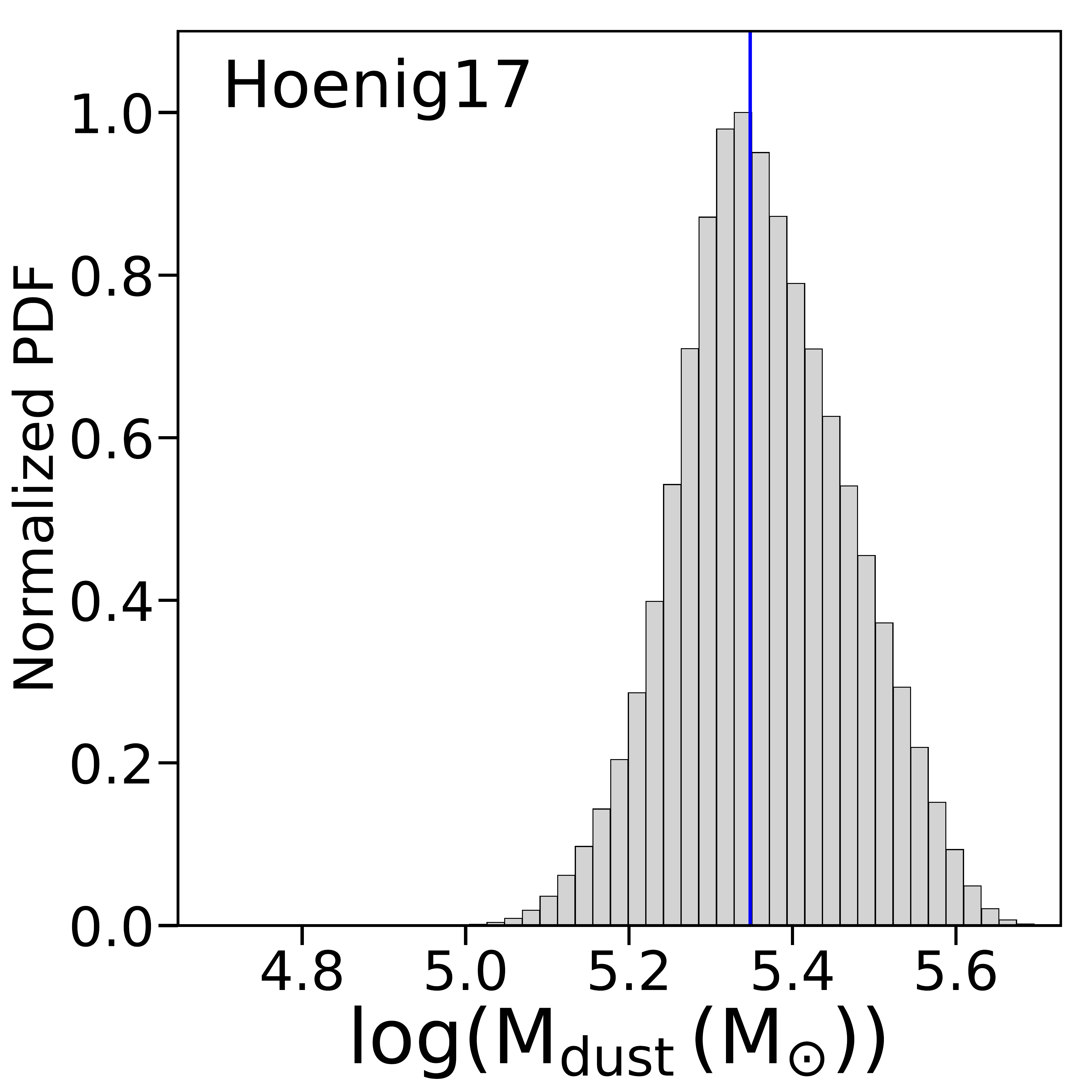}&\includegraphics[width=0.3\columnwidth]{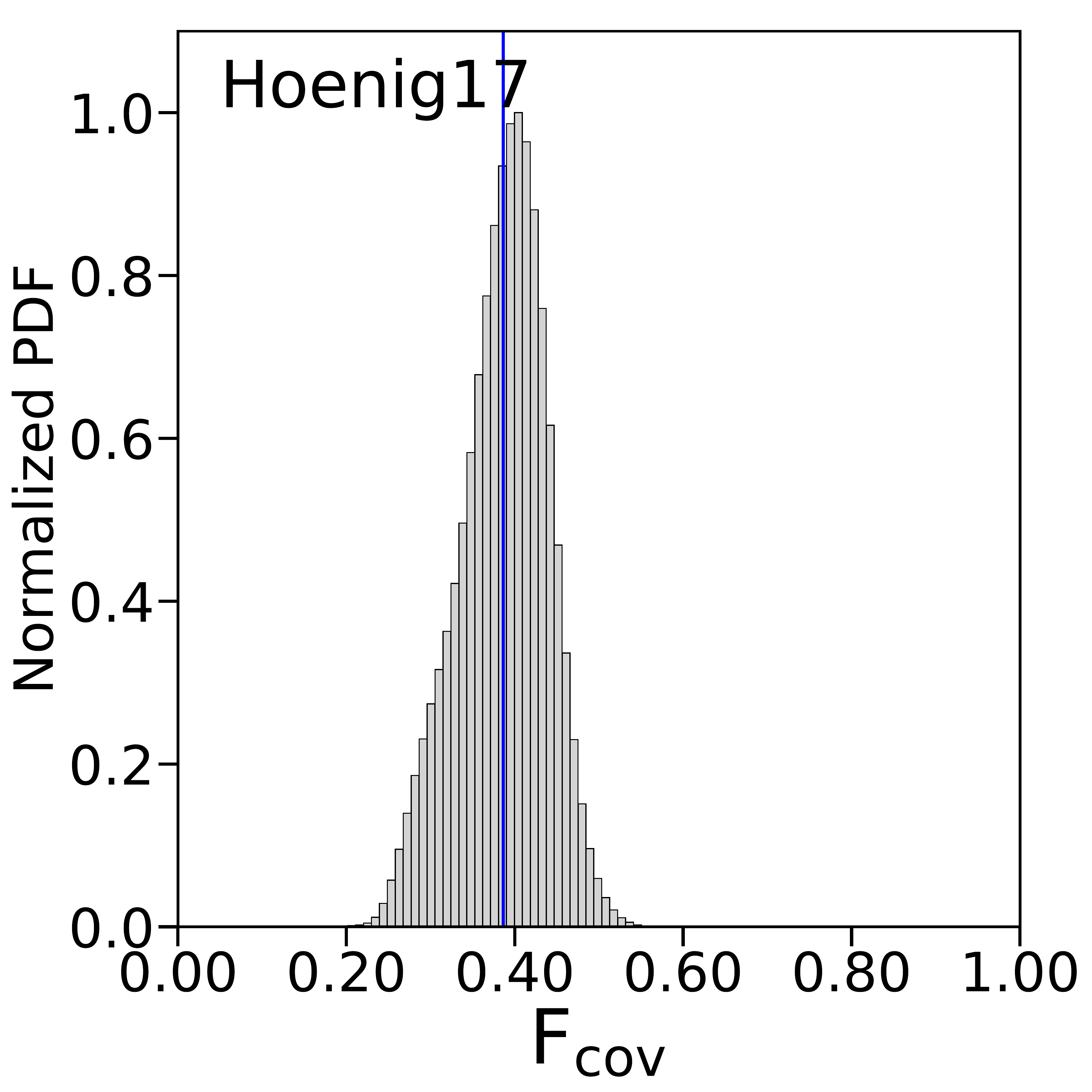}\\
\end{tabular}
\caption{{\bf The Disk+Wind H17 model parameters derived for Mrk~509}. {\bf Upper panels}: normal probability distribution function of the eight free parameters. In grey we plot the parameters derived from fit LSR spectrum, while in dark blue the distribution of the parameters obtained from fit FSR spectrum. {\bf Bottom panels}: normal PDF of the derived parameters. The blue vertical line indicates the mean value.\label{parameters_Mrk509}}
\end{figure*}

\begin{figure*}
\centering
\includegraphics[width=1\columnwidth]{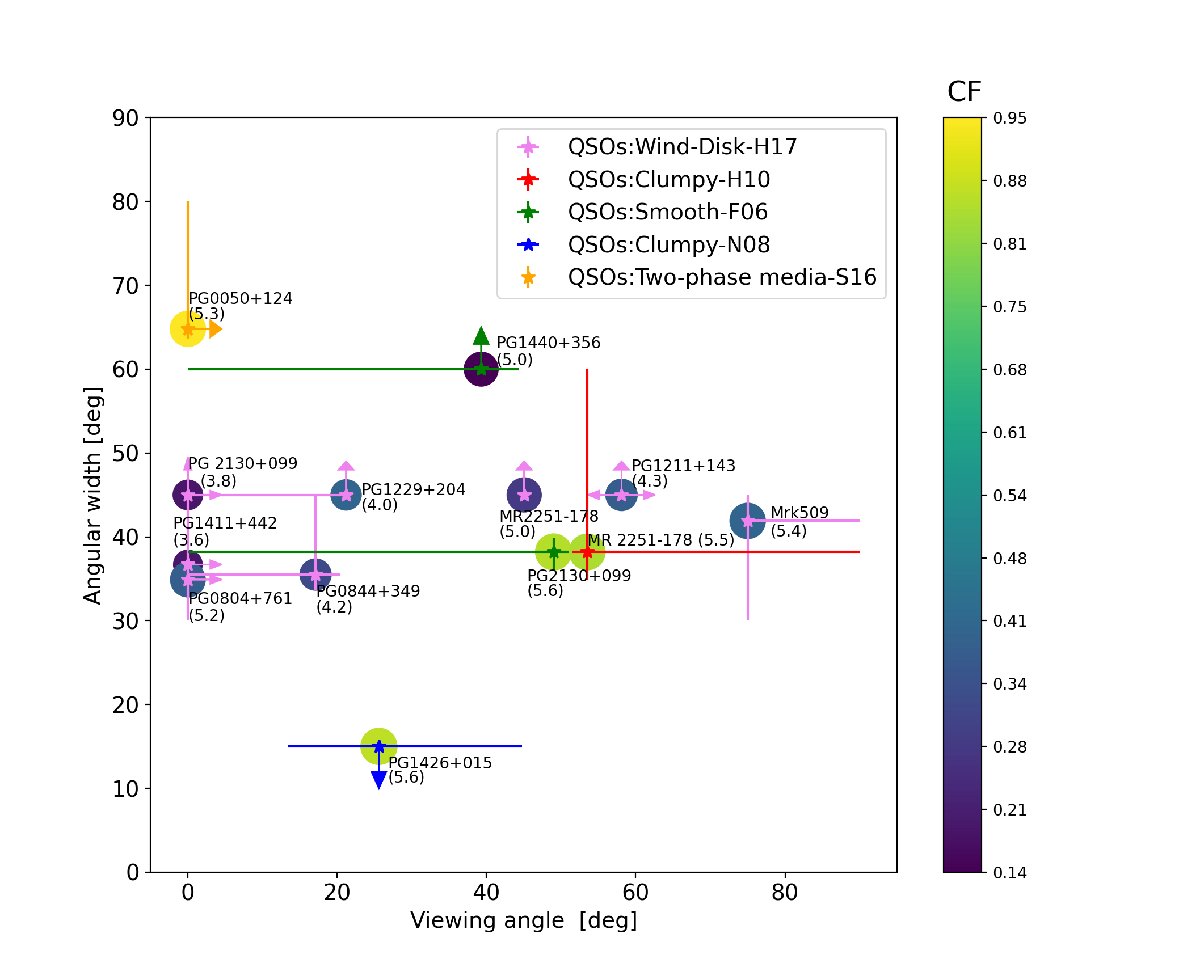}
\caption{{\bf Angular width versus viewing angle $i$ obtained for the good fits of all models}. The coloured stars indicate the model that best fits the object, the arrows an upper or lower limit. The covering factor (CF) is represented with a colour-coded circle. The set of values are indicated in the vertical colour bar. Yellow circles indicate large covering factors, while purple circles indicate low covering factors. The value of the logarithm of the dust mass (in $M_{\odot}$) is represented with a size coded circle, and the value is indicated in parenthesis. \label{ind_par}}
\end{figure*}

\subsection{Modeling procedure}
In this work 
we take advantage of the high S/N of the {\it Spitzer}/IRS spectra, 
the high angular resolution of the CC and VISIR spectra, and the ground-based high angular resolution photometry at NIR and MIR to build 
a well-sampled set of SEDs that allows us to look for the model that 
better reproduces the data. To model these data we used the interactive spectral fitting program {\sc XSPEC} \citep[][]{Arnaud96} from the {\sc HEASOFT}\footnote{\url{https://heasarc.gsfc.nasa.gov}} package. XSPEC uploads the dust models as additive tables, which have been previously created using the {\sc flx2tab} task within {\sc HEASOFT}. Briefly, a table model is composed of an N-dimensional grid of model spectra, which is calculated for a particular set of values within the N parameters space of the model \citep[for more details on the model inclusion and modelling procedure within XSPEC see][]{Gonzalez-Martin19a}. Additionally, we also included the stellar libraries of \citet{Bruzual03}, the starburst templates from \citet{Smith07}, and foreground extinction \citep{Pei92}. 

For the data, we use the {\sc flx2xsp} task within {\sc HEASOFT} to read the text file that contains the photometry and the spectroscopy plus their uncertainties, and to write a standard {\sc XSPEC} pulse height amplitude\footnote{Engineering unit used to describe the integrated charge per pixel produced by a detector.}.

We use $\rm{\chi^2}$ statistic to converge to the best fit. However, this approach requires a particular treatment of the data when using different datasets. The different spectral resolution could lead to an overestimation of the spectra compared to the photometry. Simultaneously, the spectral features (e.g., silicate features) are useful to understand the best model for each object. 
Thus, we develop a procedure able to use both the spectral and photometric data simultaneously. We create spectra, for both \emph{Spitzer} and ground-based observations, degraded to the spectral resolution of the photometry. To do so, we require that the spectrum matches the average bandpass of the photometric points in the MIR. From now on, we call this degraded spectra the low spectral resolution (LSR) \emph{Spitzer} and ground-based spectra. 

We start fitting the LSR ground-based spectrum and the high angular resolution photometry to the dust models attenuated by foreground extinction. In the case of the photometric set of data, the parameters of the dust models are tied to those of the LSR ground-based spectrum to try to find a single SED reproducing both sets of data. We assume that circumnuclear contributors do not contaminate this data set. After obtaining the best fit to the ground-based high spatial resolution data, we added the LSR {\it Spitzer}/IRS spectrum to these data set. We used the parameters derived from fitting the LSR {\it Spitzer}/IRS spectrum as an initial guess for the fit of the LSR ground-based spectra/photometry. In this case, and when necessary, we add a stellar and/or starburst component to the LSR {\it Spitzer}/IRS spectrum to consider plausible circumnuclear contributors due to the lower spatial resolution of {\it Spitzer}/IRS. 
Once the LSR data are fitted, we computed the 3$\rm{\sigma}$ errors for each parameter involved. 

We used these errors to fix the minimum, and maximum allowed range to those  parameters obtained from the spectral fit to the LSR data set. Then we replace the LSR spectra (ground-based and \emph{Spitzer}) for the full spectral resolution (FSR) version of the spectra. Therefore, we imposed these values as the initial guess of the parameters to fit the N band high angular resolution spectrum plus the photometric data. Next, besides the parameters of the models, we also calculate the dust mass and covering factor for all the models. For F06, N08, and S16, we also calculate the radial extension of the torus. To derive the dust mass and covering factor, we used the equations (1) to (6) from \citet{Esparza-Arredondo19}. The outer radius $R_{out}$ is calculated as a function of the radial extent $Y$. 

We consider as acceptable fits those with a reduced $\chi_{red}^{2}<2$ for both LSR and FSR spectral fits, see Table~\ref{tab:stat}. Then, to look for the model preferred by the data, we estimate the Akaike Information Criterion \citep[AIC][]{Akaike74} for both LSR and FSR spectral fits. The AIC criterion gives the quality of a model by estimating the likelihood of the model to predict futures values. We compute the AIC value in terms of the $\chi^{2}$, the degree of freedom (dof), and the data size (N). The dof is calculated as the total number of points in the spectrum minus the total number of free parameters of the model. In general, a good model is the one with the minimum AIC. However, to better discriminate between the models, we estimate the likelihood that the model with the minimum AIC minimizes the loss of information. 

\subsection{Spectral fitting}

\begin{figure*}
\centering
\includegraphics[width=0.7\columnwidth]{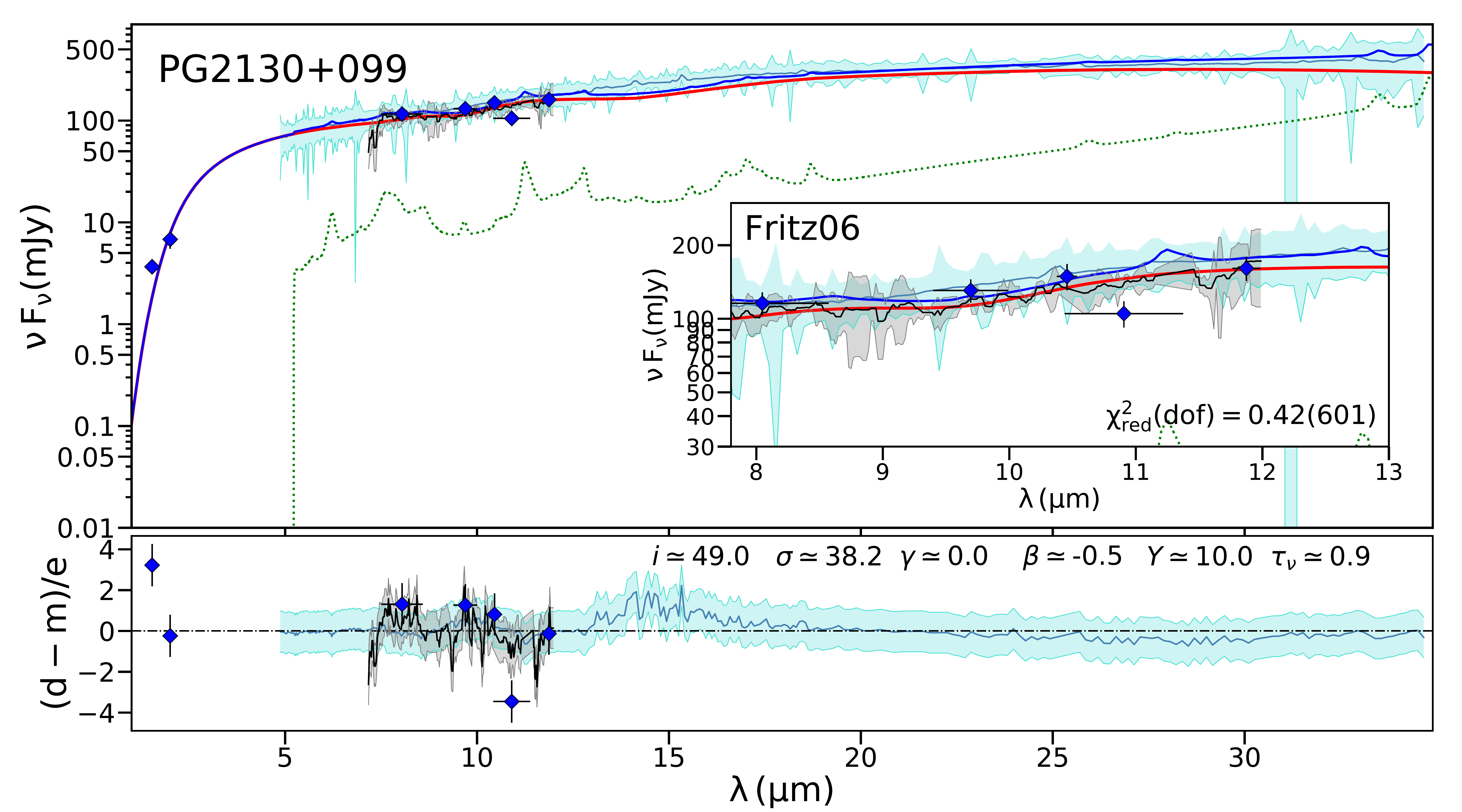}\\\includegraphics[width=0.7\columnwidth]{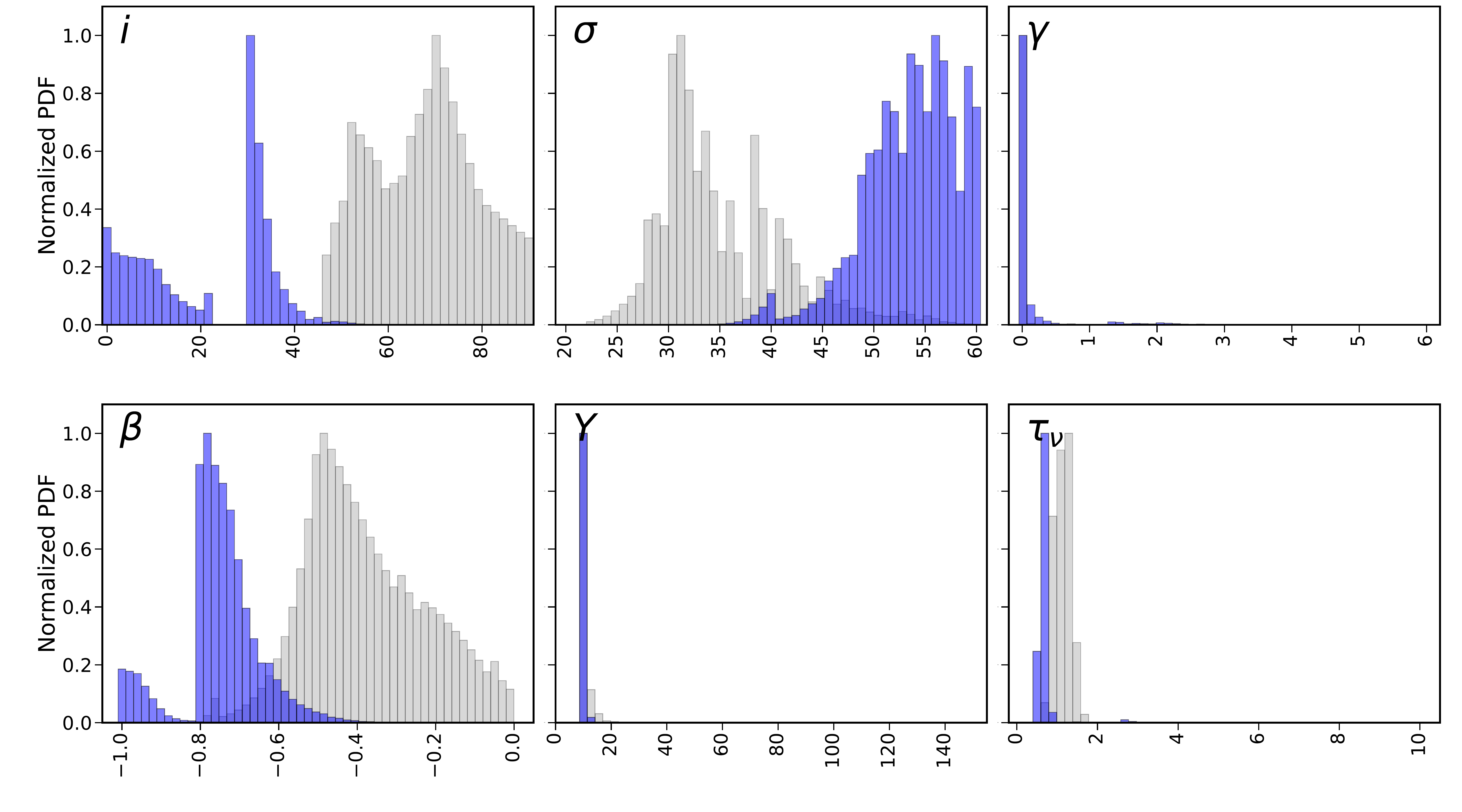}\\
\begin{tabular}{ccc}
\includegraphics[width=0.25\columnwidth]{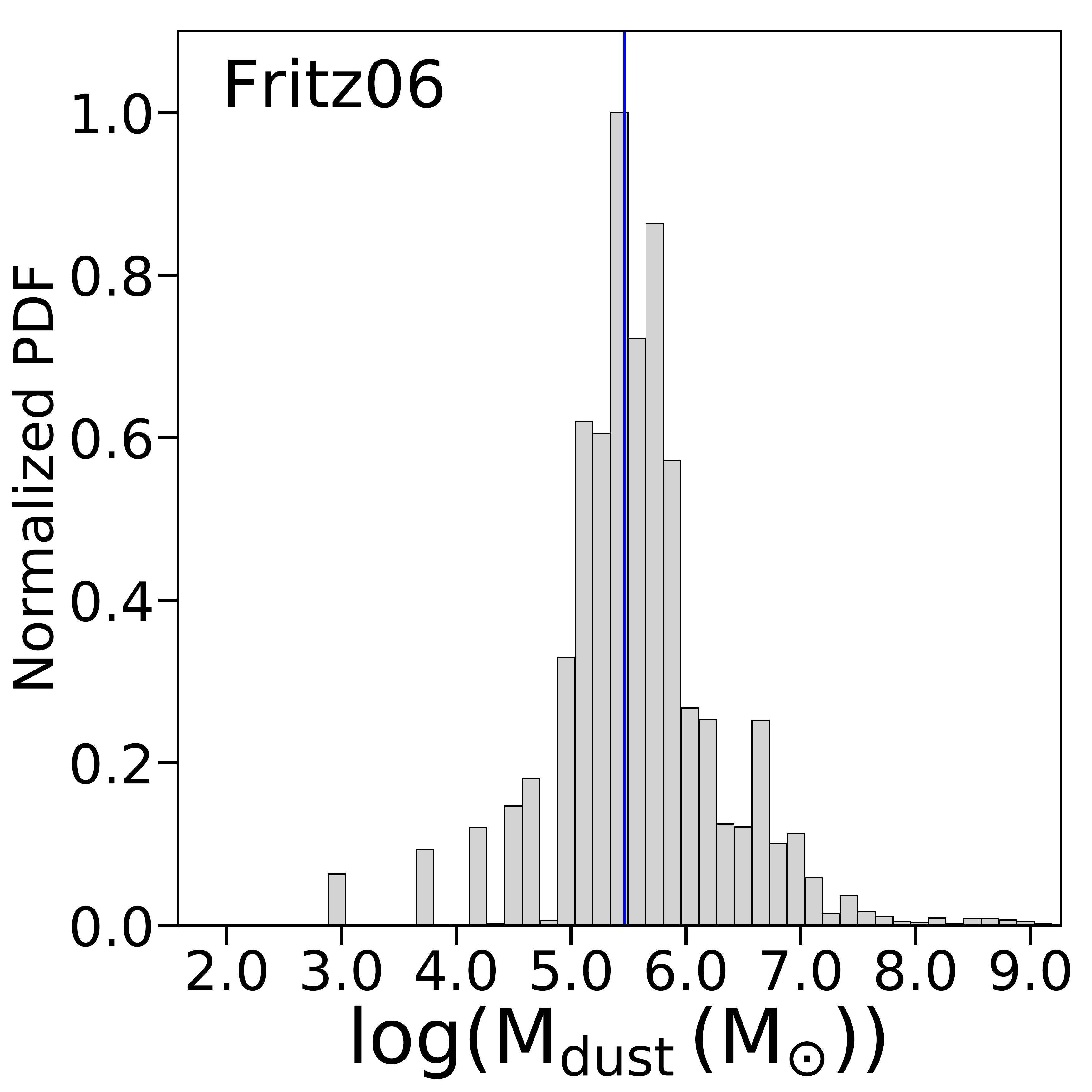}&\includegraphics[width=0.25\columnwidth]{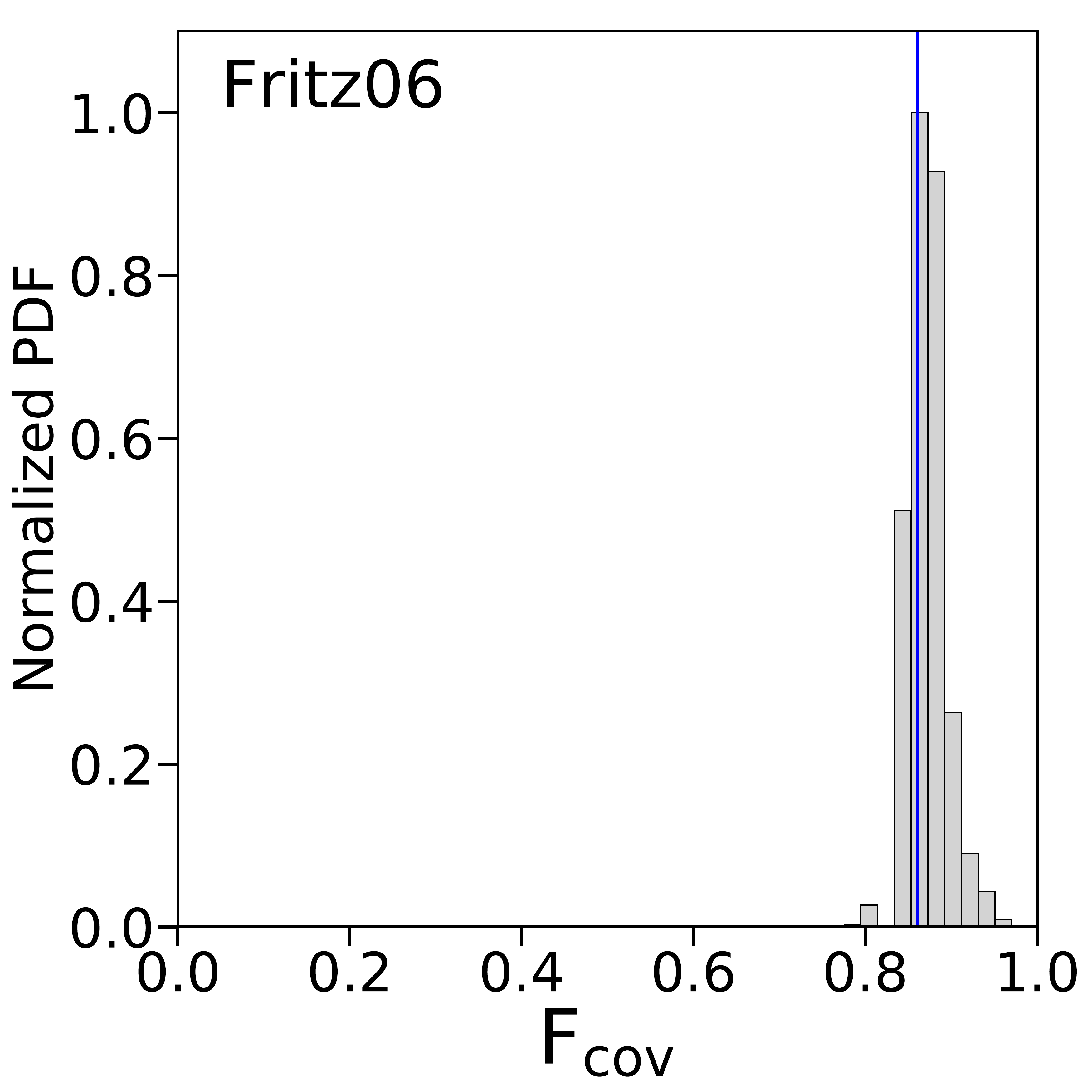}&\includegraphics[width=0.25\columnwidth]{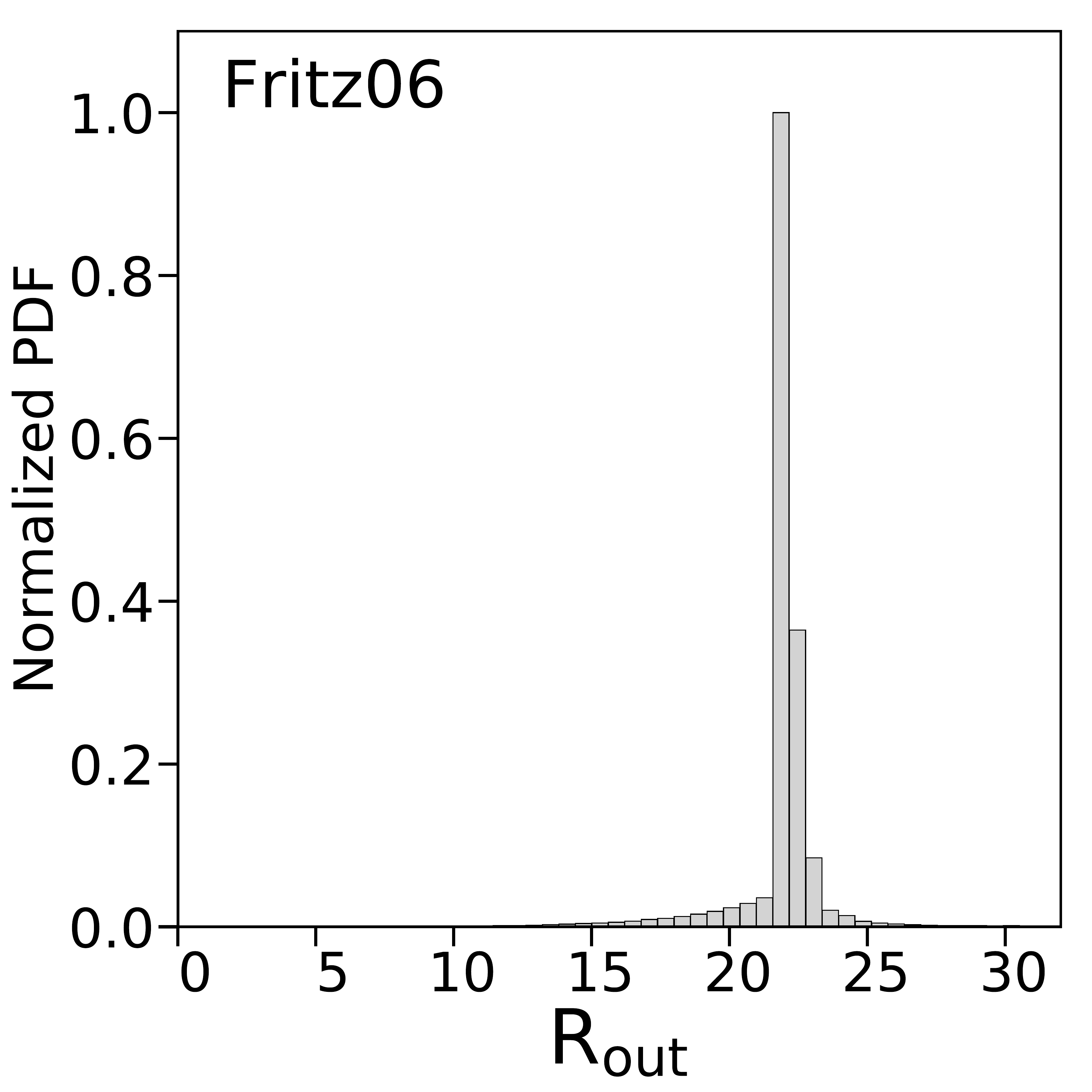}\\
\end{tabular}
\caption{{\bf The SED of PG~2130+099 best modeled by the Smooth F06 model}. {\bf Upper panel}: the high angular resolution photometric points are plotted as blue points with their error bars, the black arrows are upper limits. The high angular resolution spectrum is plotted with a black solid line, the grey shaded region represent the errors. The {\it Spitzer}/IRS spectrum is plotted with a dark cyan solid line and its error with a cyan shaded region. The red solid line is the best model resulting from 
fit the high angular resolution data. The green and yellow dotted lines are the starburst and stellar components, respectively. The blue solid line represents the sum of the stellar, starburst and torus components that best fit the {\it Spitzer}/IRS spectrum.  {\bf Middle panel}. {\bf The model parameters derived}: normal probability distribution function of the free parameters. In grey we plot the parameters derived from fit the LSR spectrum, while in dark blue the distribution of the parameters derived from fit the FSR spectrum. {\bf Bottom panel}: normal PDF of the derived parameters. The blue vertical line indicate the mean.\label{fit1}}
\end{figure*}

\begin{figure*}
\centering
\includegraphics[width=0.7\columnwidth]{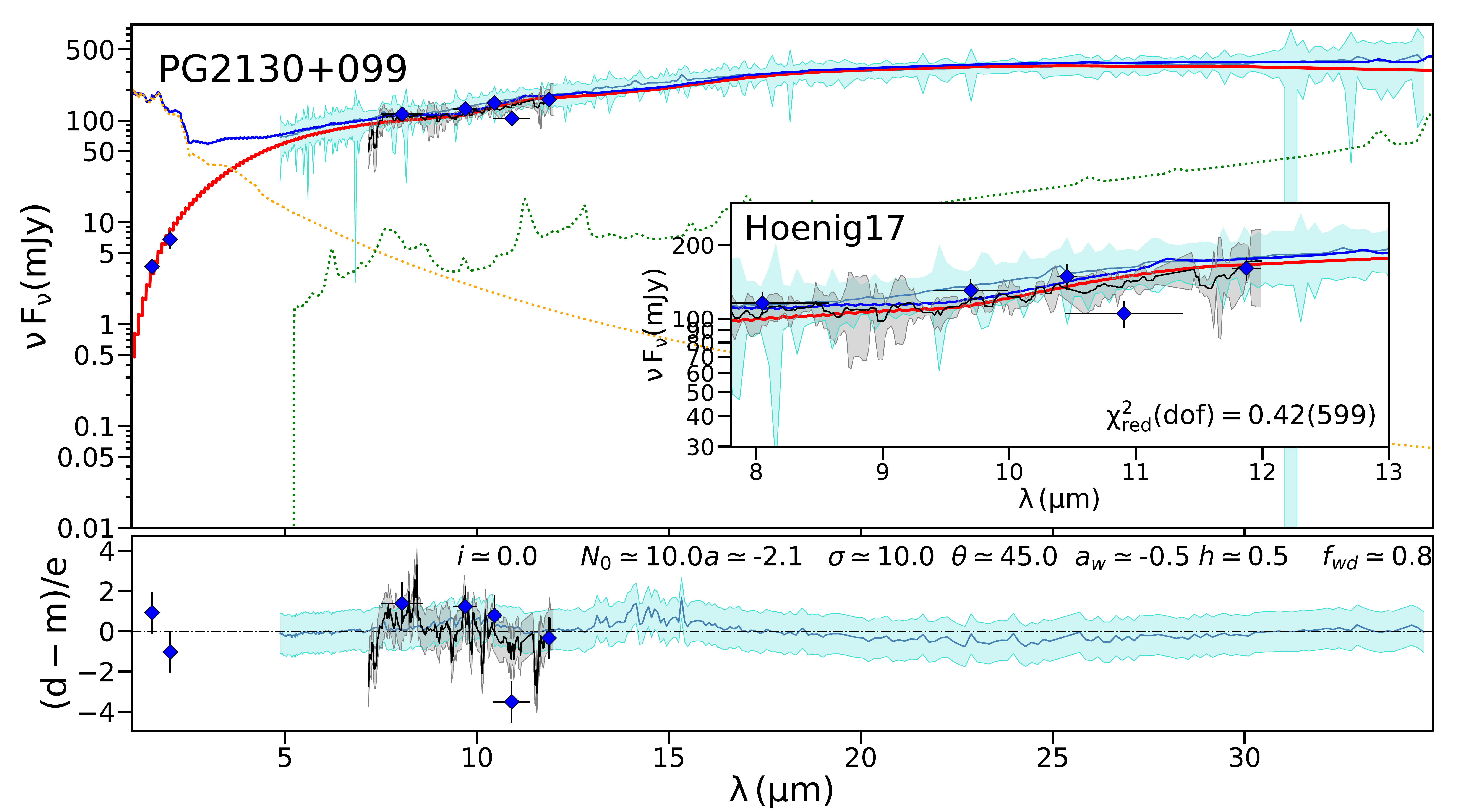}\\\includegraphics[width=0.7\columnwidth]{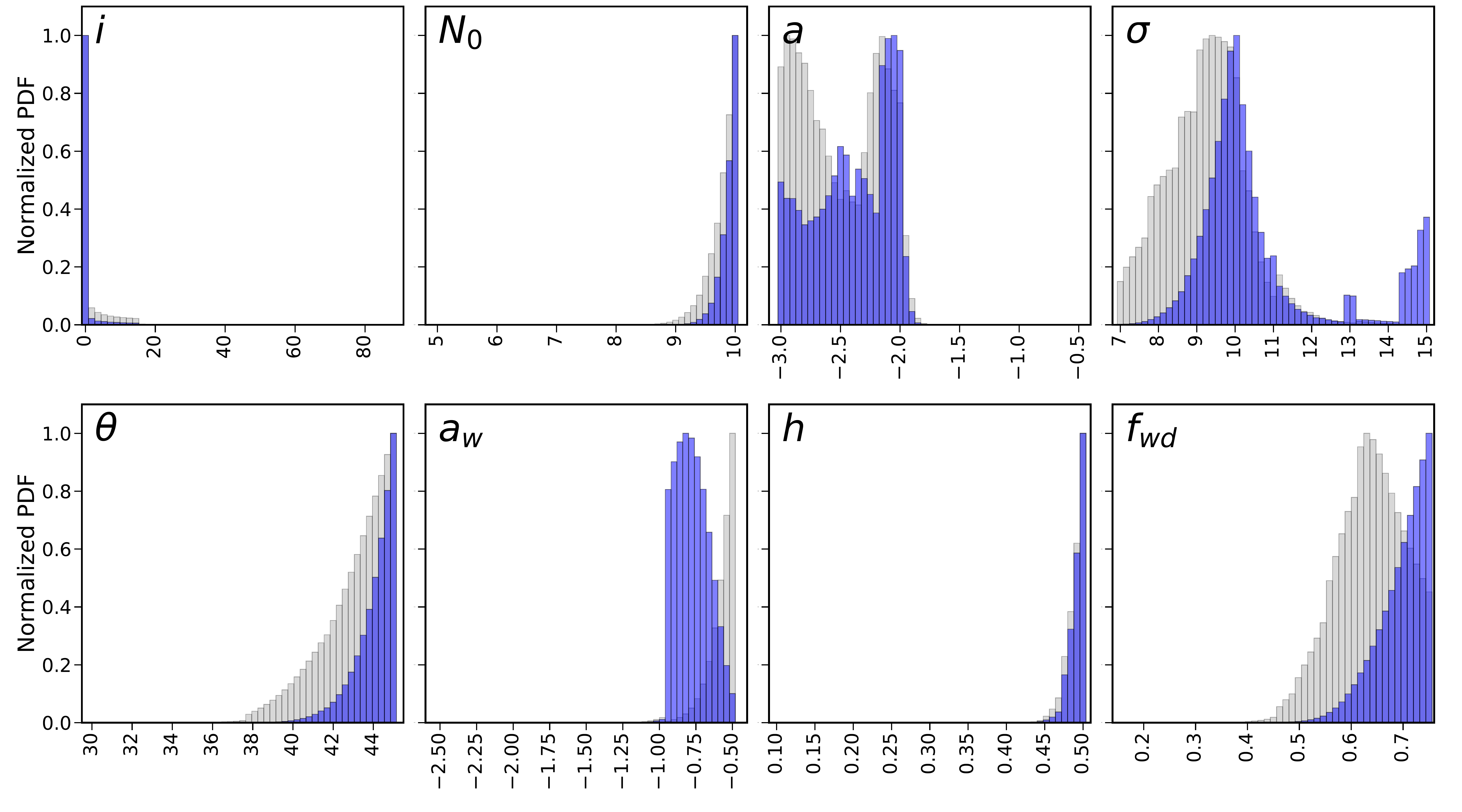}\\
\begin{tabular}{ccc}
\includegraphics[width=0.25\columnwidth]{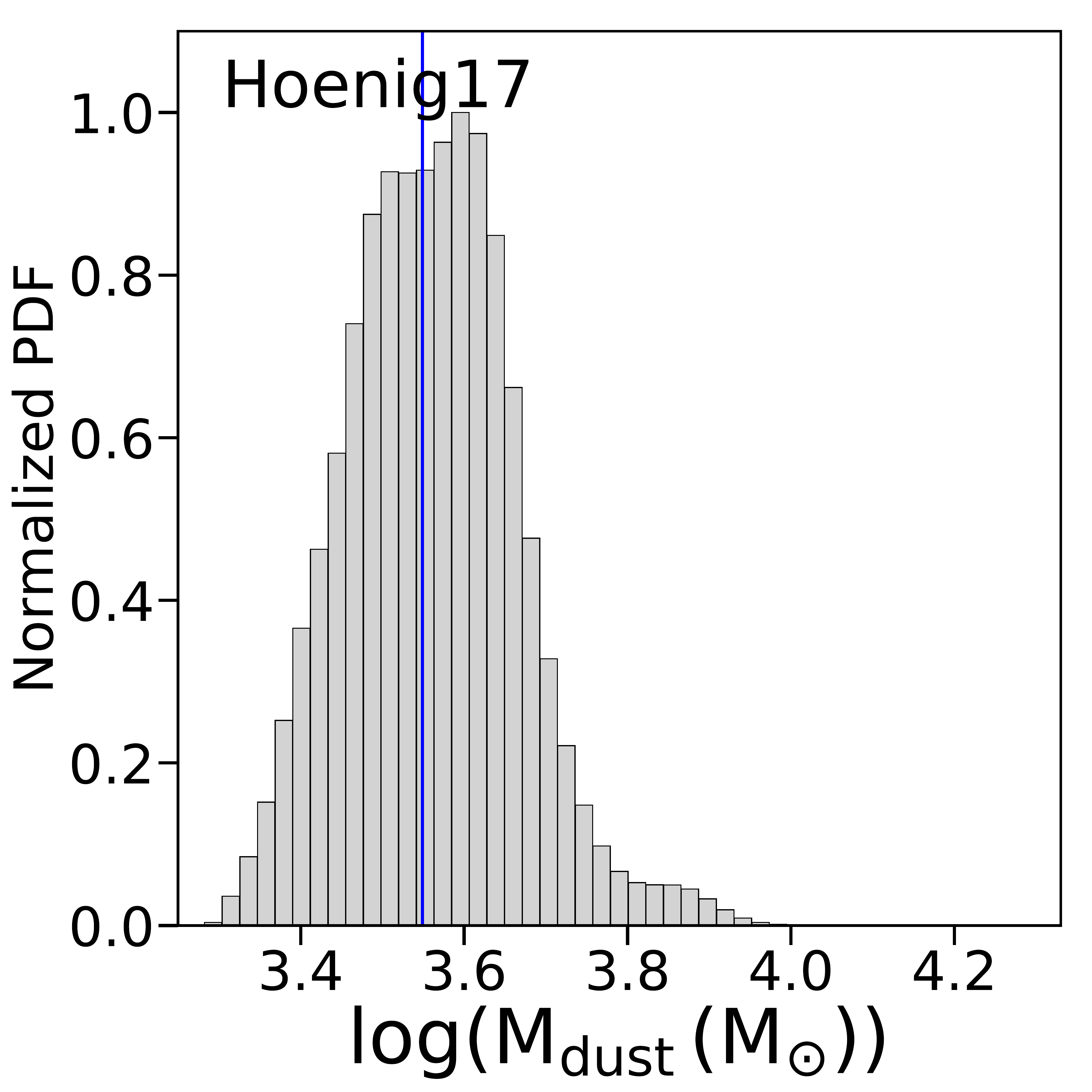}&\includegraphics[width=0.25\columnwidth]{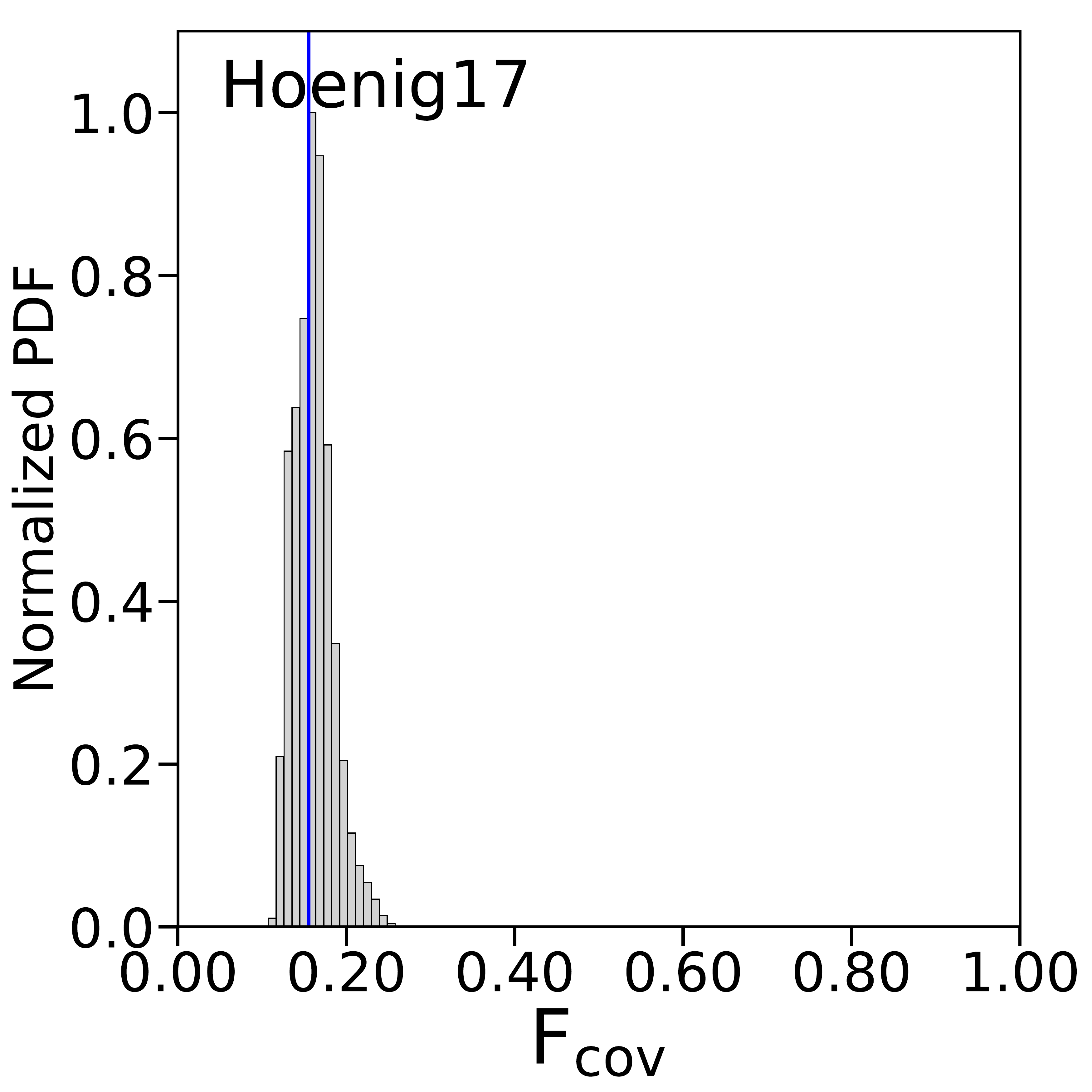}&\\
\end{tabular}
\caption{As Figure \ref{fit1} but for the Disk+Wind H17 model.\label{fit2}}
\end{figure*}

\begin{figure*}
\centering
\includegraphics[width=0.7\columnwidth]{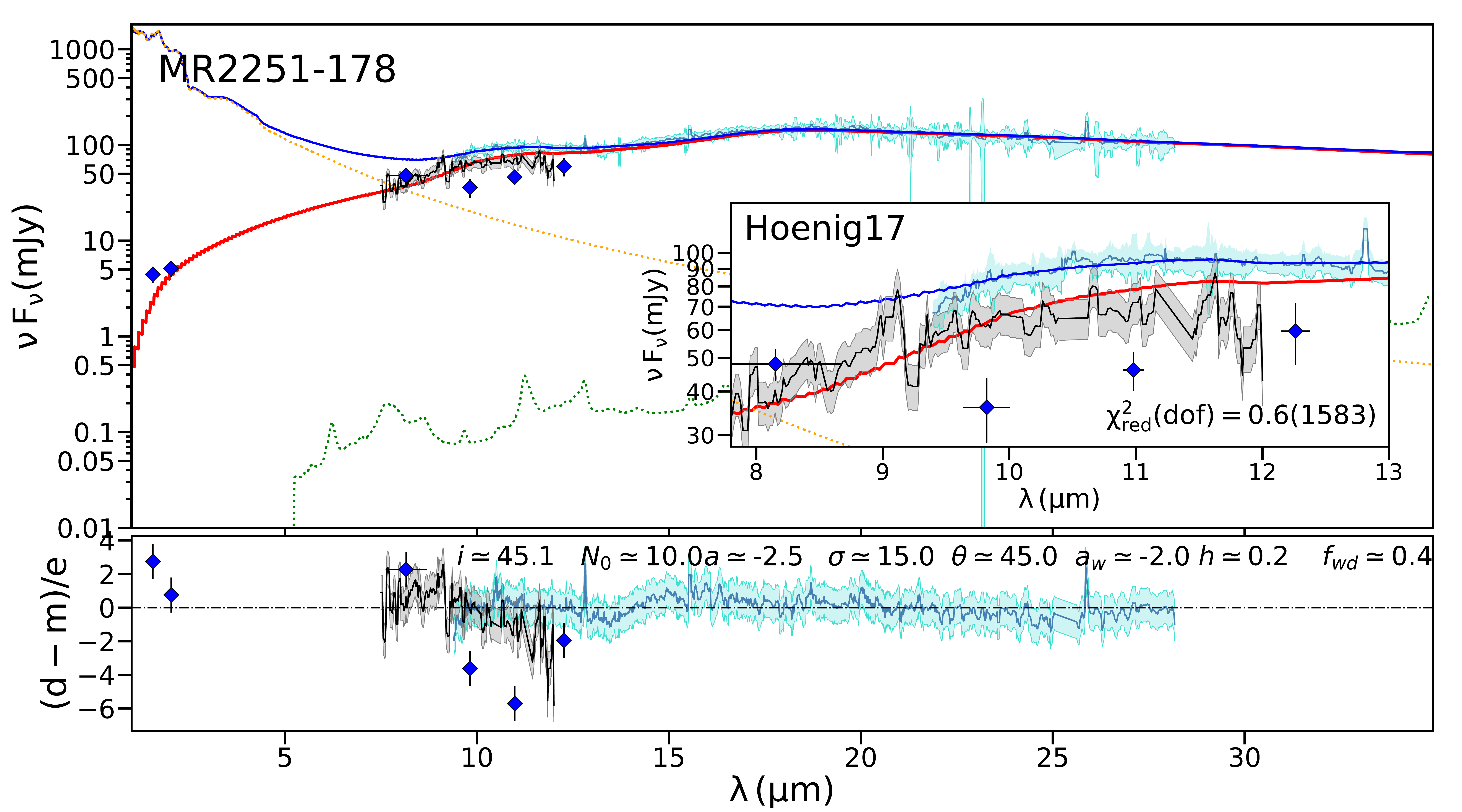}\\
\includegraphics[width=0.7\columnwidth]{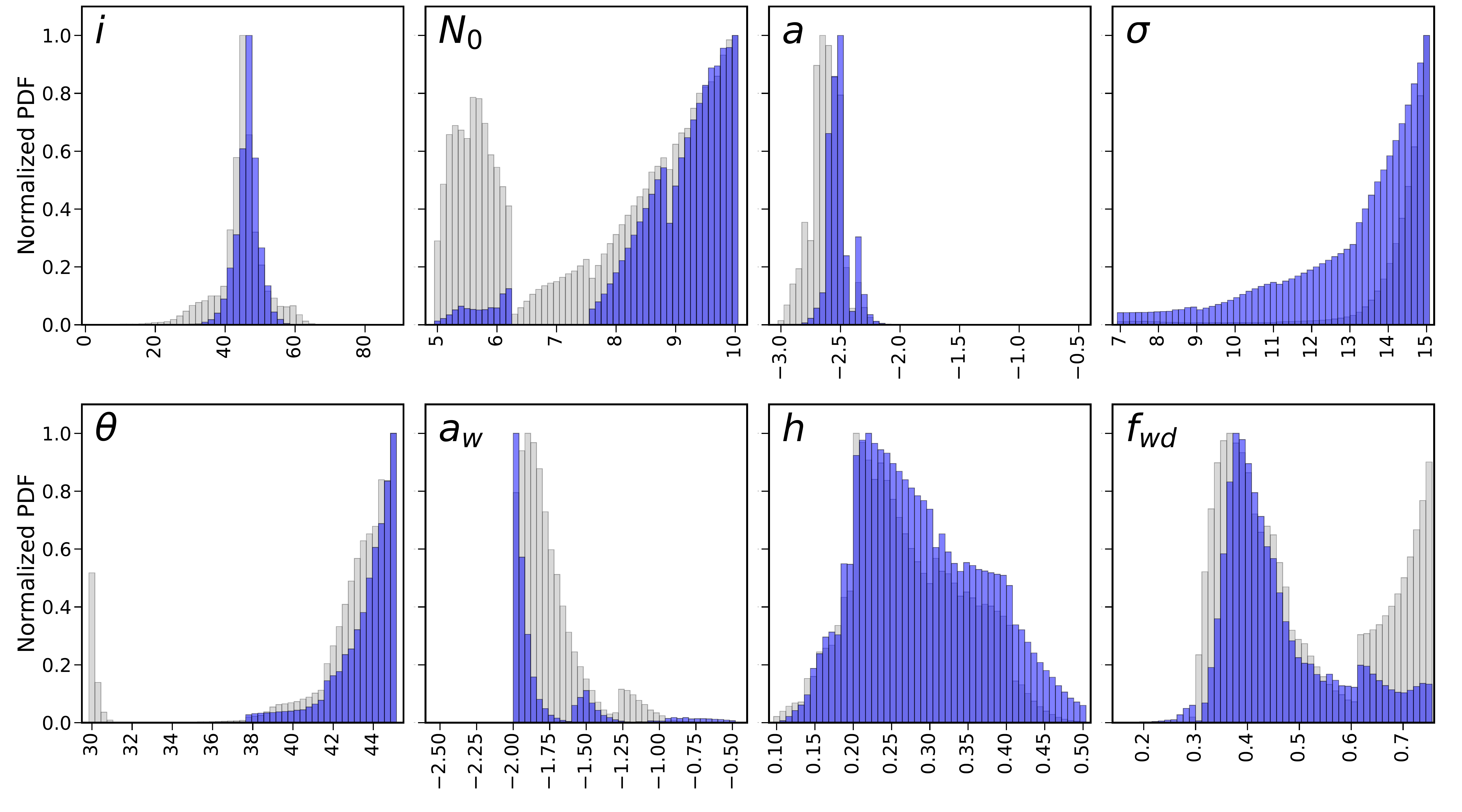}\\
\begin{tabular}{cc}
\includegraphics[width=0.3\columnwidth]{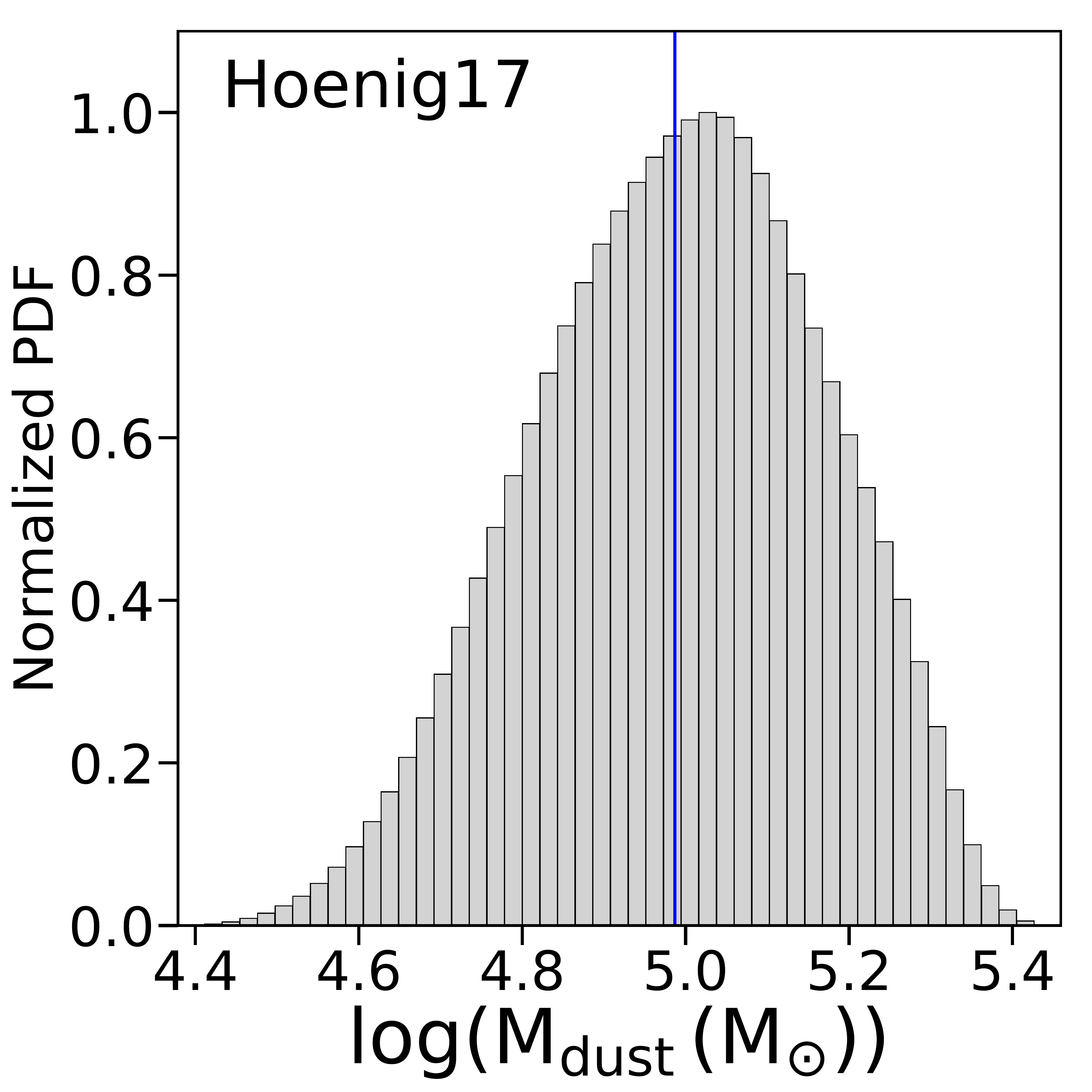}&\includegraphics[width=0.3\columnwidth]{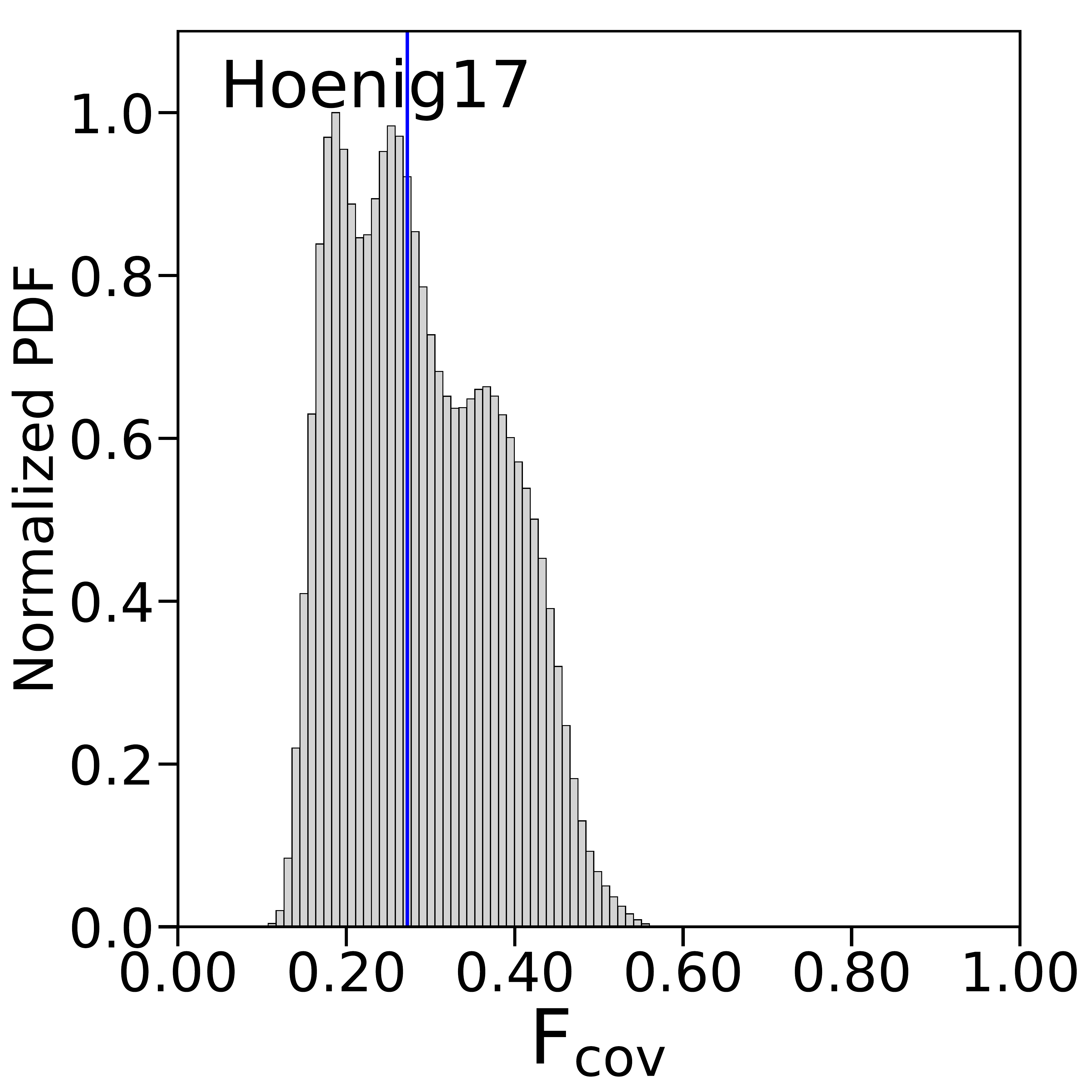}\\
\end{tabular}
\caption{{\bf The SED of MR~2251-178 best modeled by the Disk+Wind H17 model}. {\bf Upper panel}: the high angular resolution photometric points are plotted as blue points with their error bars, the black arrows are upper limits. The high angular resolution spectrum is plotted with a black solid line, the grey shaded region represent the errors. The {\it Spitzer}/IRS spectrum is plotted with a dark cyan solid line and its error with a cyan shaded region. The red solid line is the best model resulting from 
fit the high angular resolution data. The green and yellow dotted lines are the starburst and stellar components, respectively. The blue solid line represents the sum of the stellar, starburst and torus components that best fit the {\it Spitzer}/IRS spectrum.  {\bf Middle panel}. {\bf The model parameters derived}: normal probability distribution function of the free parameters. In grey we plot the parameters derived from fit the LSR spectrum, while in dark blue the distribution of the parameters derived from fit the FSR spectrum. {\bf Bottom panel}: normal PDF of the derived parameters. The blue vertical line indicate the mean.\label{fit3}}
\end{figure*}

\begin{figure*}
\centering
\includegraphics[width=0.7\columnwidth]{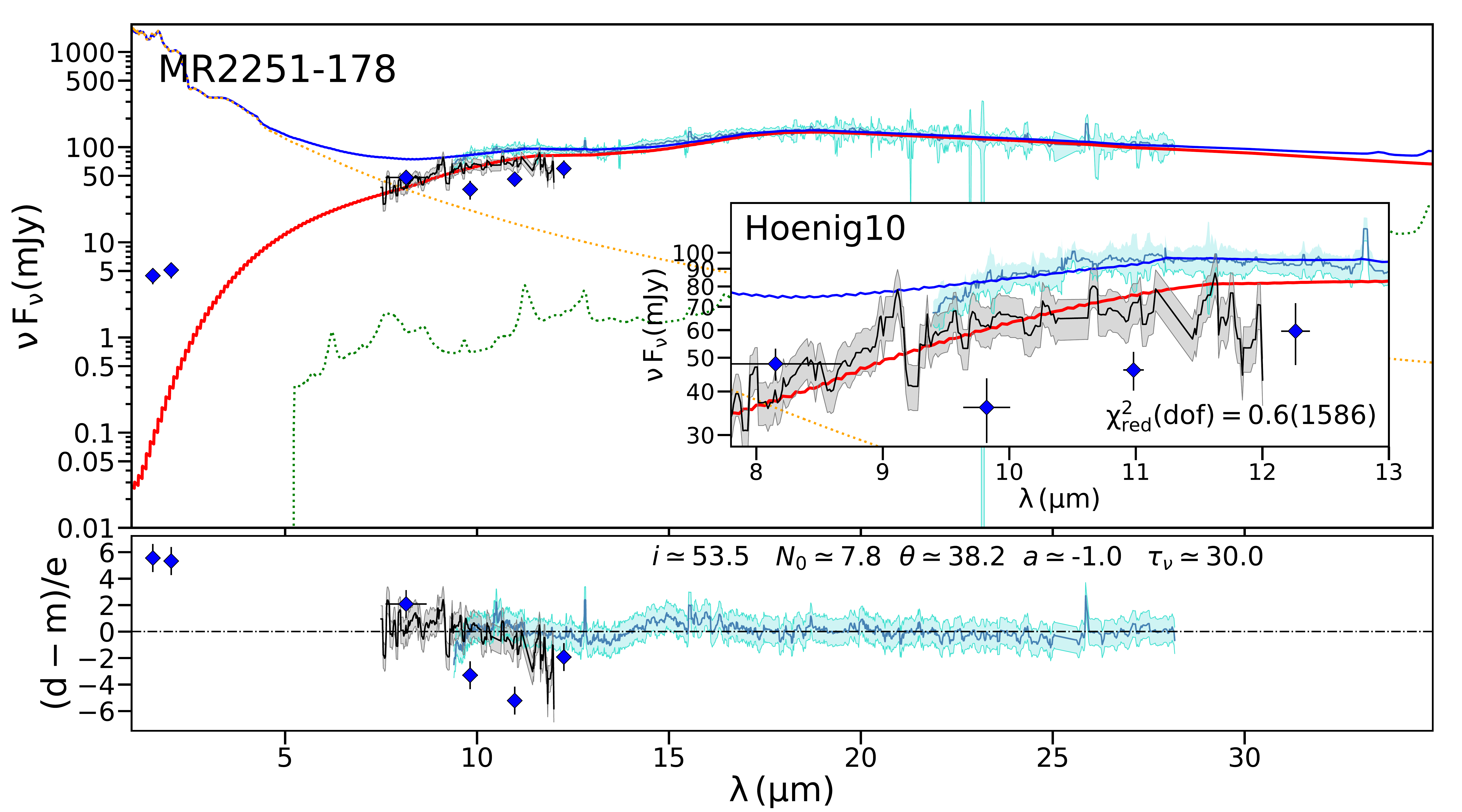}\\
\includegraphics[width=0.7\columnwidth]{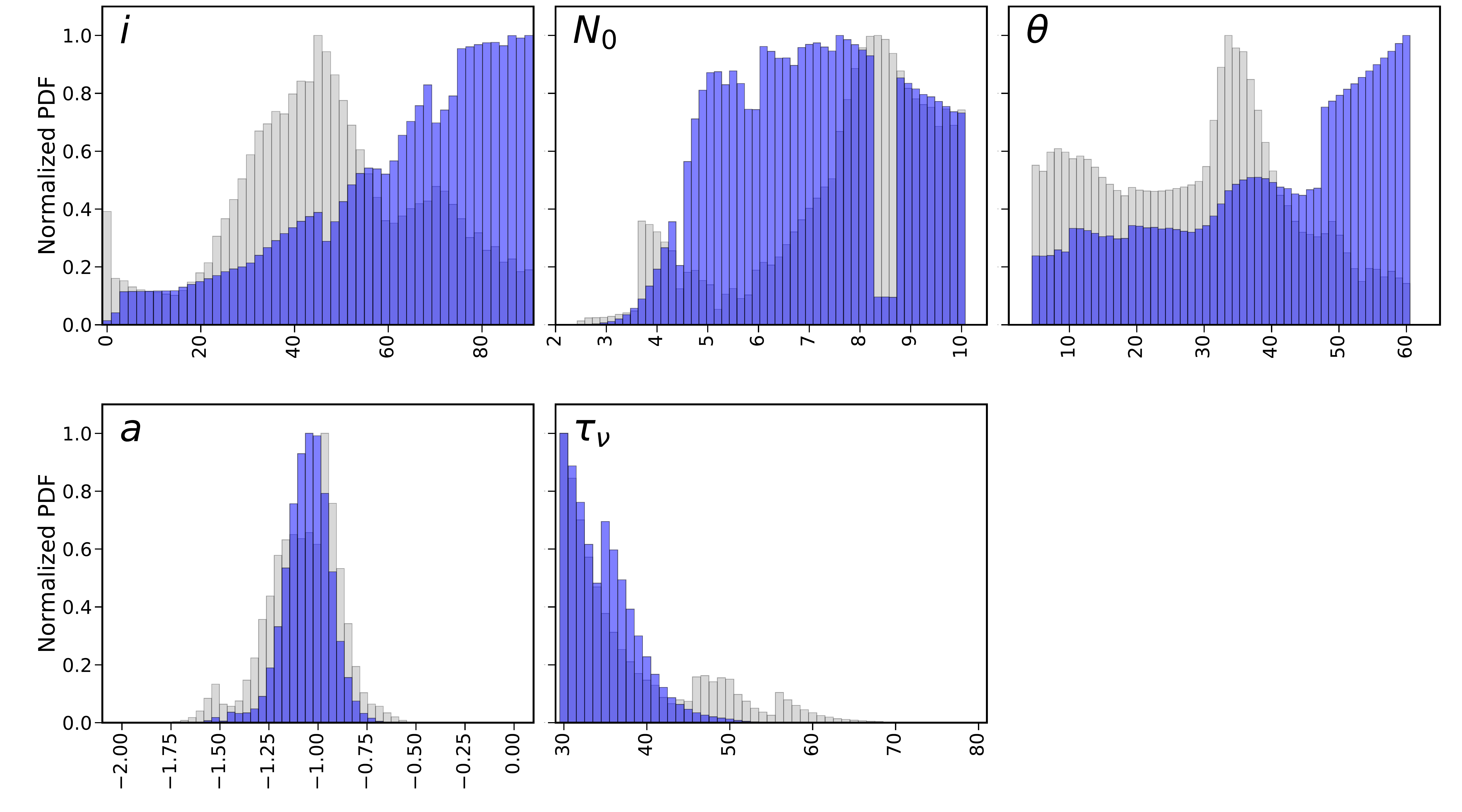}\\
\begin{tabular}{cc}
\includegraphics[width=0.3\columnwidth]{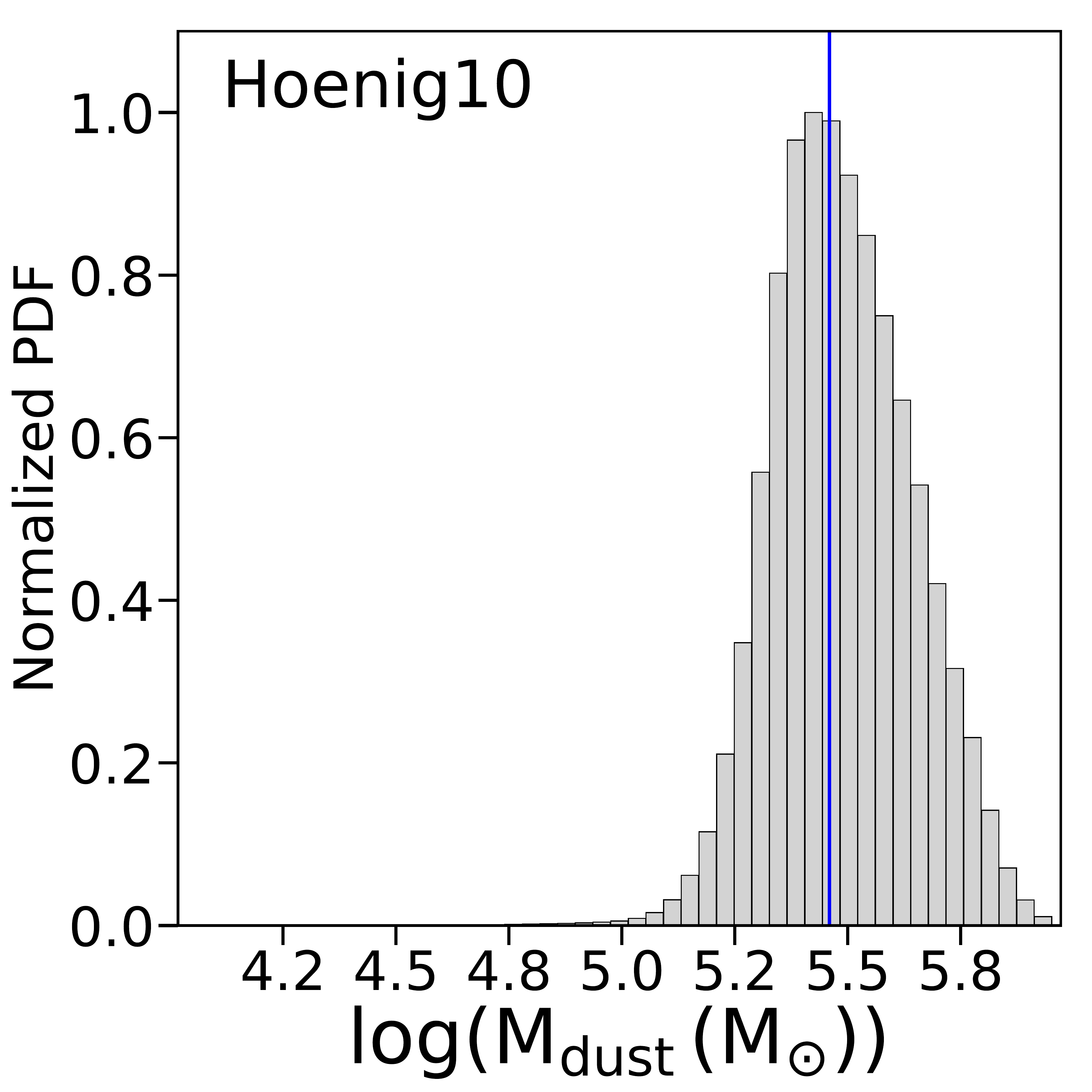}&\includegraphics[width=0.3\columnwidth]{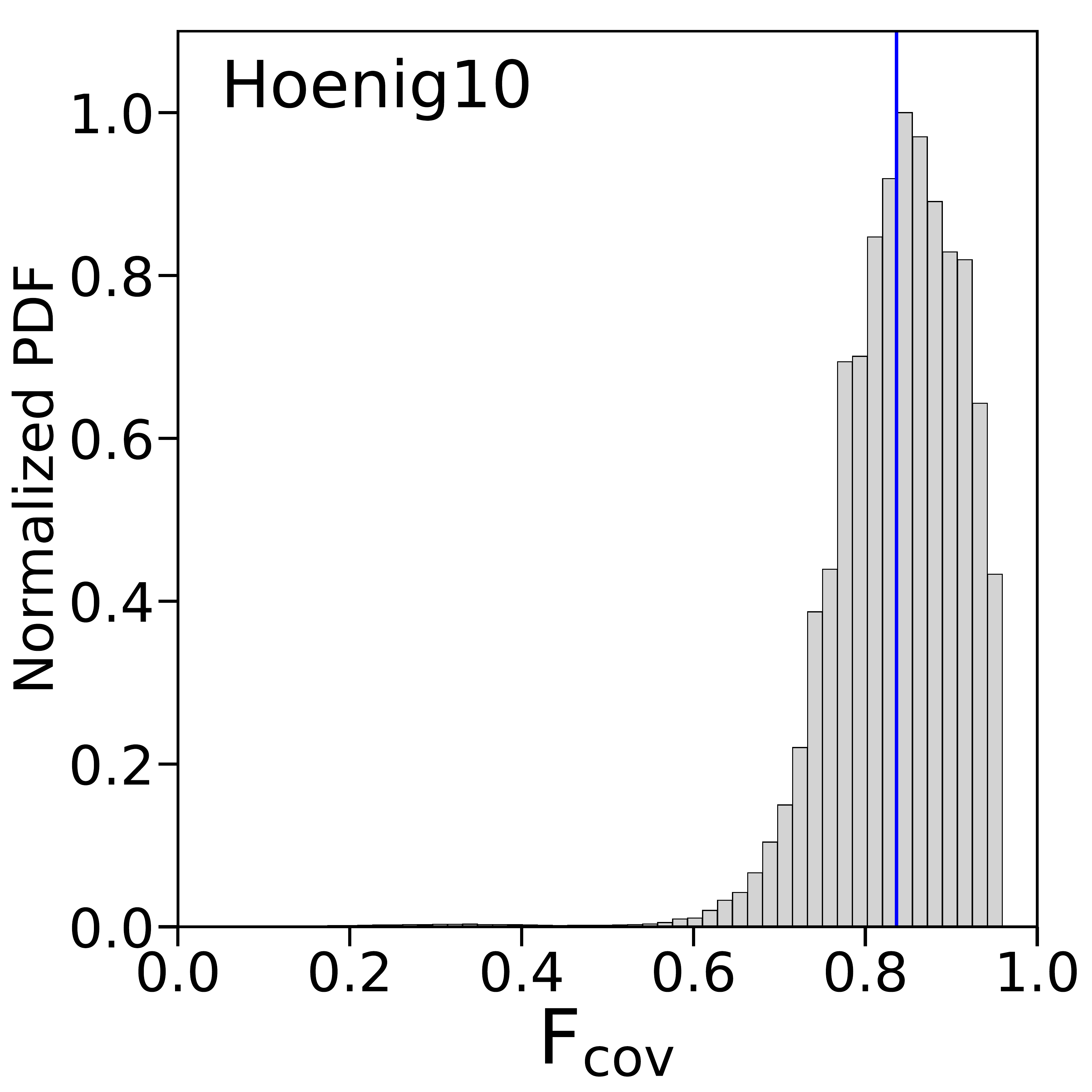}\\
\end{tabular}
\caption{As Figure~\ref{fit3} but for the Clumpy H10 model.\label{fit4}}
\end{figure*}

\begin{figure*}
\centering
\includegraphics[width=0.7\columnwidth]{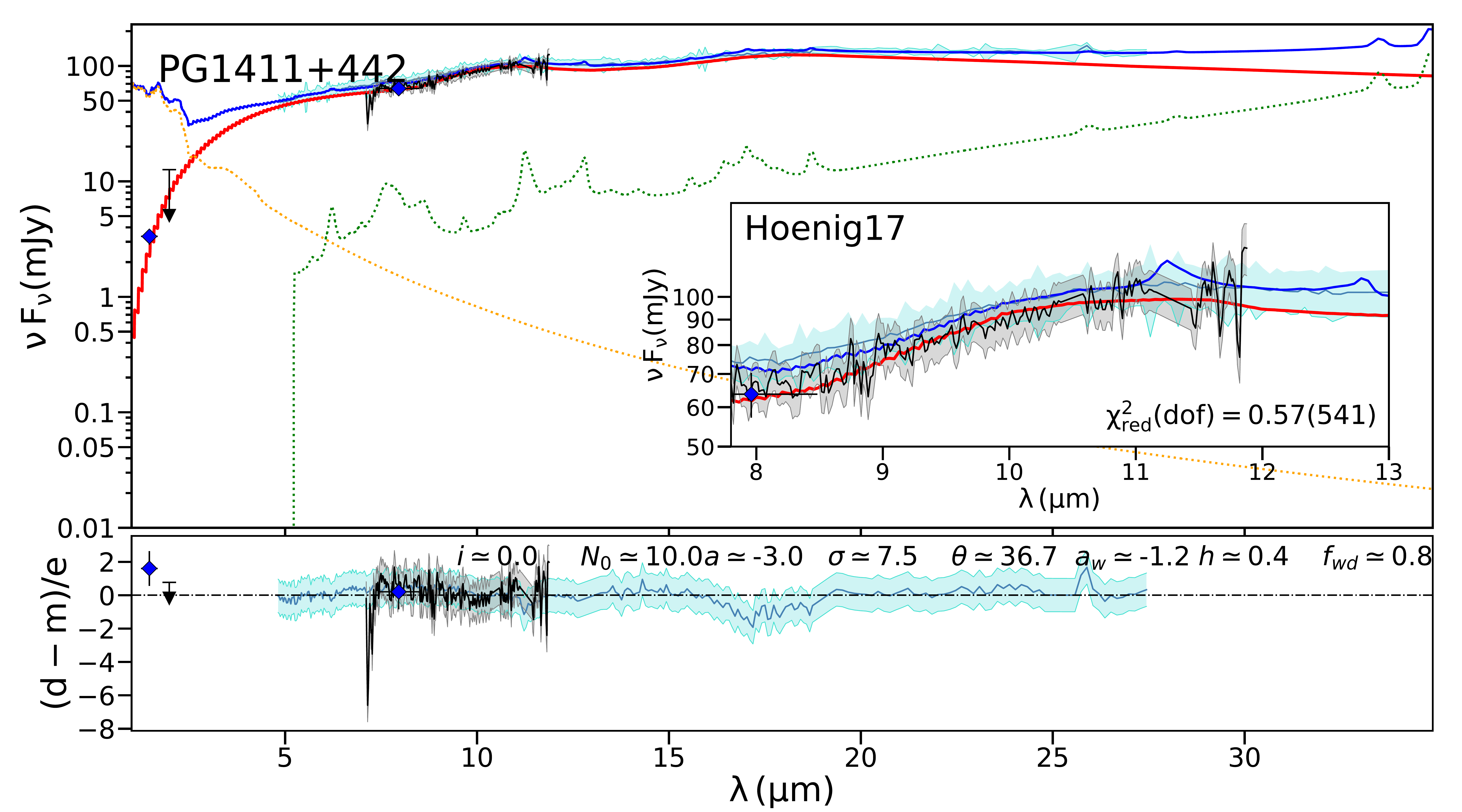}\\
\includegraphics[width=0.7\columnwidth]{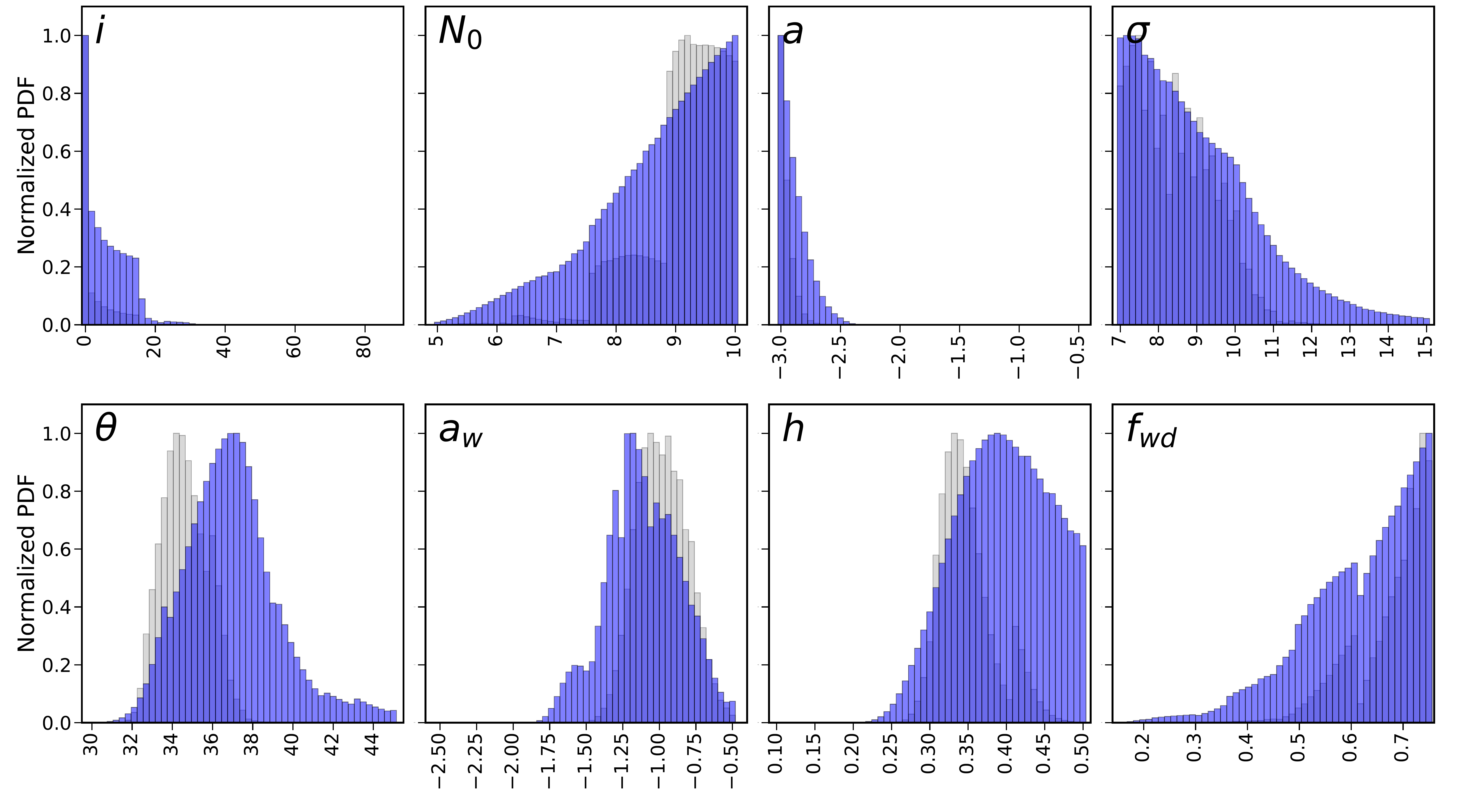}\\
\begin{tabular}{cc}
\includegraphics[width=0.3\columnwidth]{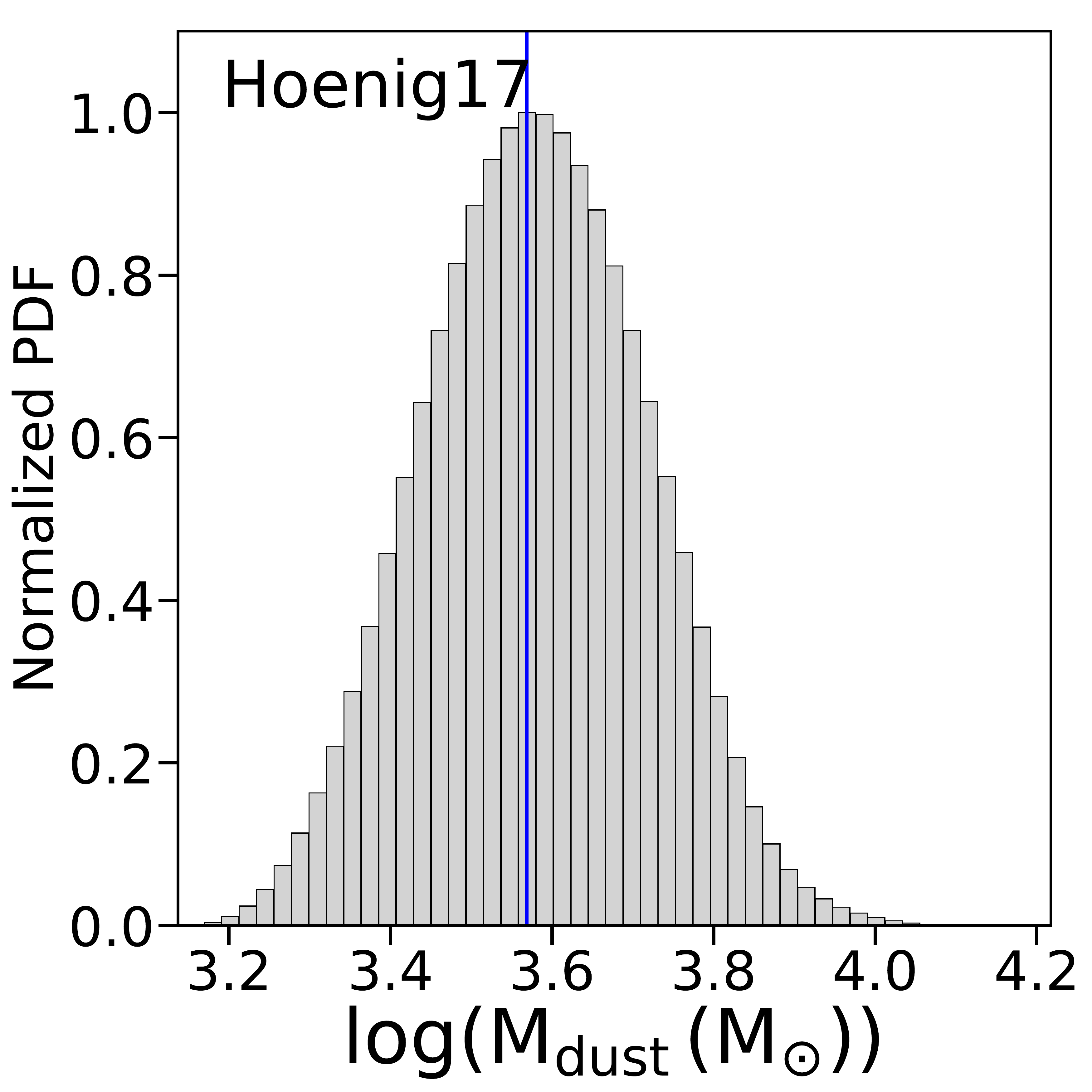}&\includegraphics[width=0.3\columnwidth]{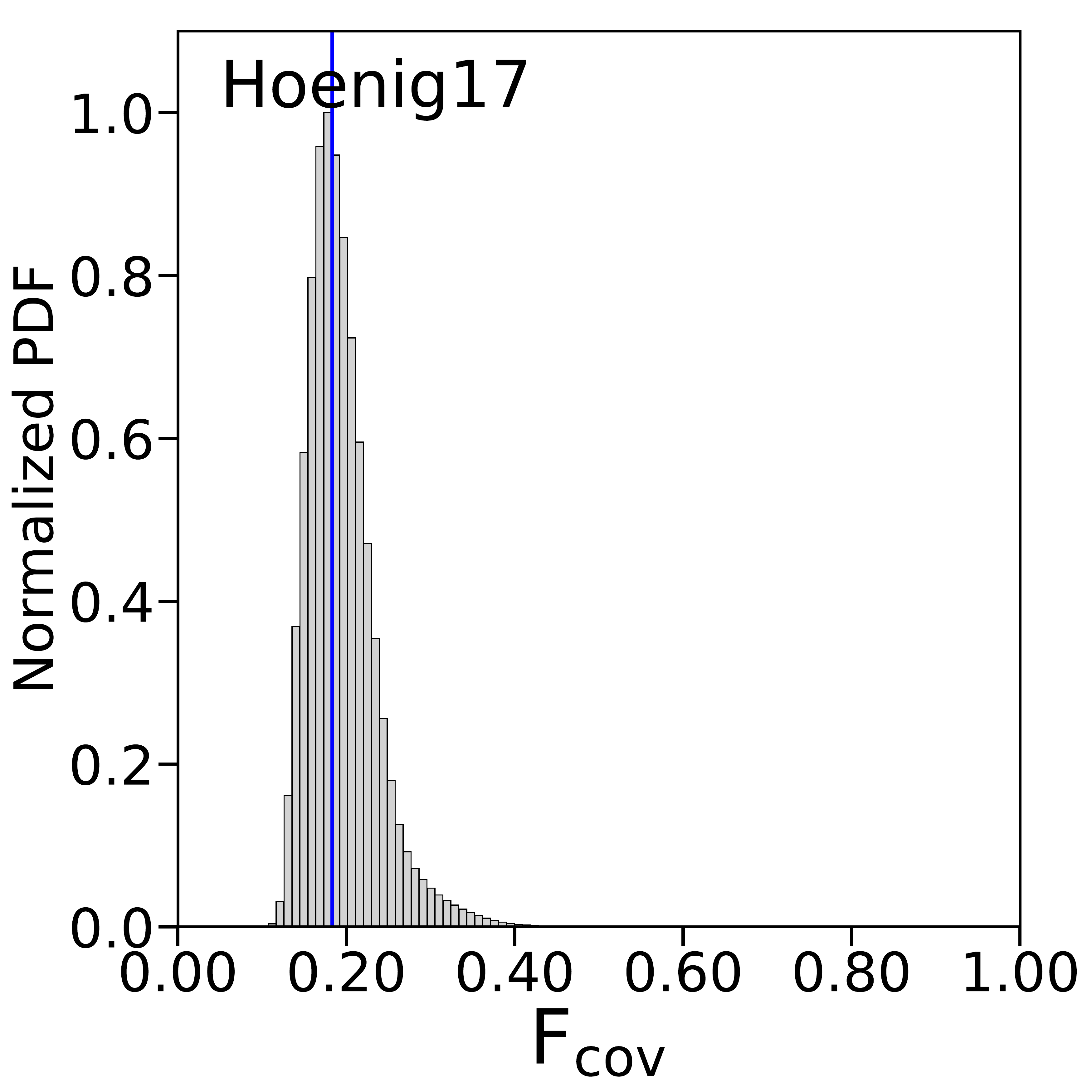}\\
\end{tabular}
\caption{{\bf The SED of PG~1411+442 best modeled by the Disk+Wind H17 model}. {\bf Upper panel}: the high angular resolution photometric points are plotted as blue points with their error bars, the black arrows are upper limits. The high angular resolution spectrum is plotted with a black solid line, the grey shaded region represent the errors. The {\it Spitzer}/IRS spectrum is plotted with a dark cyan solid line and its error with a cyan shaded region. The red solid line is the best model resulting from 
fit the high angular resolution data. The green and yellow dotted lines are the starburst and stellar components, respectively. The blue solid line represents the sum of the stellar, starburst and torus components that best fit the {\it Spitzer}/IRS spectrum. {\bf Middle panel}. {\bf The model parameters derived}: normal probability distribution function of the free parameters. In grey we plot the parameters derived from fit the LSR spectrum, while in dark blue the distribution of the parameters derived from fit the FSR spectrum. {\bf Bottom panel}: normal PDF of the derived parameters. The blue vertical line indicate the mean. \label{fit5}}
\end{figure*}

\begin{figure*}
\centering
\includegraphics[width=0.7\columnwidth]{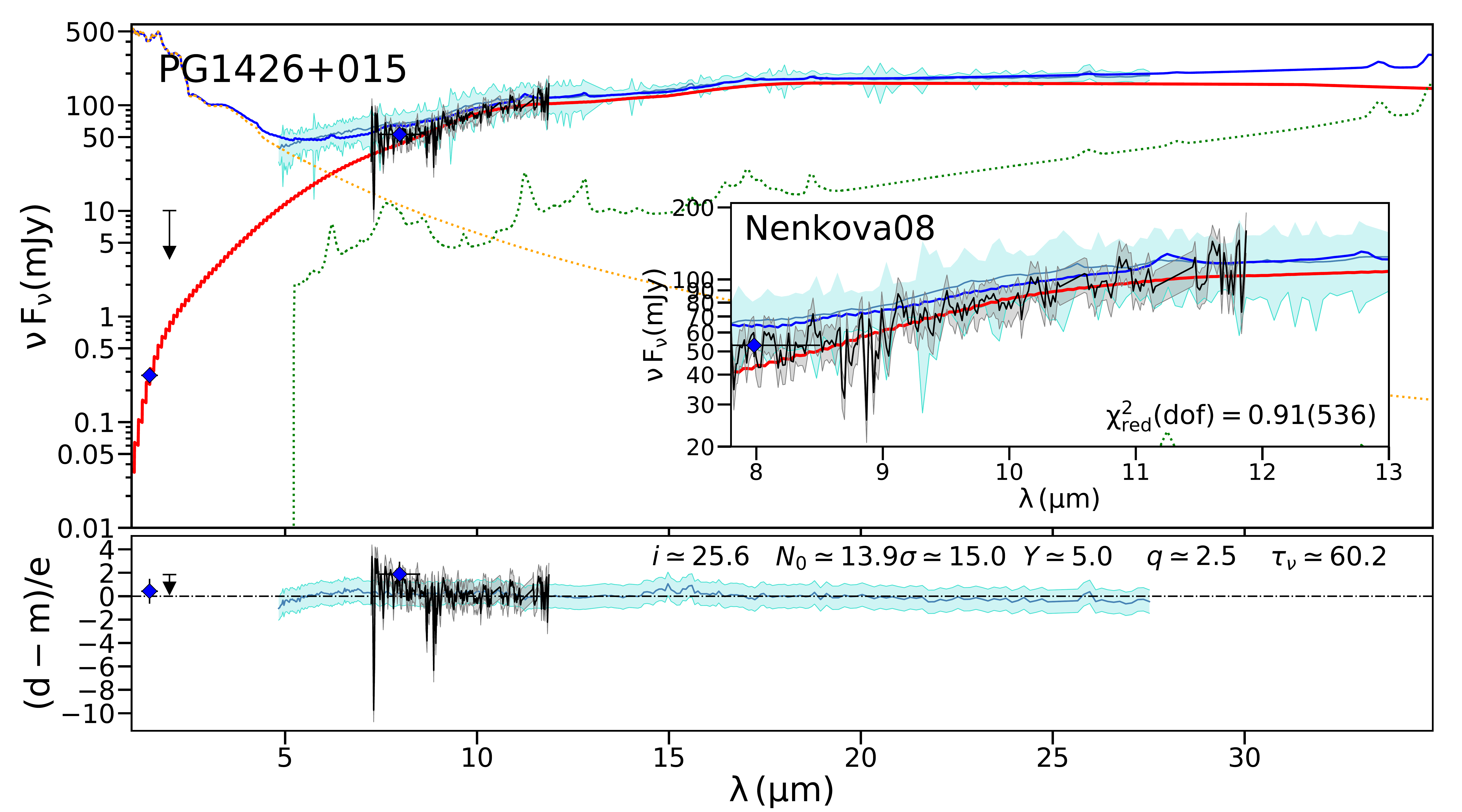}\\
\includegraphics[width=0.7\columnwidth]{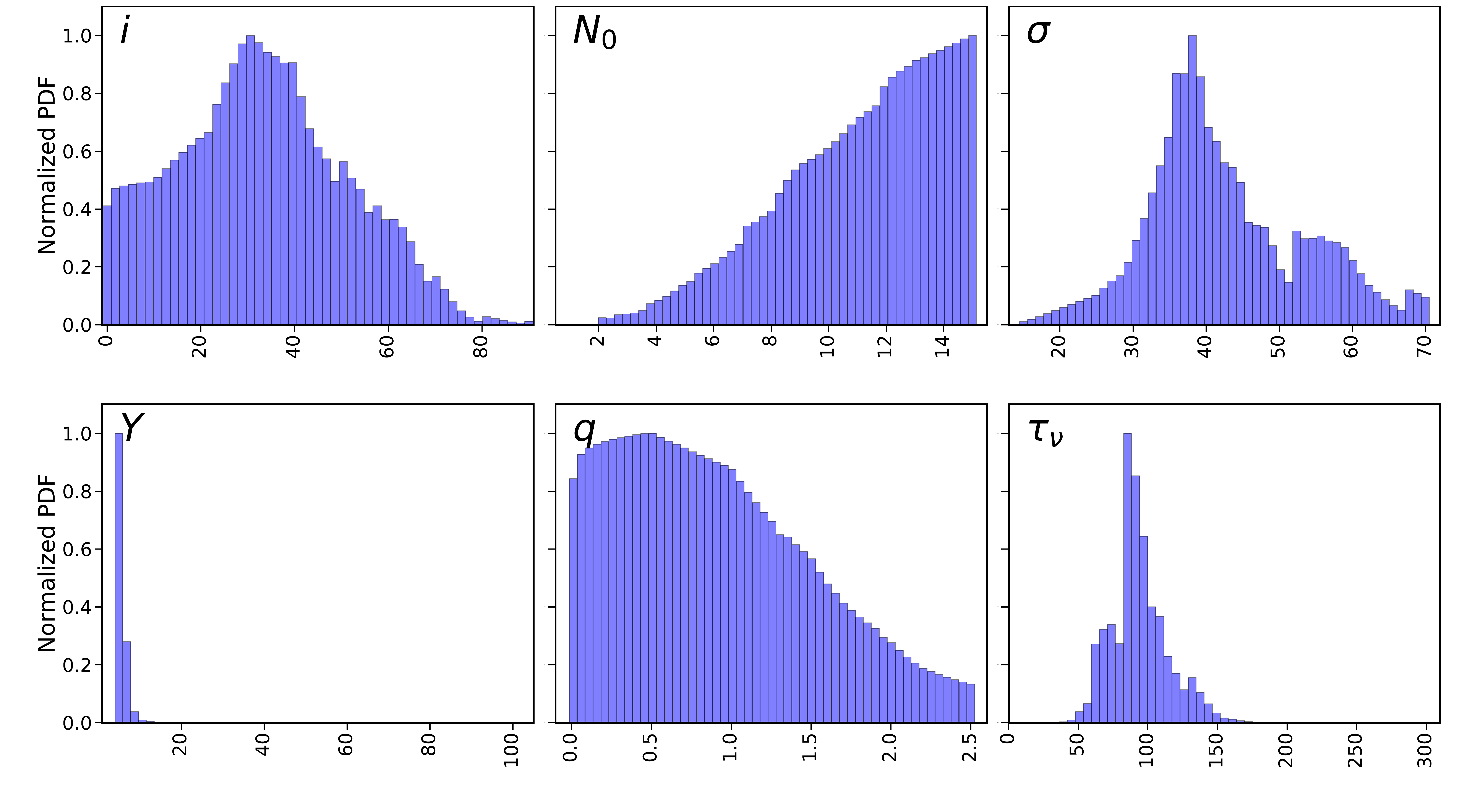}\\
\begin{tabular}{ccc}
\includegraphics[width=0.3\columnwidth]{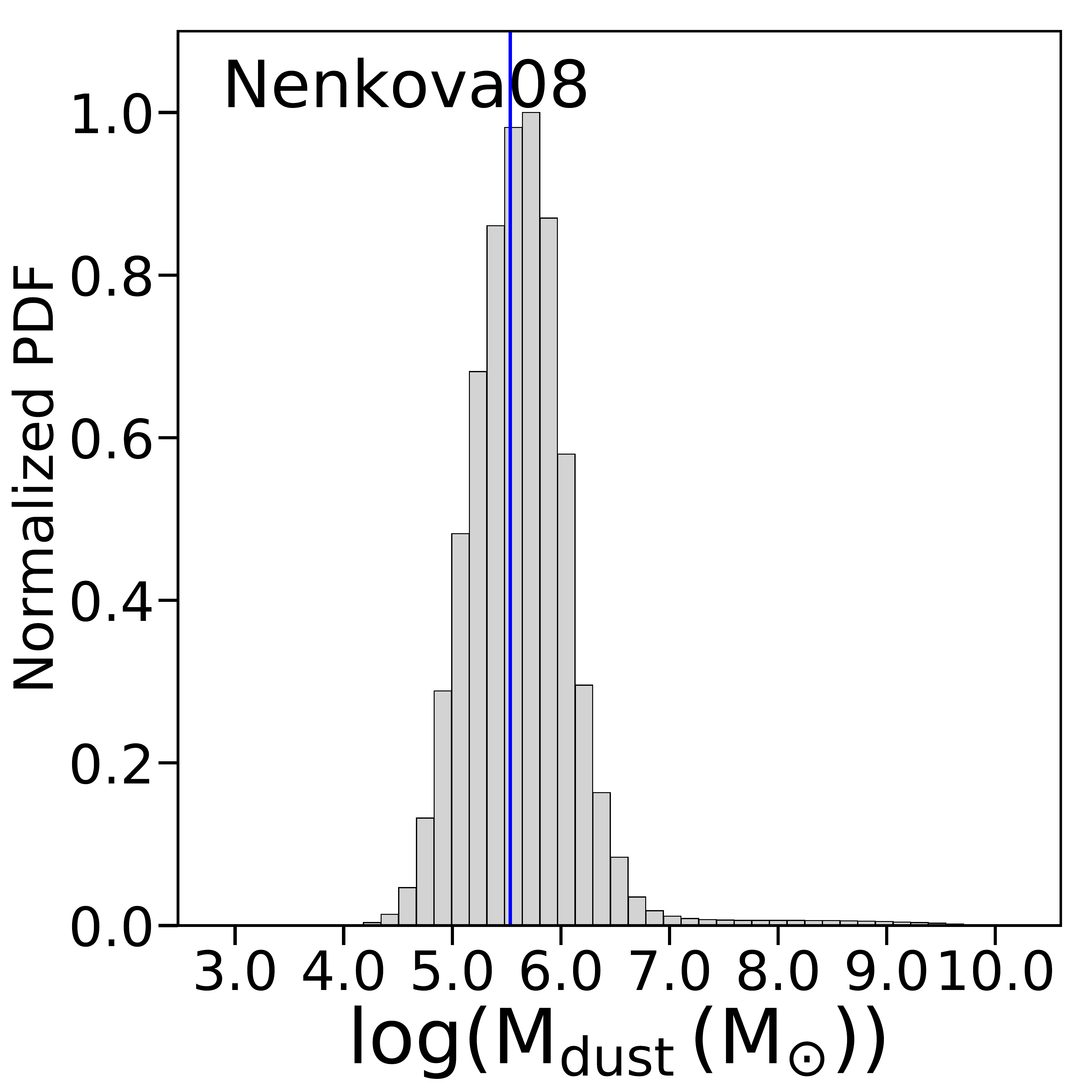}&\includegraphics[width=0.3\columnwidth]{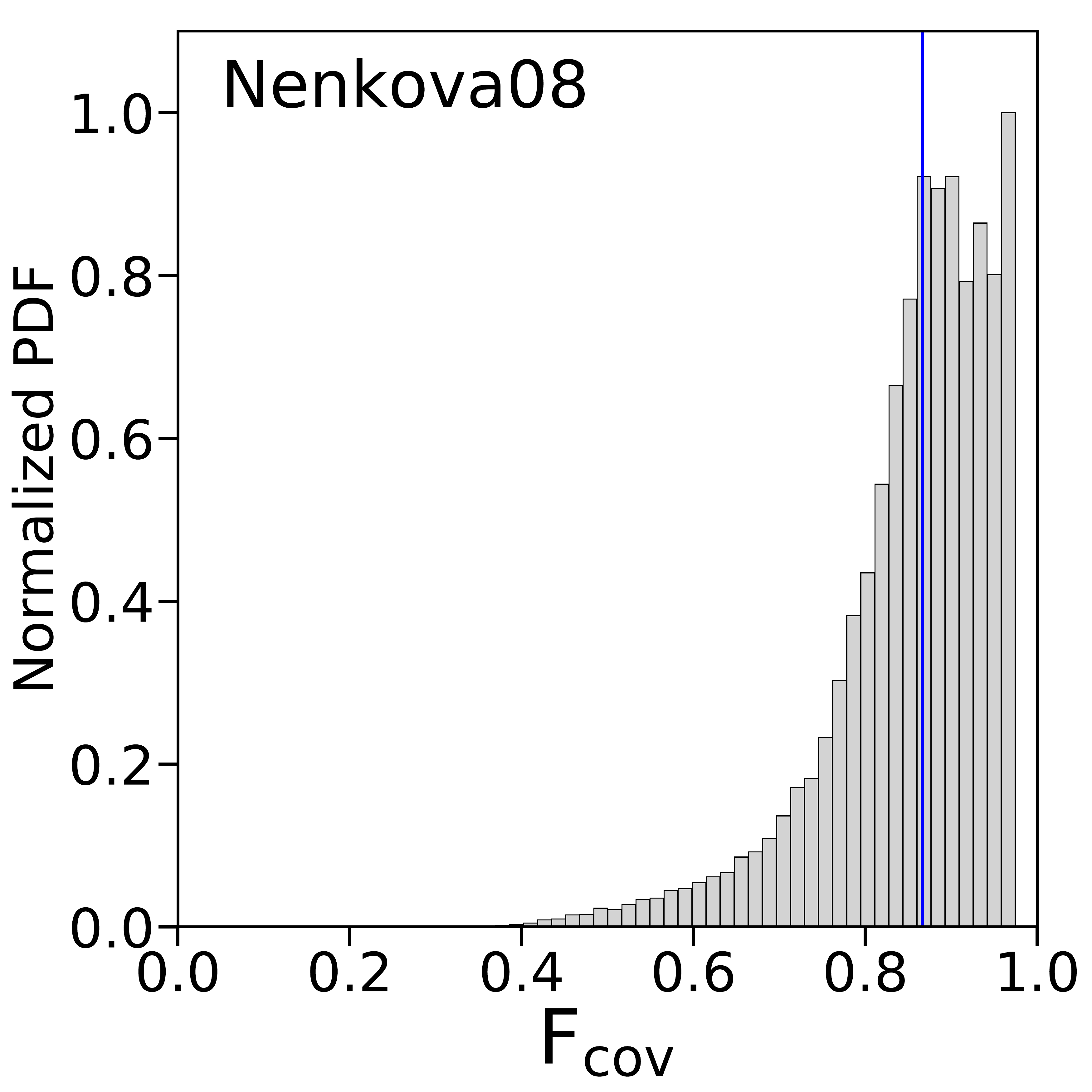}&
\includegraphics[width=0.3\columnwidth]{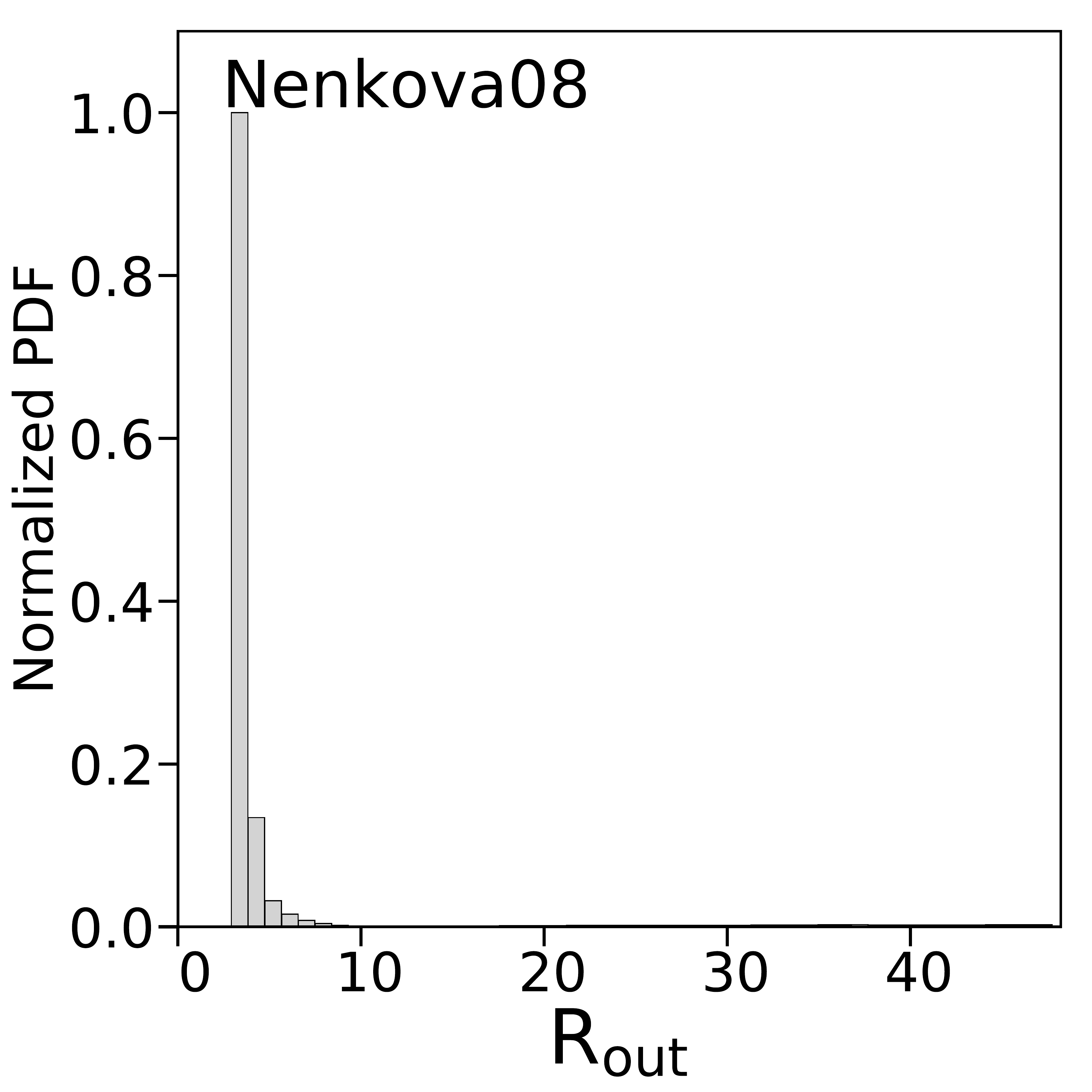}\\
\end{tabular}
\caption{{\bf The SED of PG~1426+015 best modeled by the Clumpy N08 model}. {\bf Upper panel}: the high angular resolution photometric points are plotted as blue points with their error bars, the black arrows are upper limits. The high angular resolution spectrum is plotted with a black solid line, the grey shaded region represent the errors. The {\it Spitzer}/IRS spectrum is plotted with a dark cyan solid line and its error with a cyan shaded region. The red solid line is the best model resulting from 
fit the high angular resolution data. The green and yellow dotted lines are the starburst and stellar components, respectively. The blue solid line represents the sum of the stellar, starburst and torus components that best fit the {\it Spitzer}/IRS spectrum.  {\bf Middle panel}. {\bf The model parameters derived}: normal probability distribution function of the free parameters. In grey we plot the parameters derived from fit the LSR spectrum, while in dark blue the distribution of the parameters derived from fit the FSR spectrum. {\bf Bottom panel}: normal PDF of the derived parameters. The blue vertical line indicate the mean.\label{fit6}}
\end{figure*}

\begin{figure*}
\centering
\includegraphics[width=0.7\columnwidth]{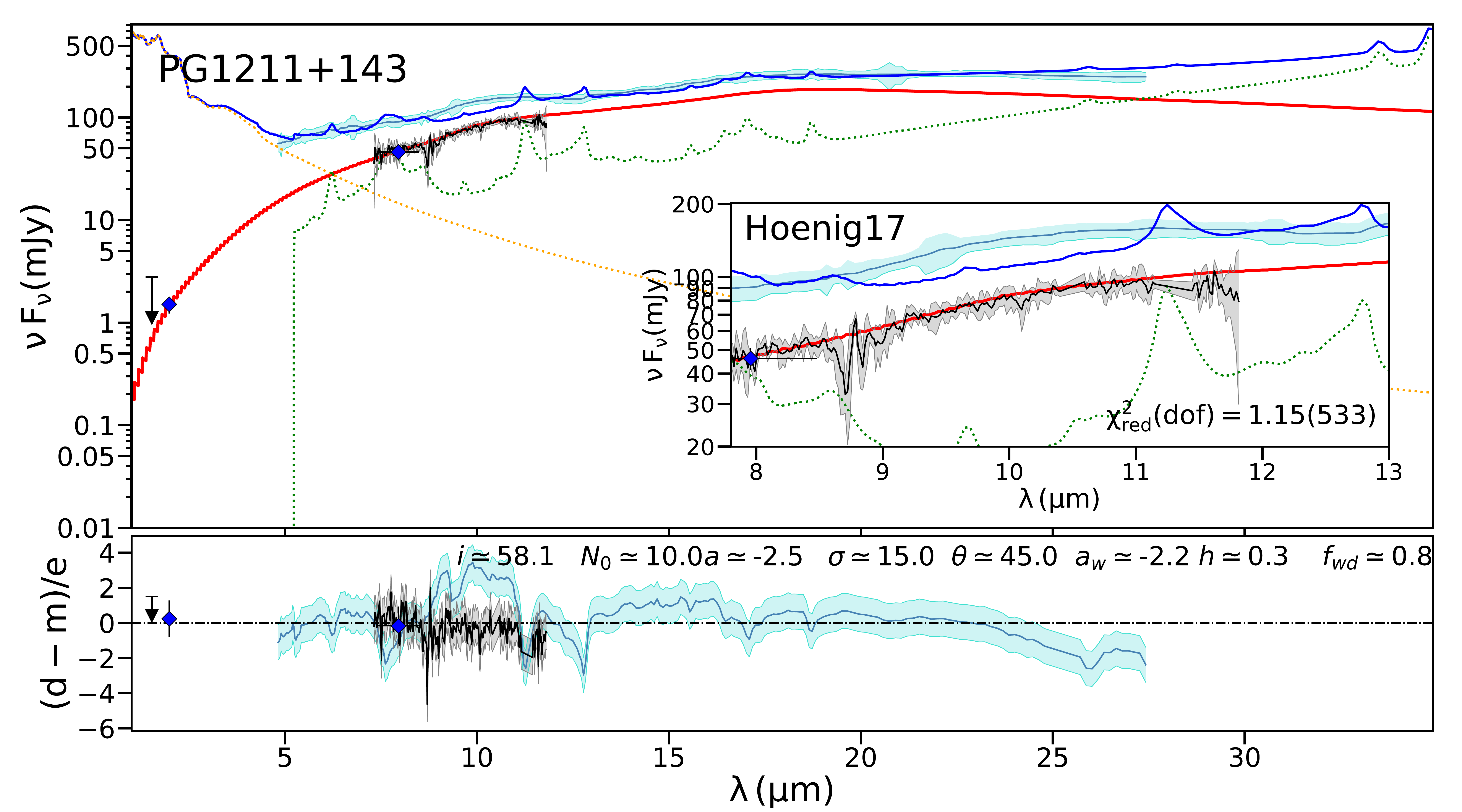}\\
\includegraphics[width=0.7\columnwidth]{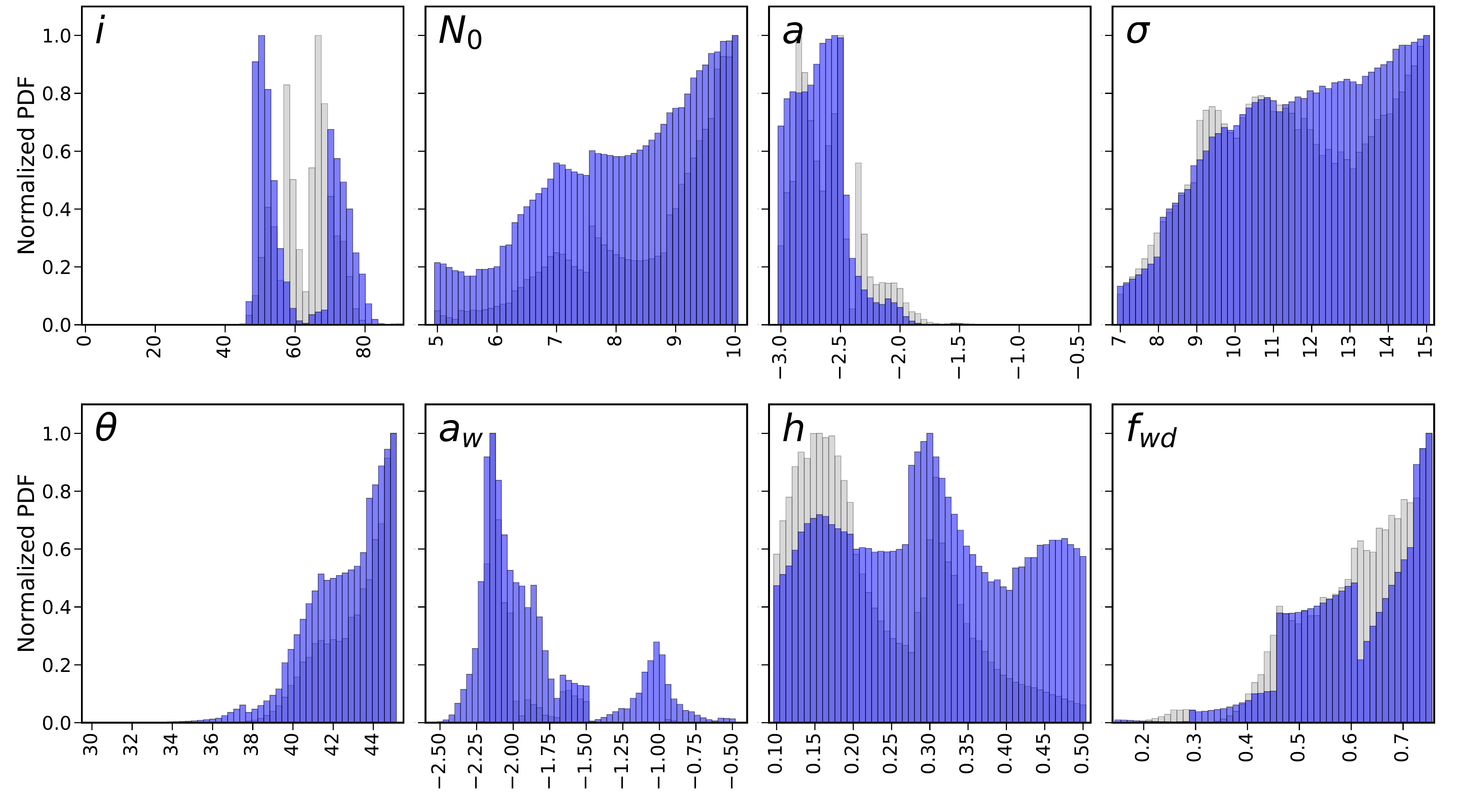}\\
\begin{tabular}{cc}
\includegraphics[width=0.3\columnwidth]{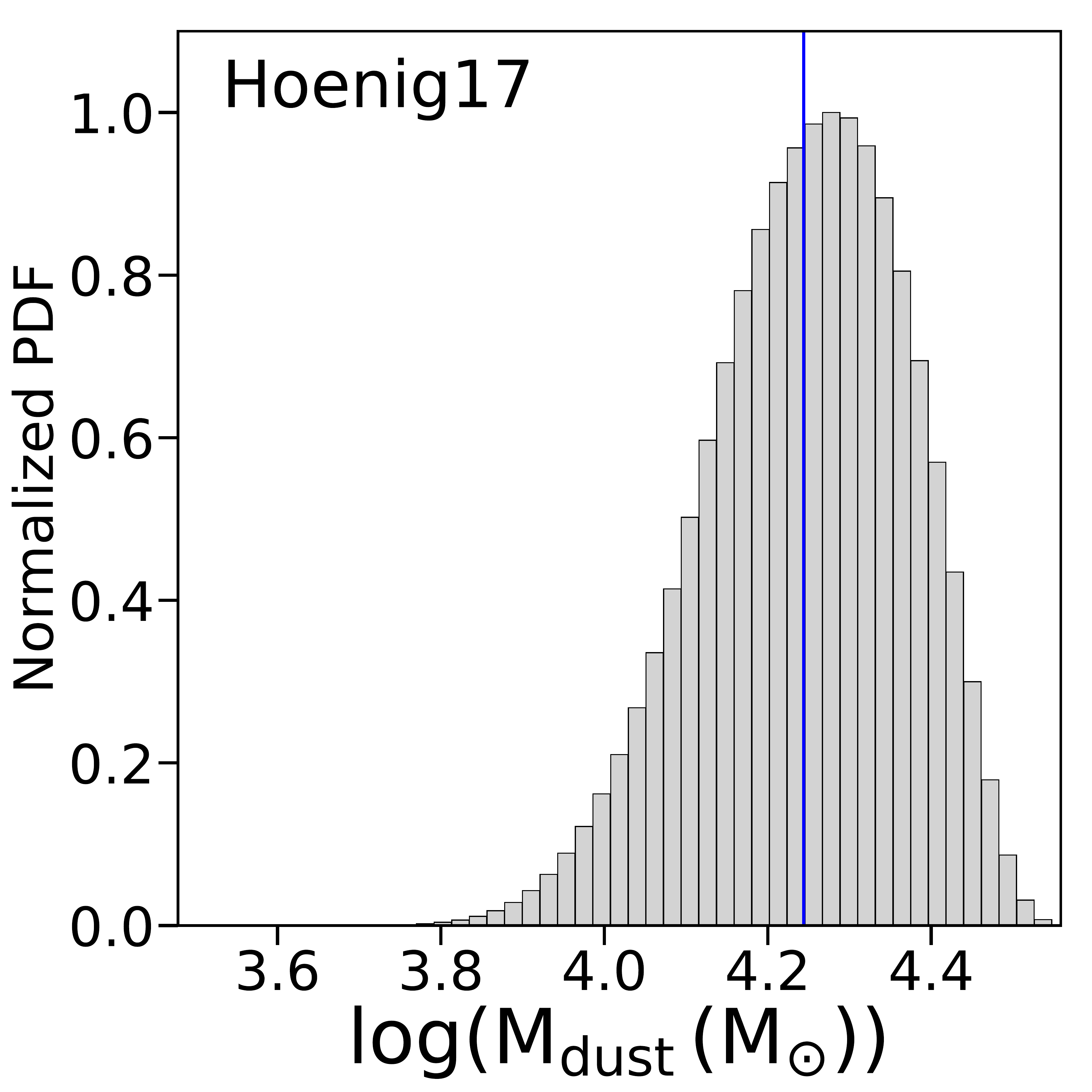}&\includegraphics[width=0.3\columnwidth]{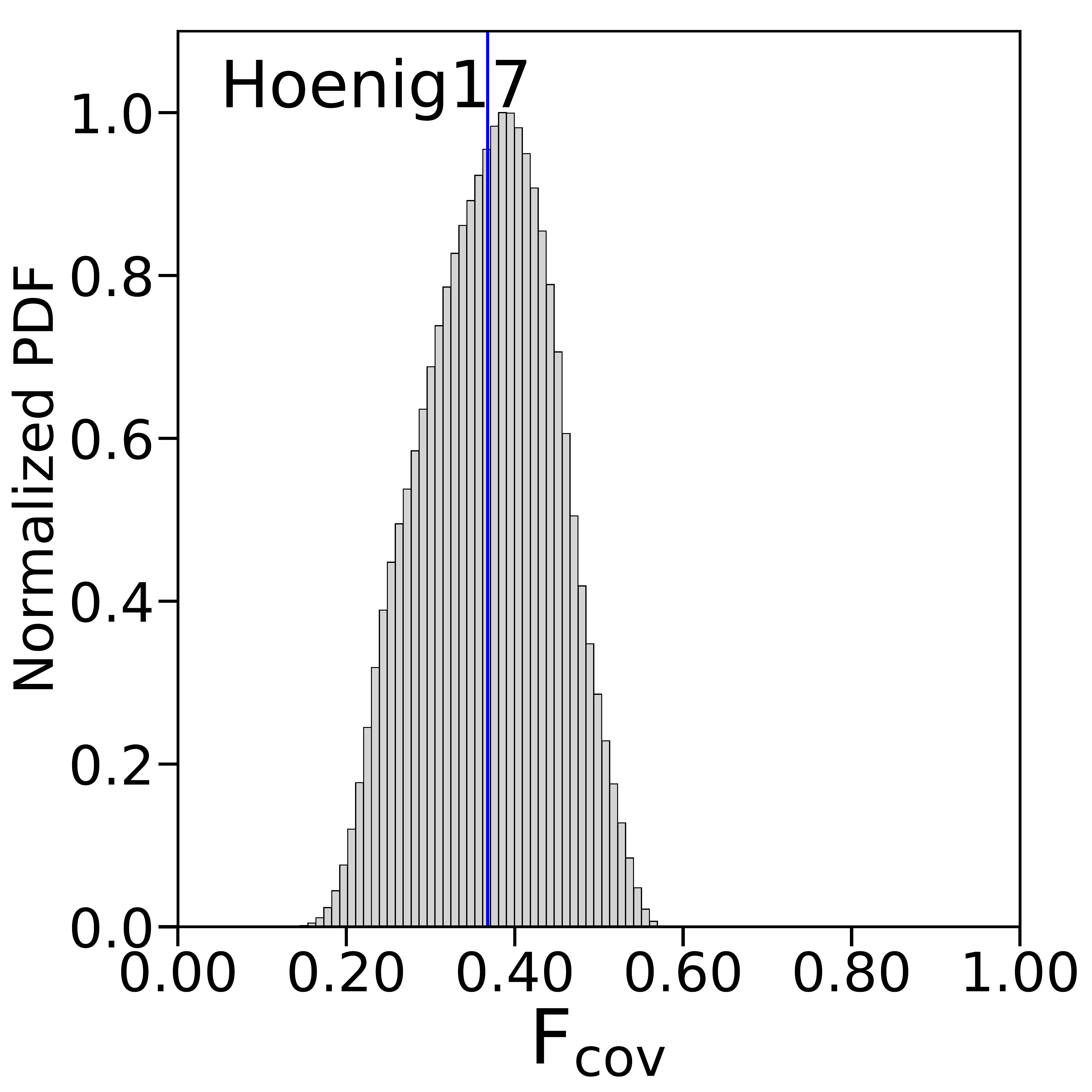}\\
\end{tabular}
\caption{As Figure \ref{fit5} but for the QSO PG~1211+143.\label{fit7}}
\end{figure*}

\begin{figure*}
\centering
\includegraphics[width=0.7\columnwidth]{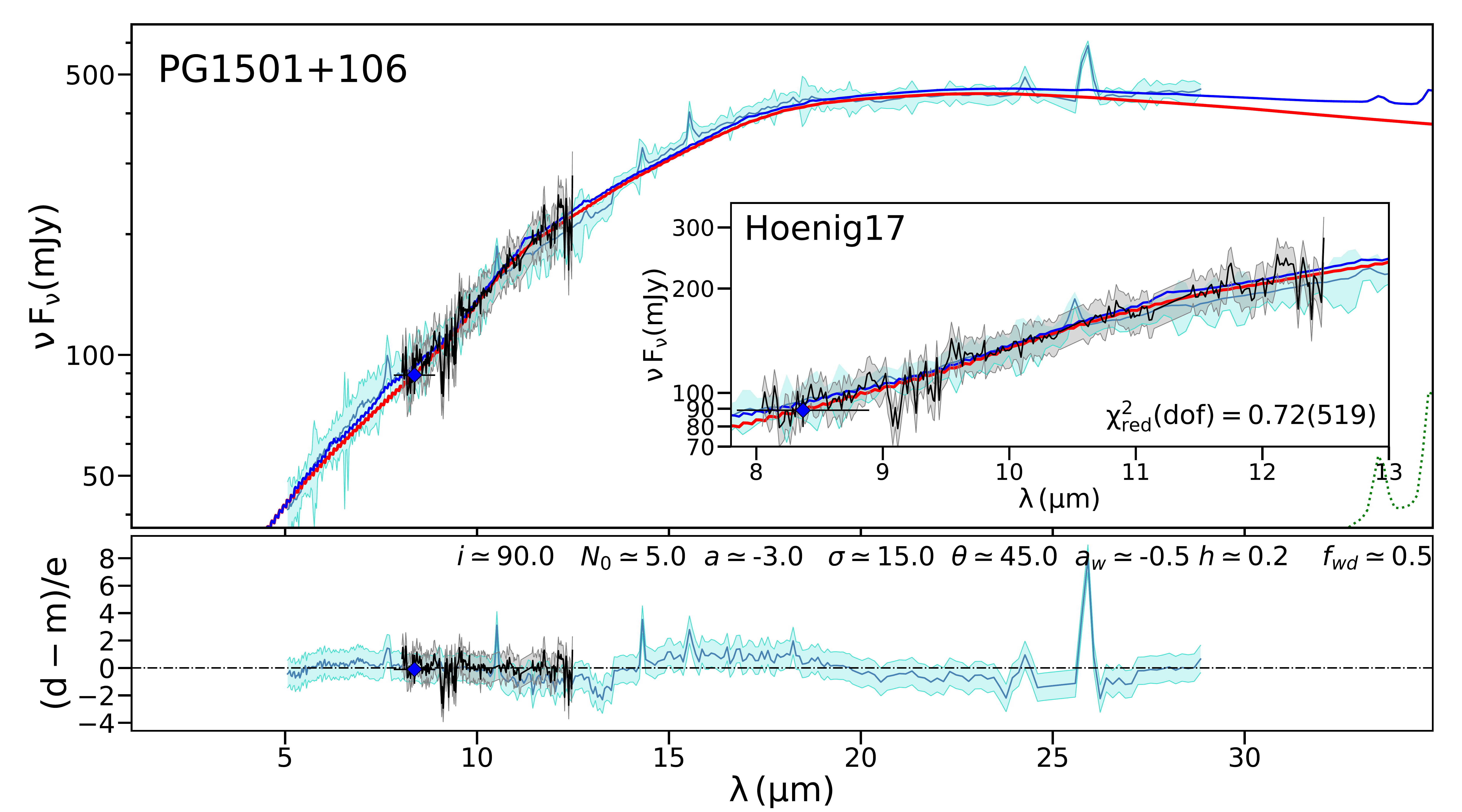}\\
\includegraphics[width=0.7\columnwidth]{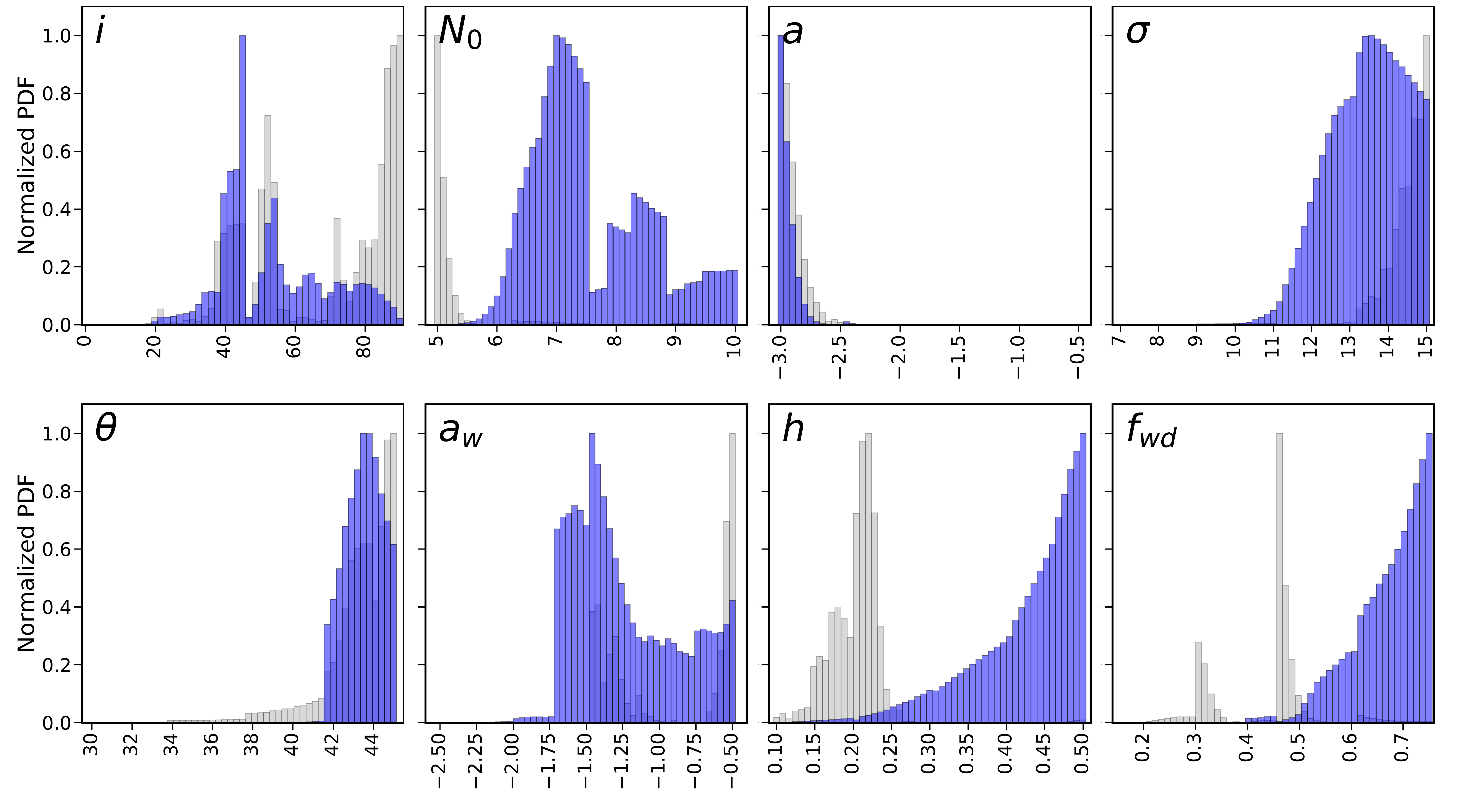}\\
\begin{tabular}{cc}
\includegraphics[width=0.3\columnwidth]{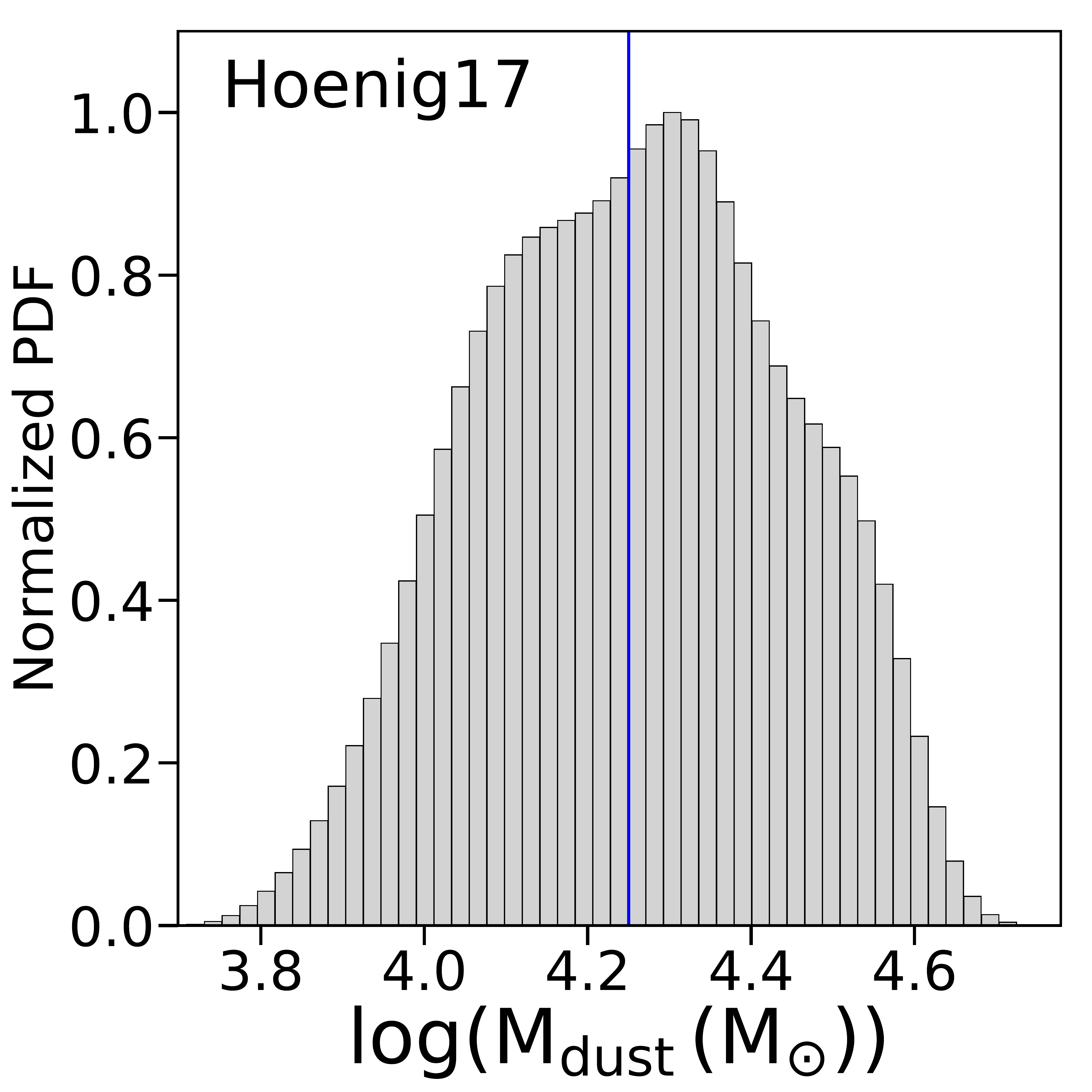}&\includegraphics[width=0.3\columnwidth]{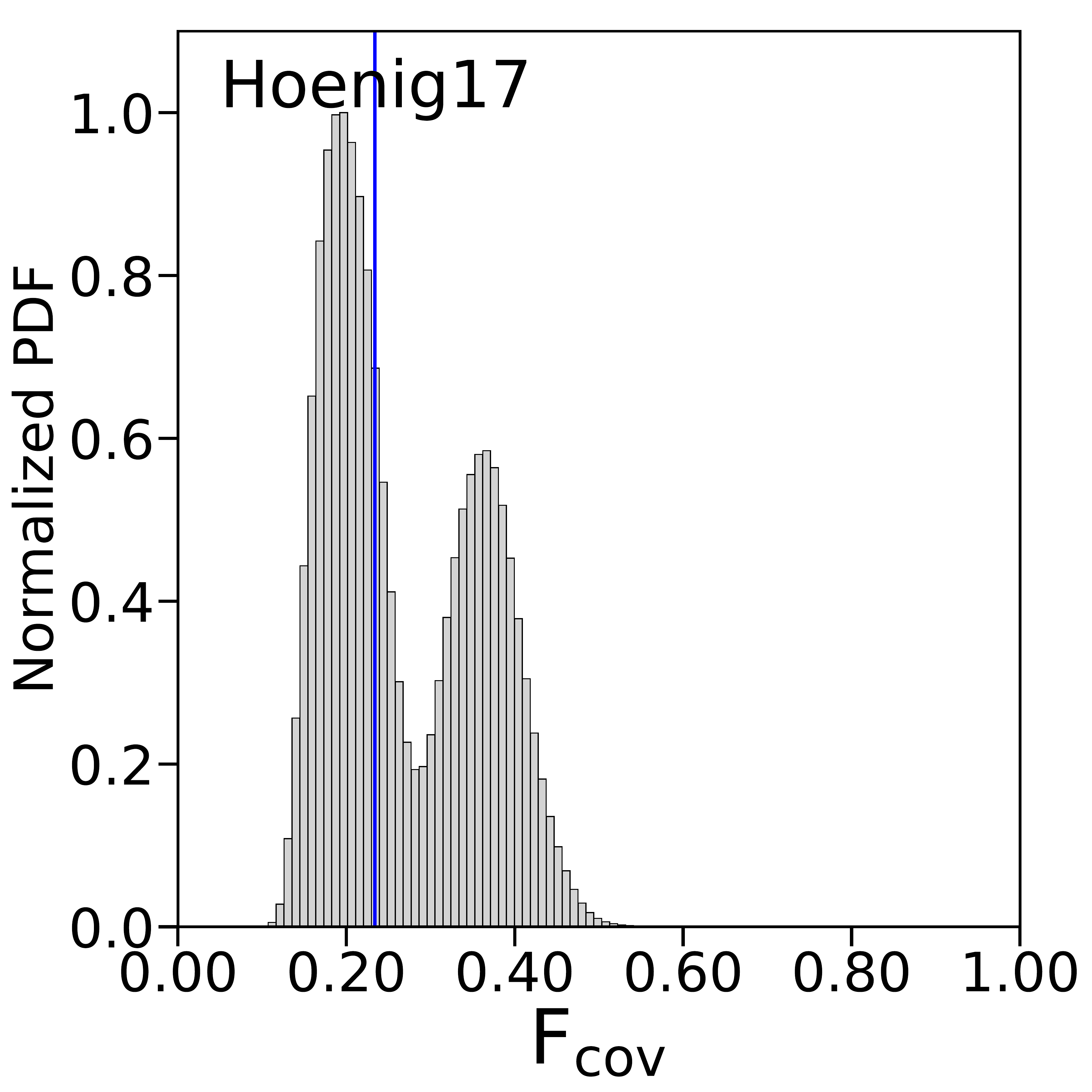}\\
\end{tabular}
\caption{As Figure \ref{fit5} but for the QSO PG~1501+106.\label{fit8}}
\end{figure*}

\begin{figure*}
\centering
\includegraphics[width=0.7\columnwidth]{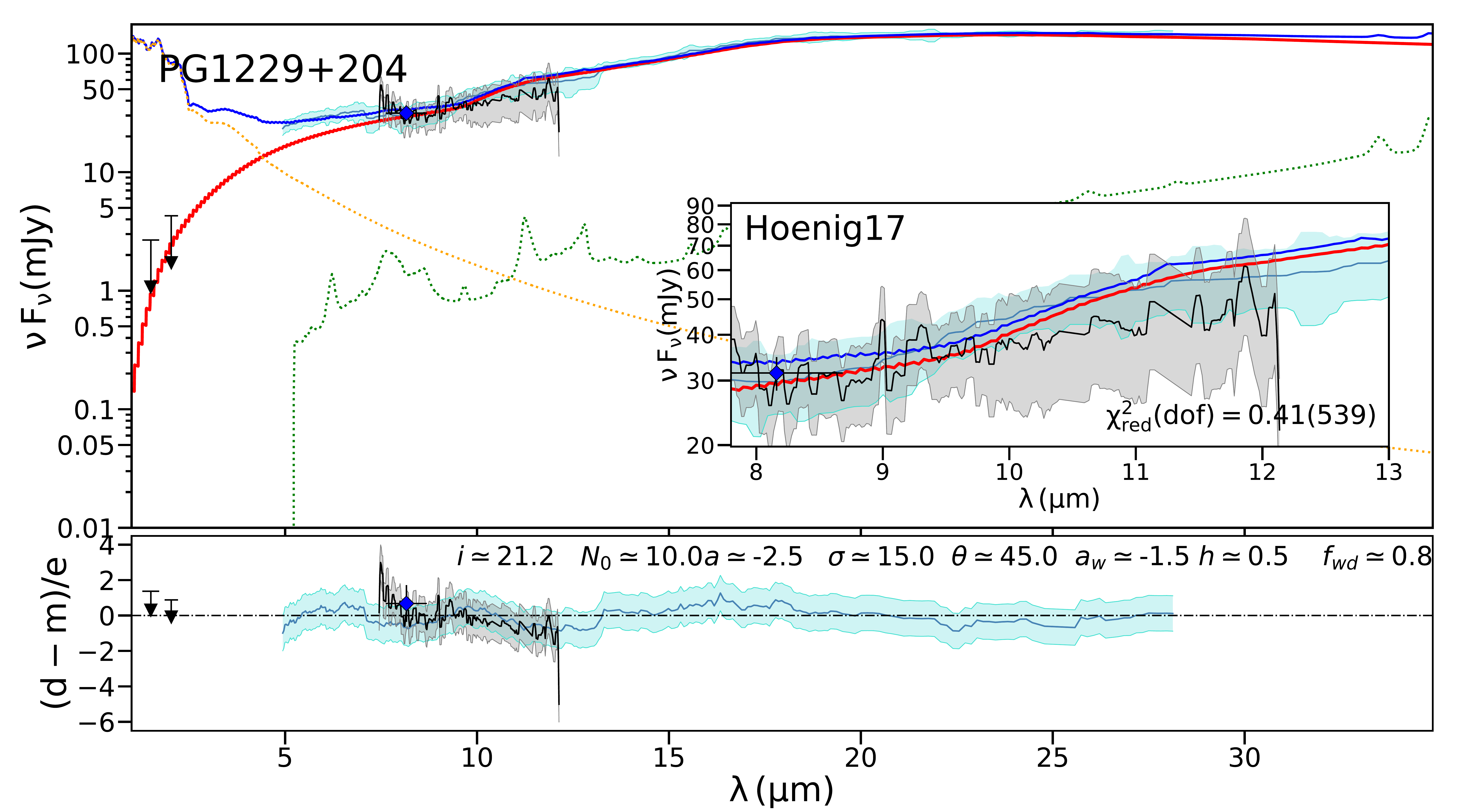}\\\includegraphics[width=0.7\columnwidth]{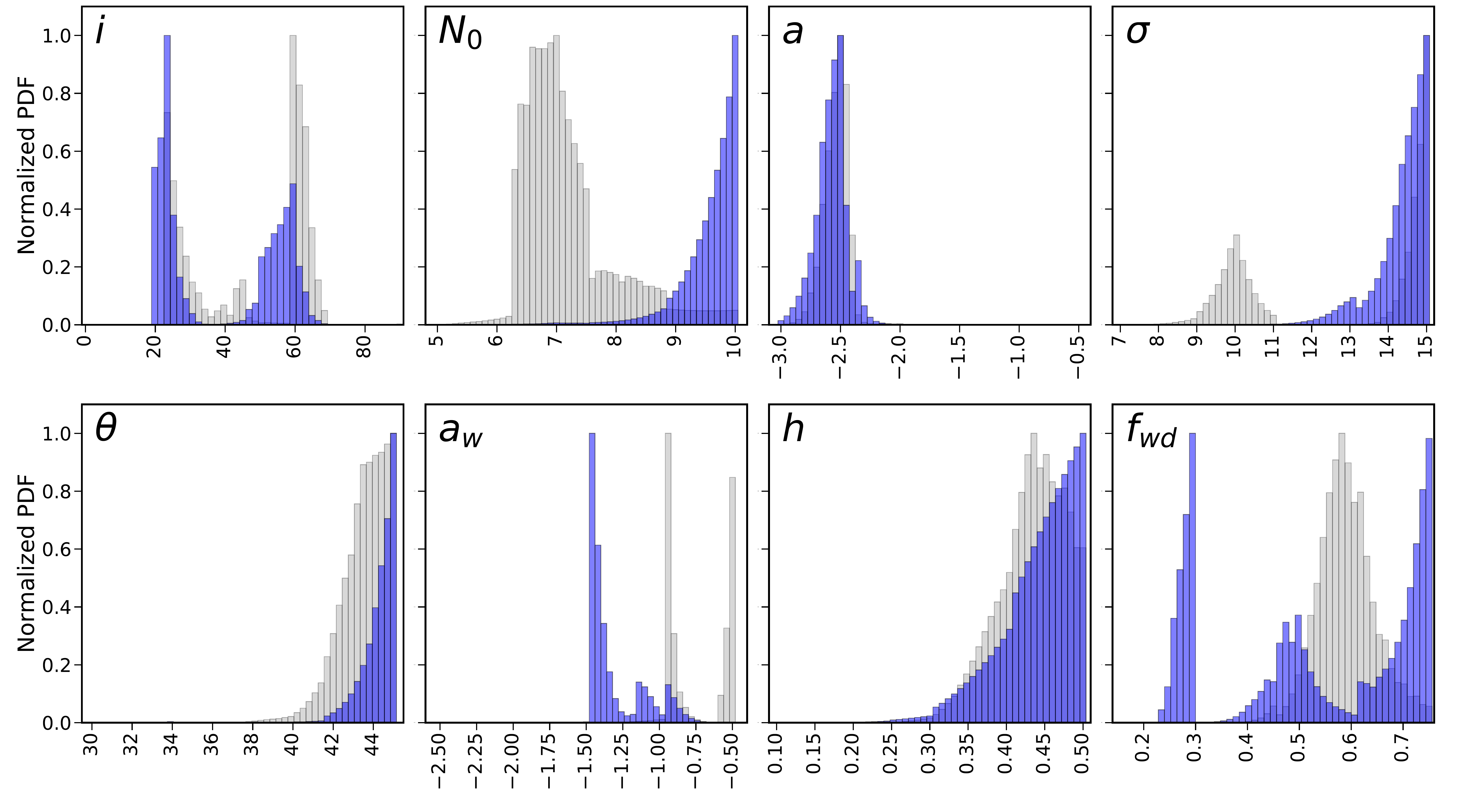}\\
\begin{tabular}{cc}
\includegraphics[width=0.3\columnwidth]{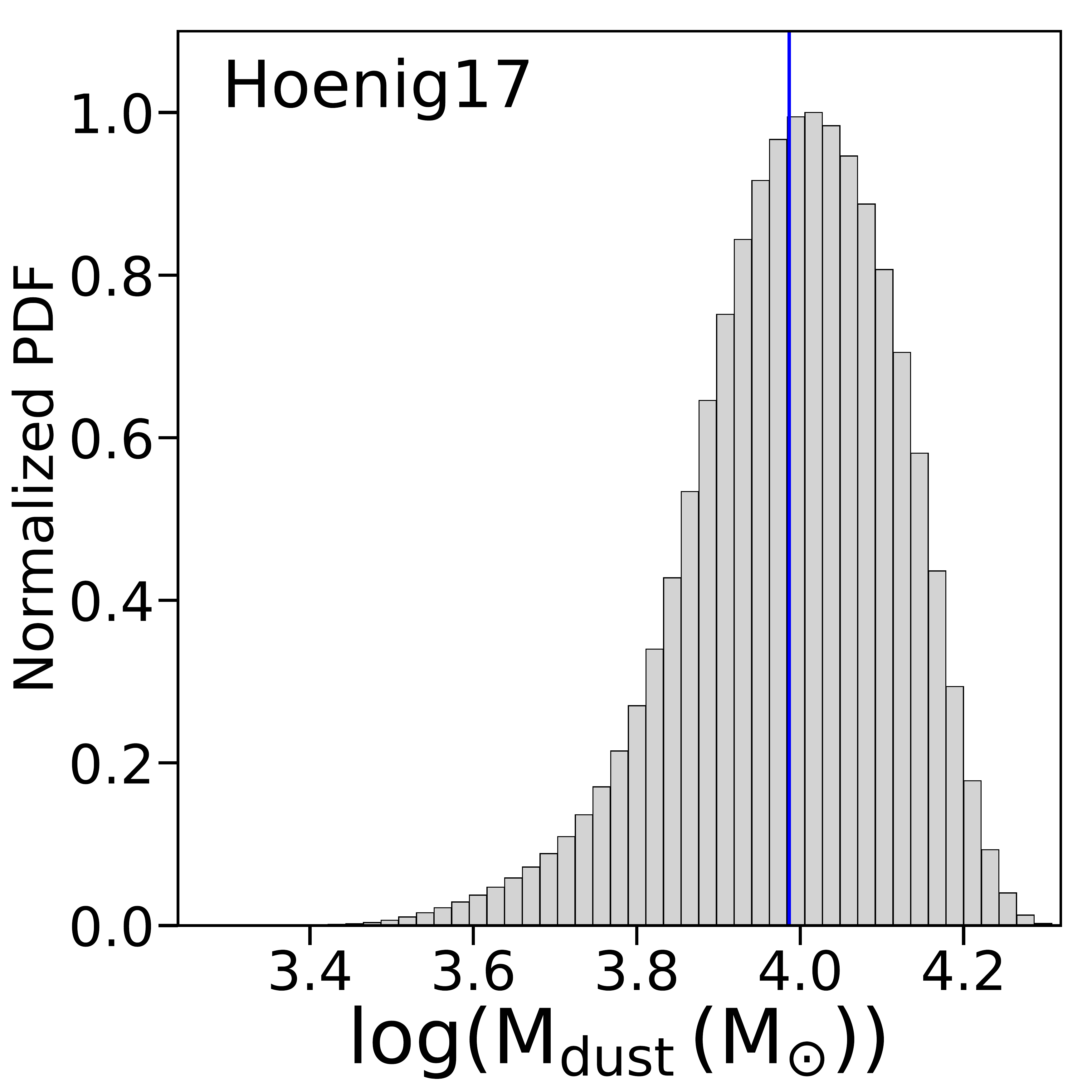}&\includegraphics[width=0.3\columnwidth]{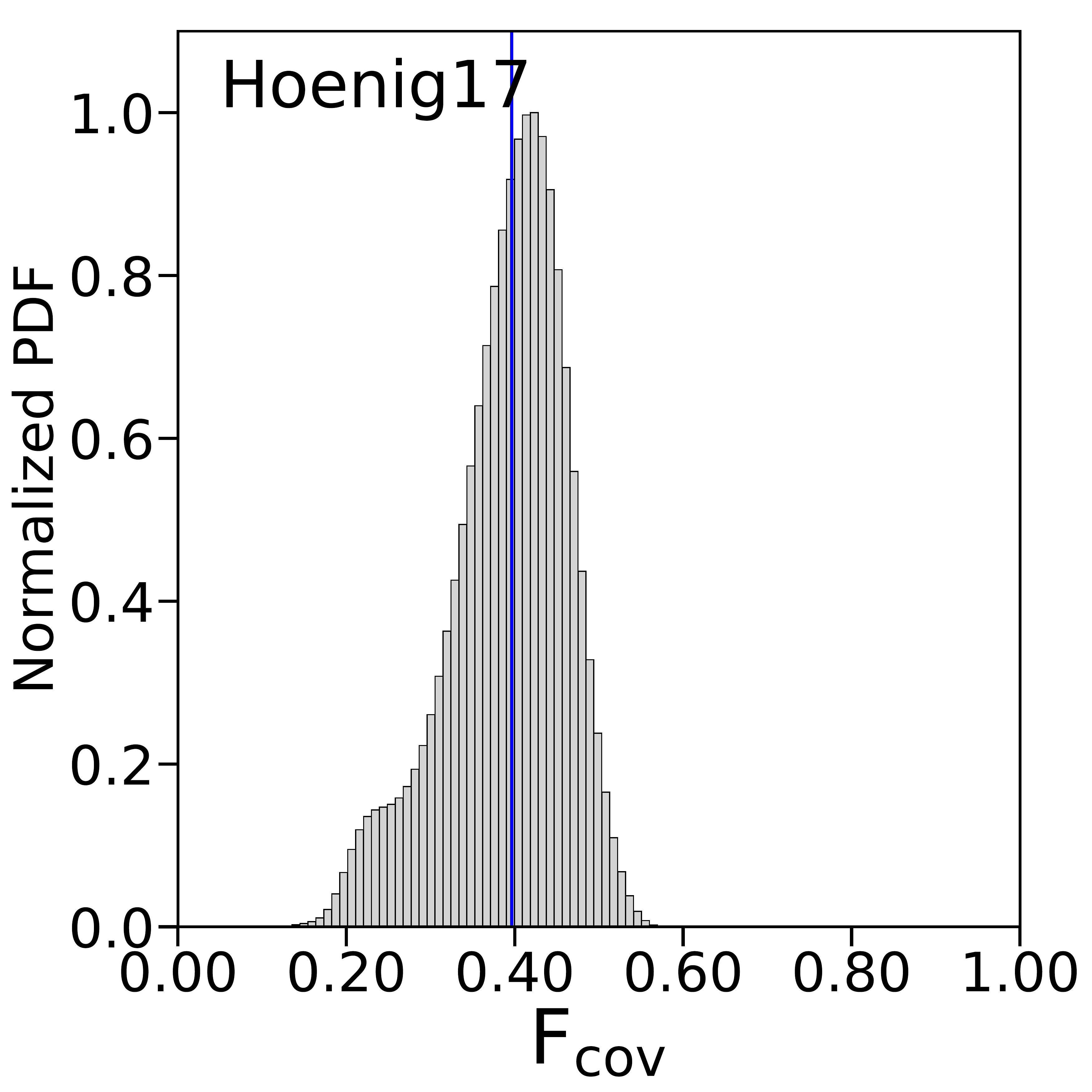}\\
\end{tabular}
\caption{As Figure \ref{fit5} but for the QSO PG~1229+204.\label{fit9}}
\end{figure*}

\begin{figure*}
\centering
\includegraphics[width=0.7\columnwidth]{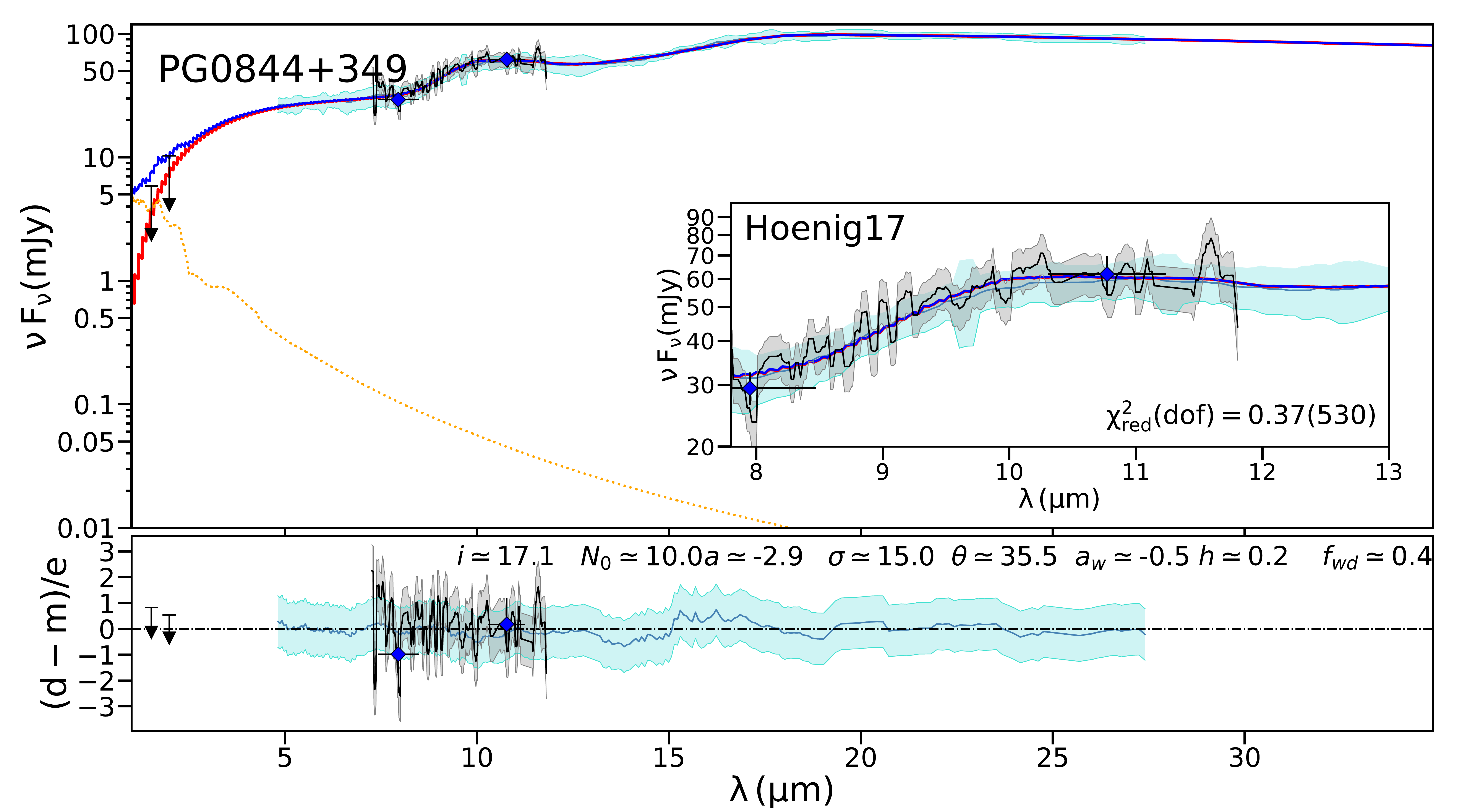}\\\includegraphics[width=0.7\columnwidth]{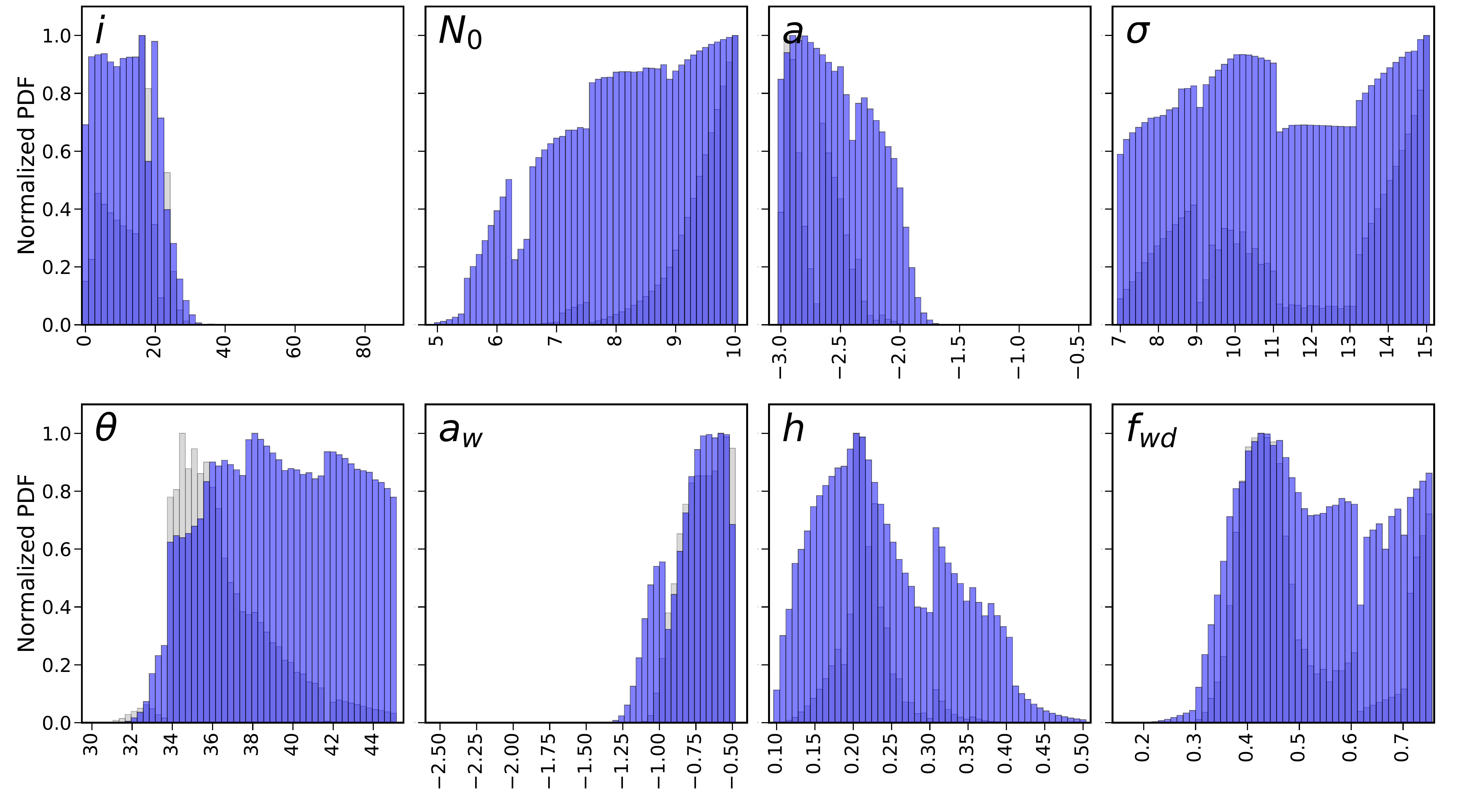}\\
\begin{tabular}{cc}
\includegraphics[width=0.3\columnwidth]{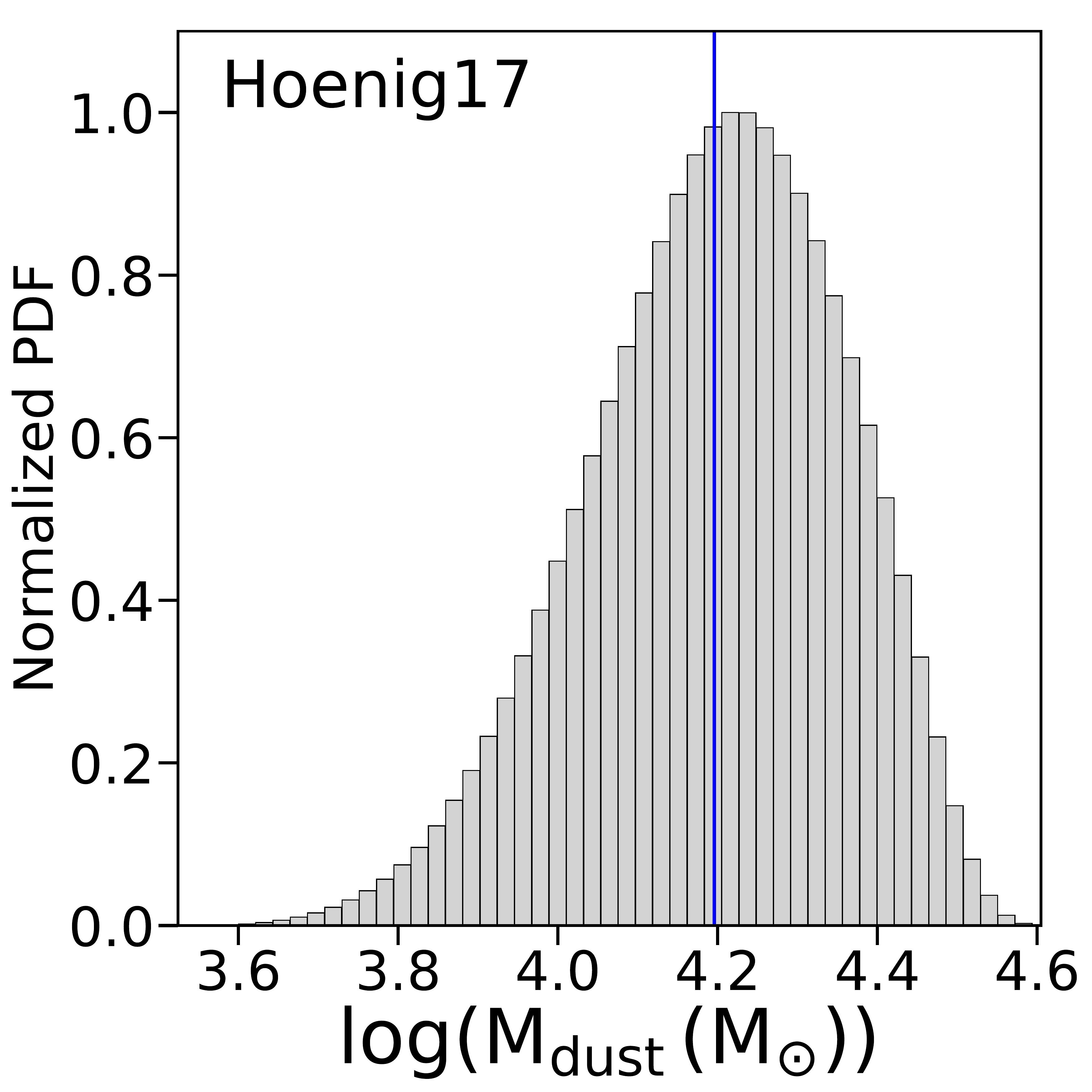}&\includegraphics[width=0.3\columnwidth]{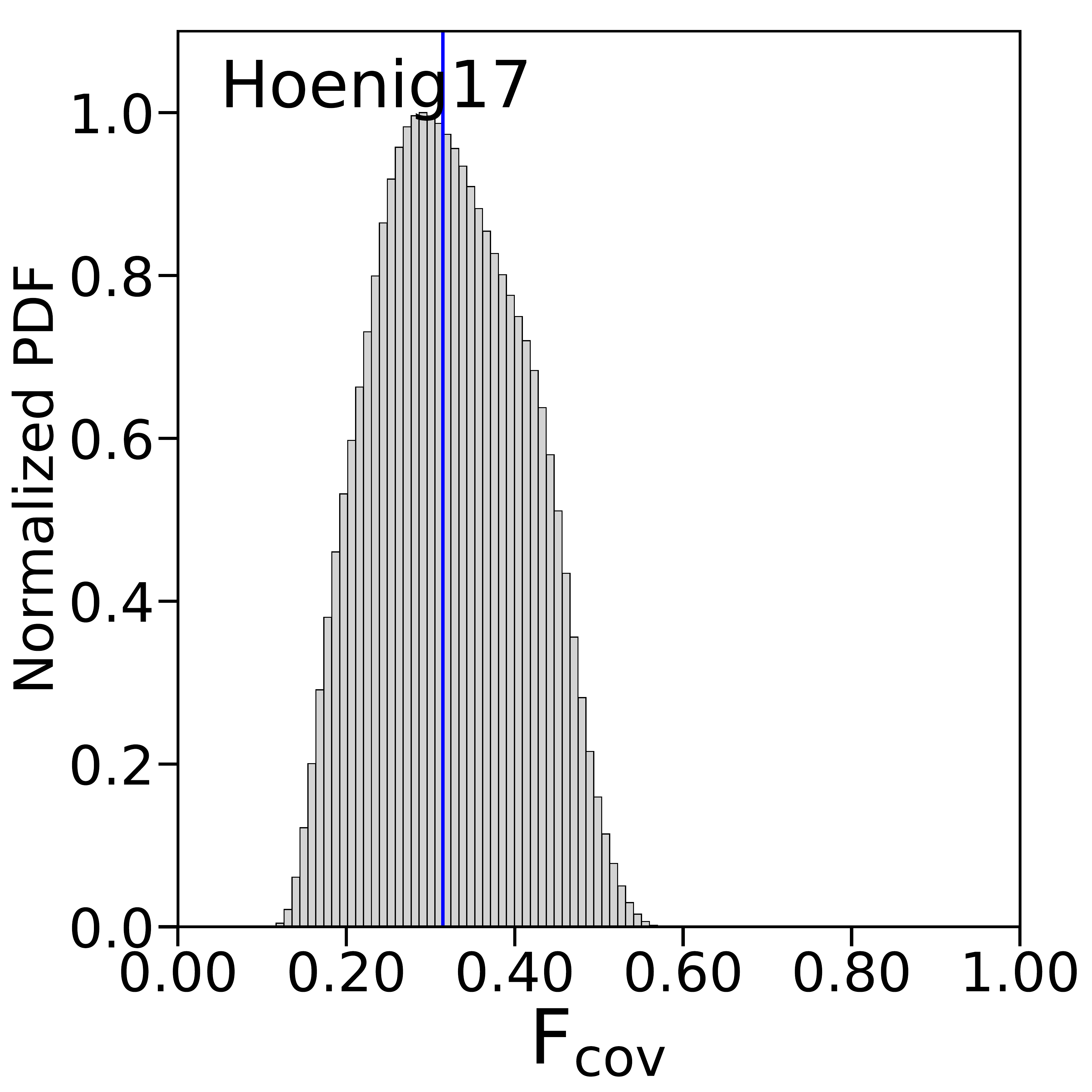}\\
\end{tabular}
\caption{As Figure \ref{fit5} but for the QSO PG~0844+349.\label{fit10}}
\end{figure*}

\begin{figure*}
\centering
\includegraphics[width=0.8\columnwidth]{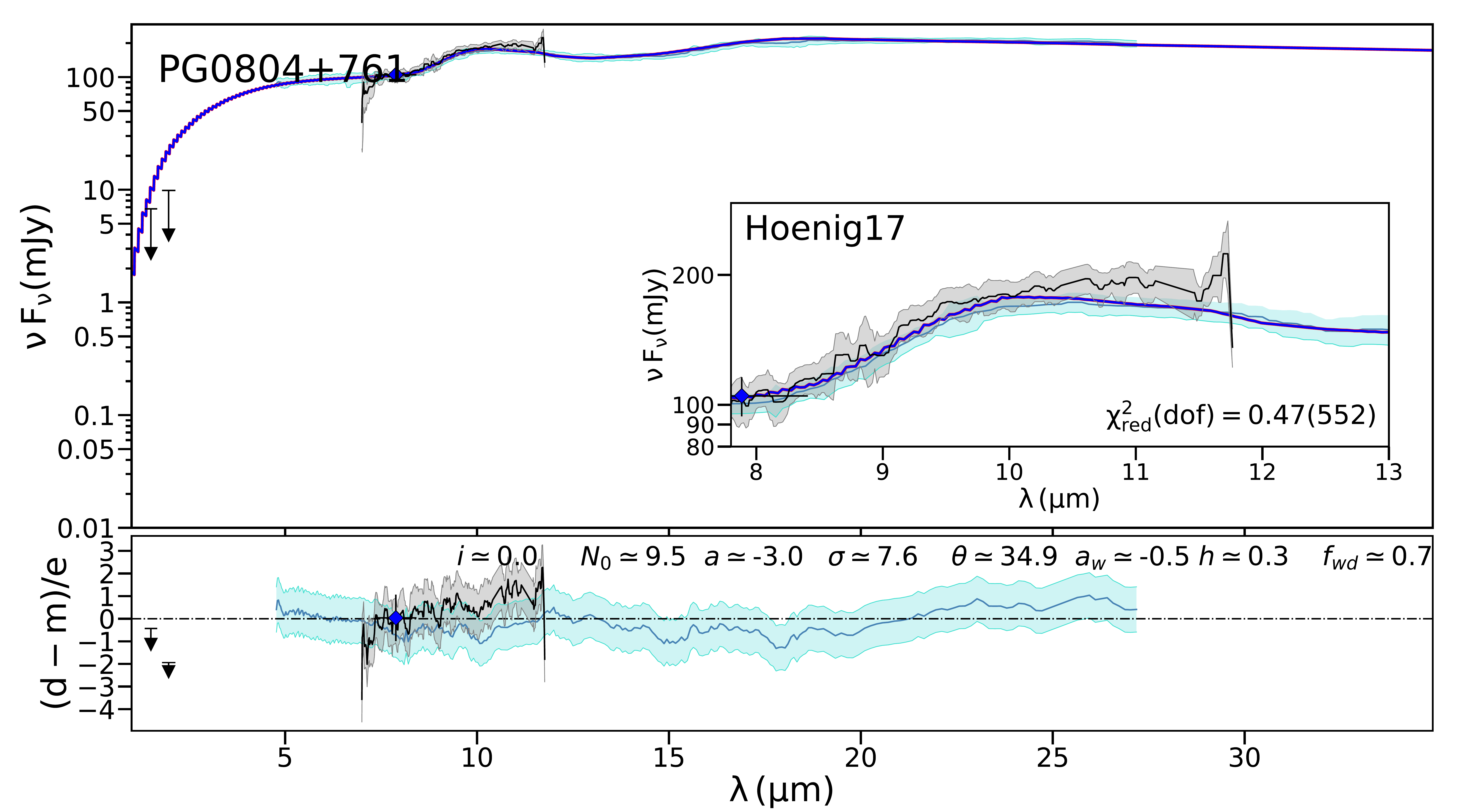}\\\includegraphics[width=0.8\columnwidth]{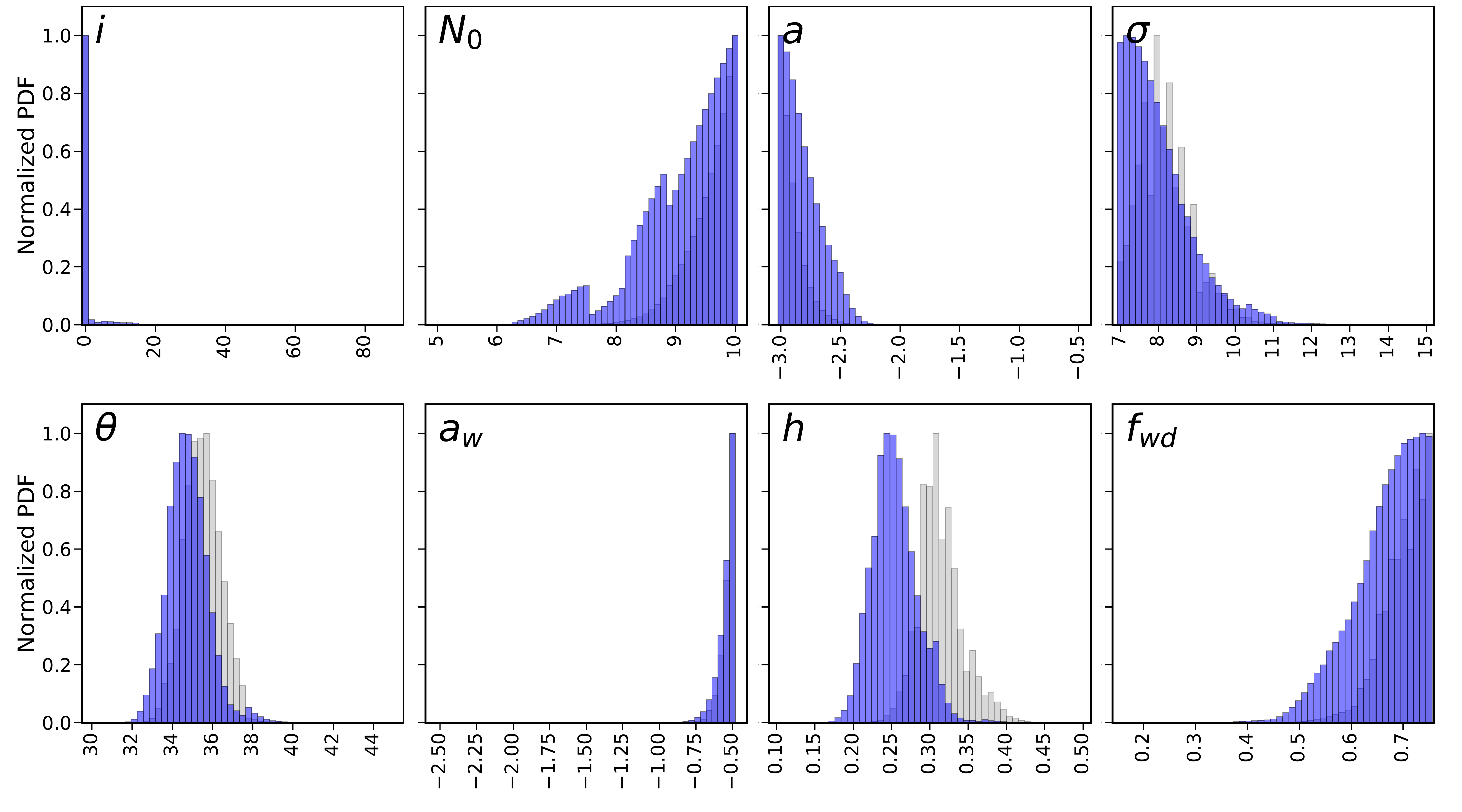}\\
\begin{tabular}{cc}
\includegraphics[width=0.3\columnwidth]{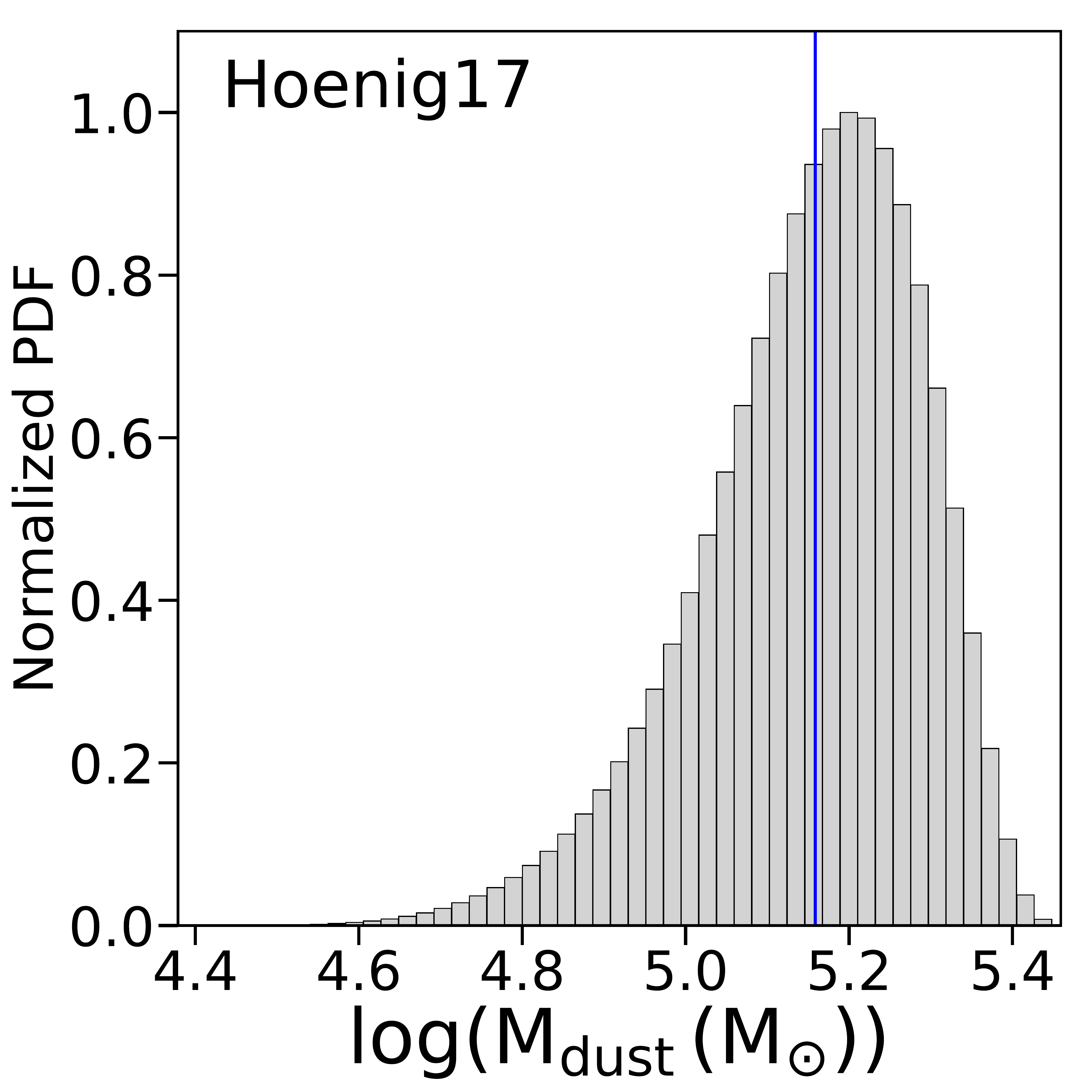}&\includegraphics[width=0.3\columnwidth]{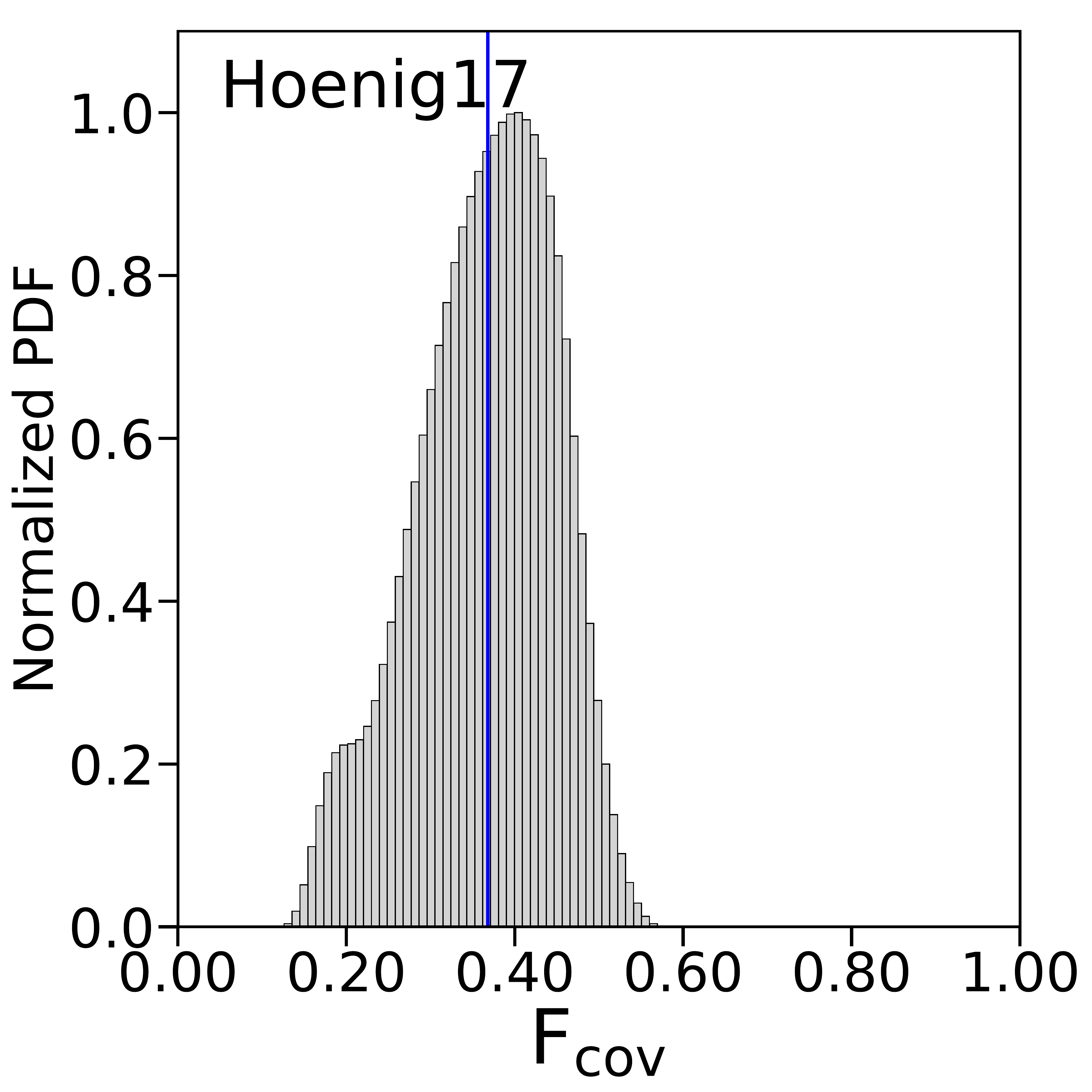}\\
\end{tabular}
\caption{As Figure \ref{fit5} but for the QSO PG~0804+761.\label{fit11}}
\end{figure*}

\begin{figure*}
\centering
\includegraphics[width=0.7\columnwidth]{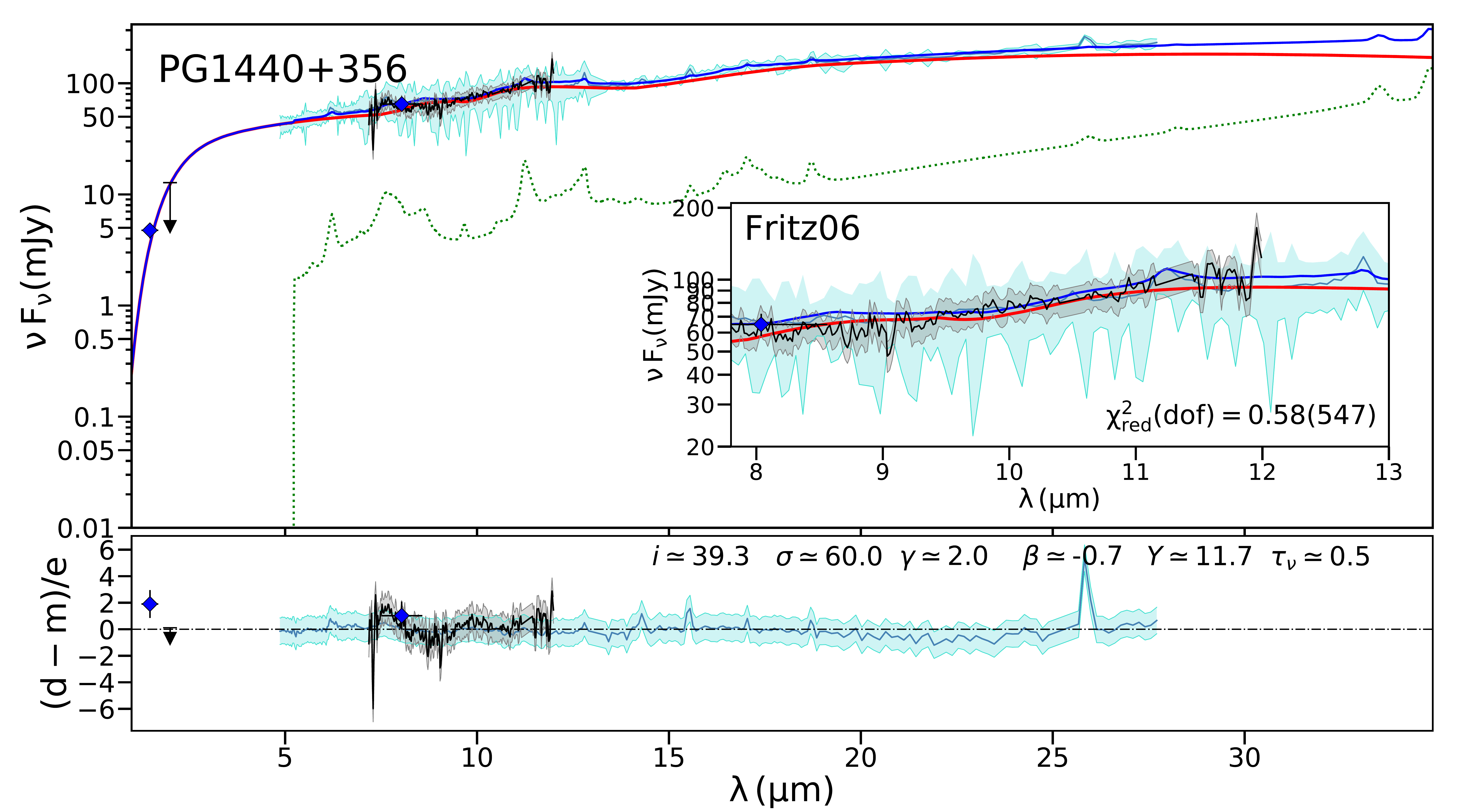}\\
\includegraphics[width=0.7\columnwidth]{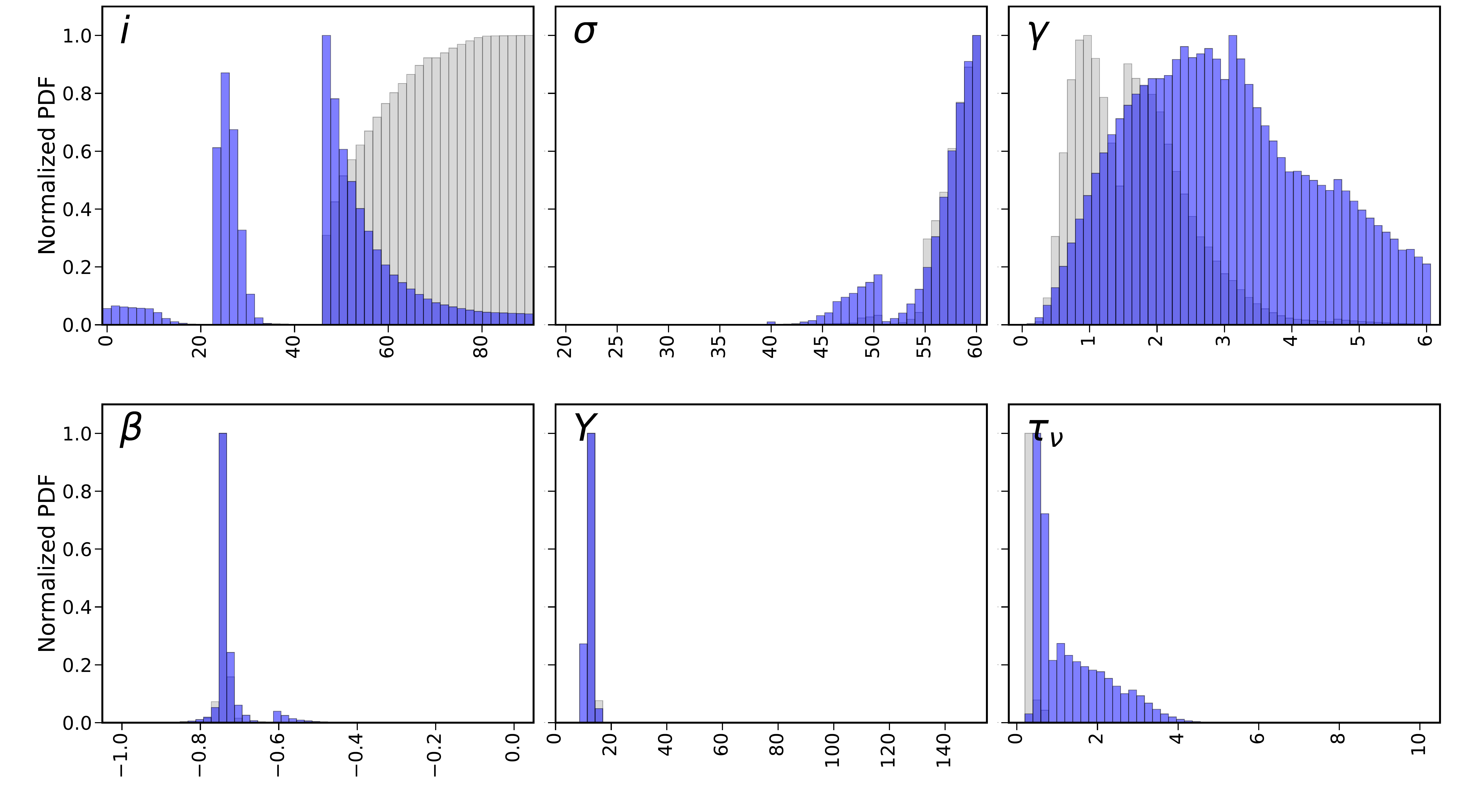}\\
\begin{tabular}{ccc}
\includegraphics[width=0.3\columnwidth]{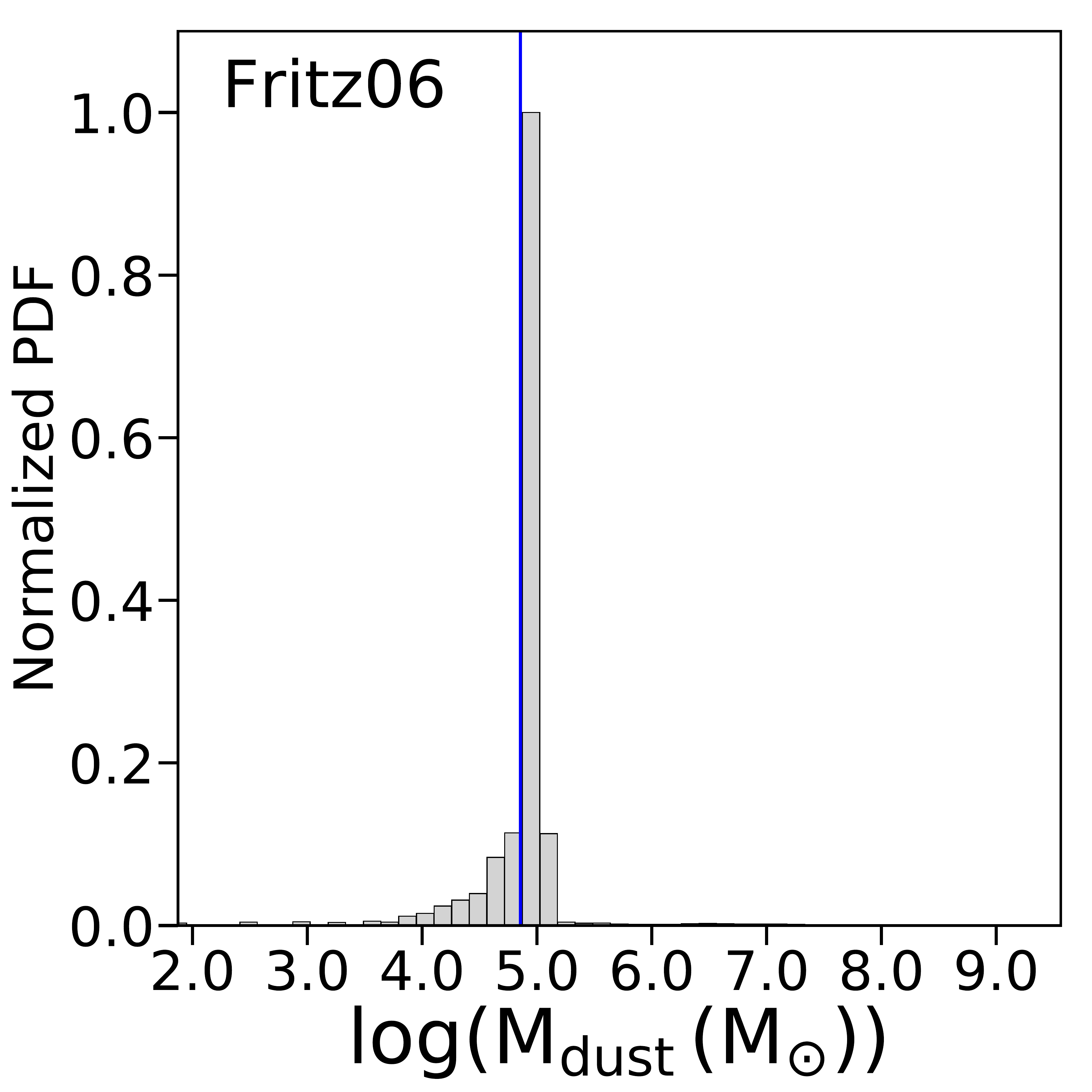}&\includegraphics[width=0.3\columnwidth]{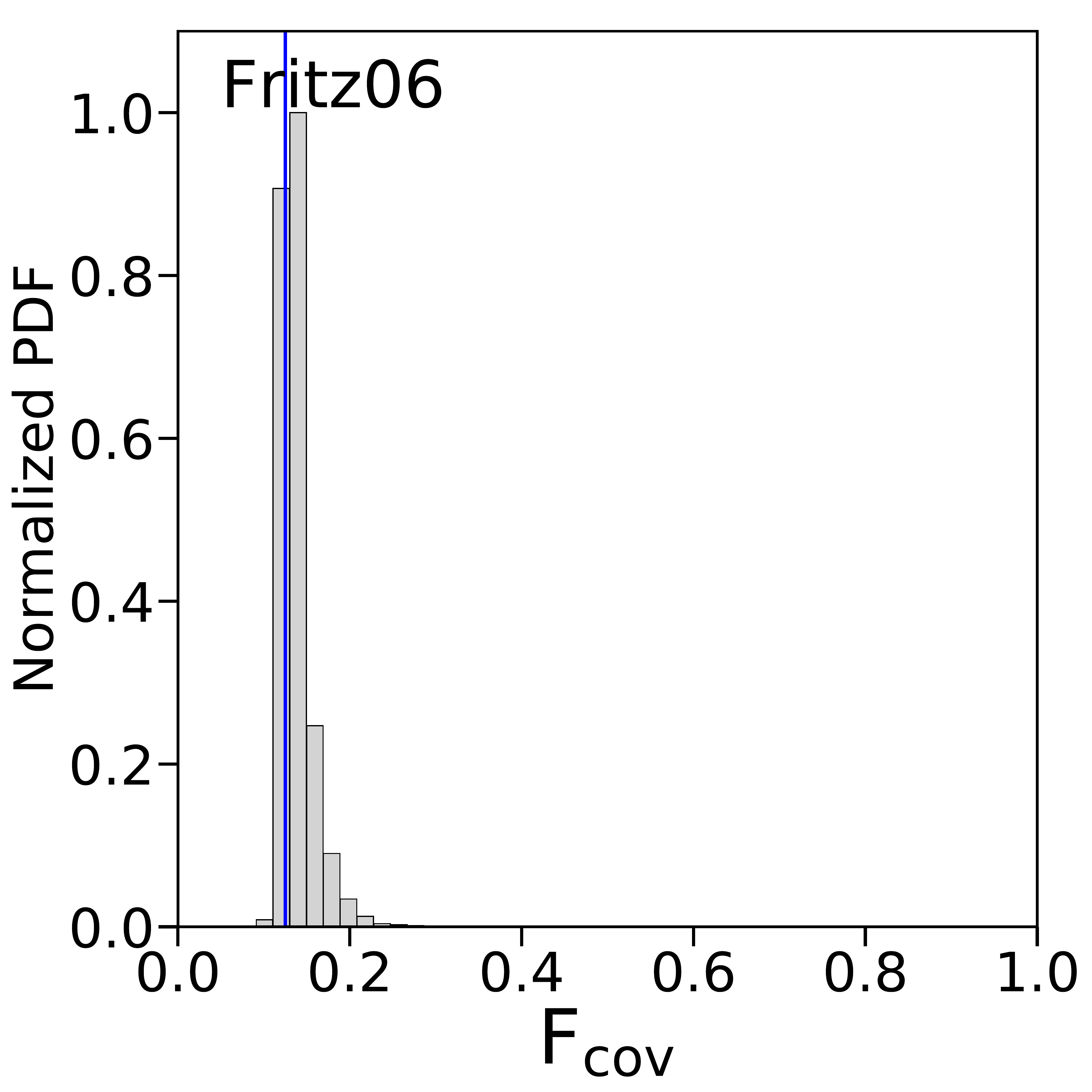}&
\includegraphics[width=0.3\columnwidth]{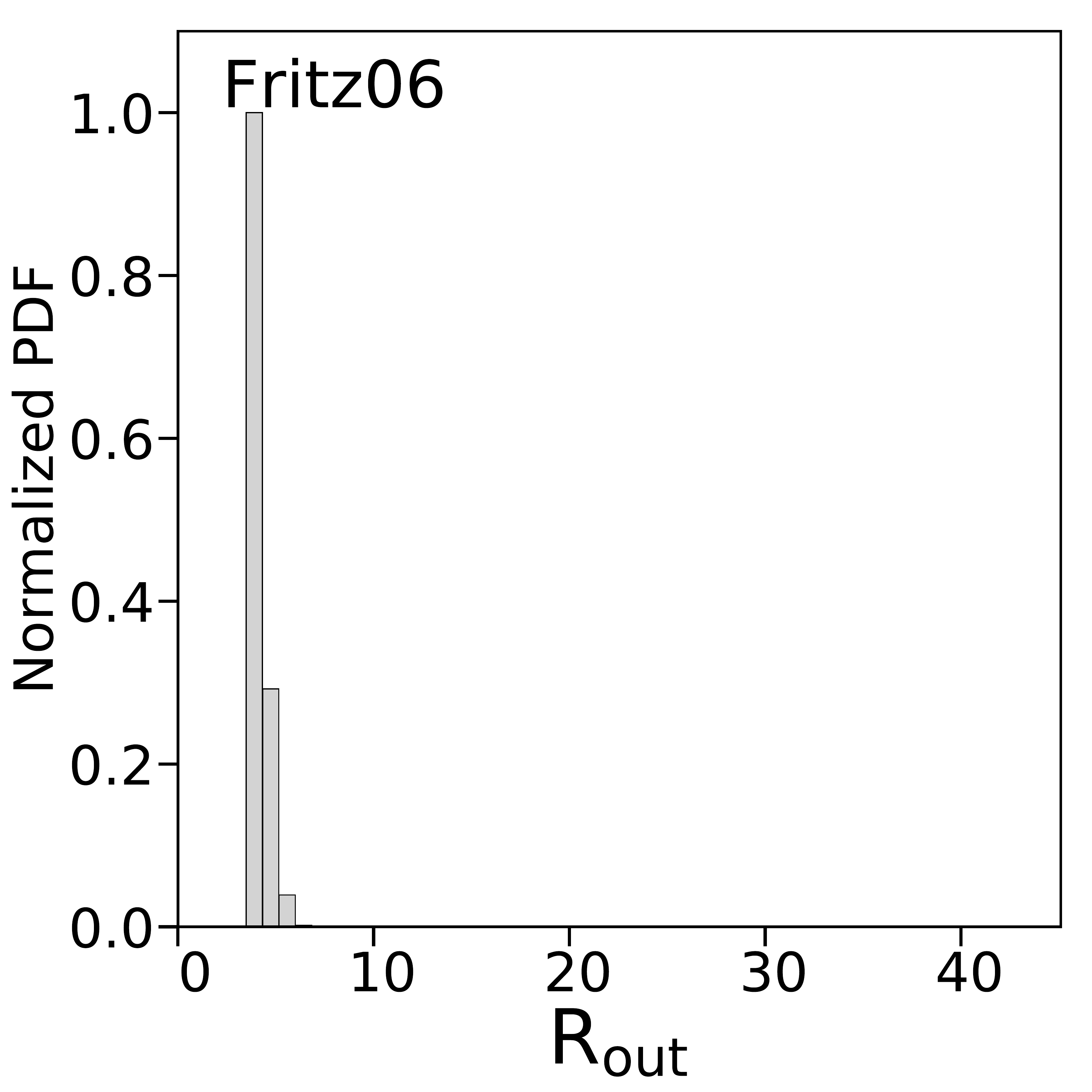} \\
\end{tabular}
\caption{{\bf The SED of PG~1440+356 best modeled by the Smooth F06 model}. {\bf Upper panel}: the high angular resolution photometric points are plotted as blue points with their error bars, the black arrows are upper limits. The high angular resolution spectrum is plotted with a black solid line, the grey shaded region represent the errors. The {\it Spitzer}/IRS spectrum is plotted with a dark cyan solid line and its error with a cyan shaded region. The red solid line is the best model resulting from 
fit the high angular resolution data. The green and yellow dotted lines are the starburst and stellar components, respectively. The blue solid line represents the sum of the stellar, starburst and torus components that best fit the {\it Spitzer}/IRS spectrum.  {\bf Middle panel}. {\bf The model parameters derived}: normal probability distribution function of the free parameters. In grey we plot the parameters derived from fit the LSR spectrum, while in dark blue the distribution of the parameters derived from fit the FSR spectrum. {\bf Bottom panel}: normal PDF of the derived parameters. The blue vertical line indicate the mean.\label{fit12}}
\end{figure*}

\begin{figure*}
\centering
\includegraphics[width=0.7\columnwidth]{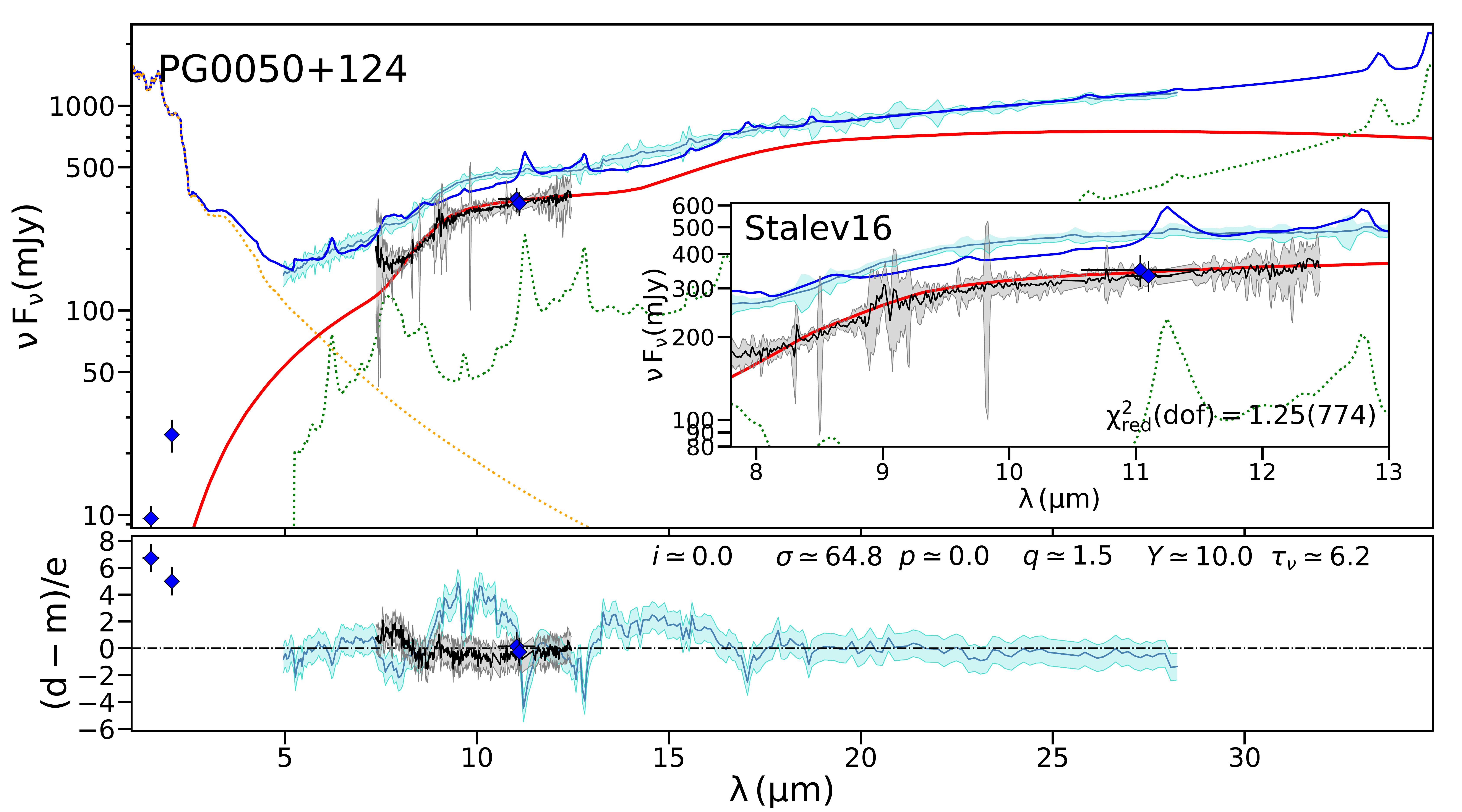}\\
\includegraphics[width=0.7\columnwidth]{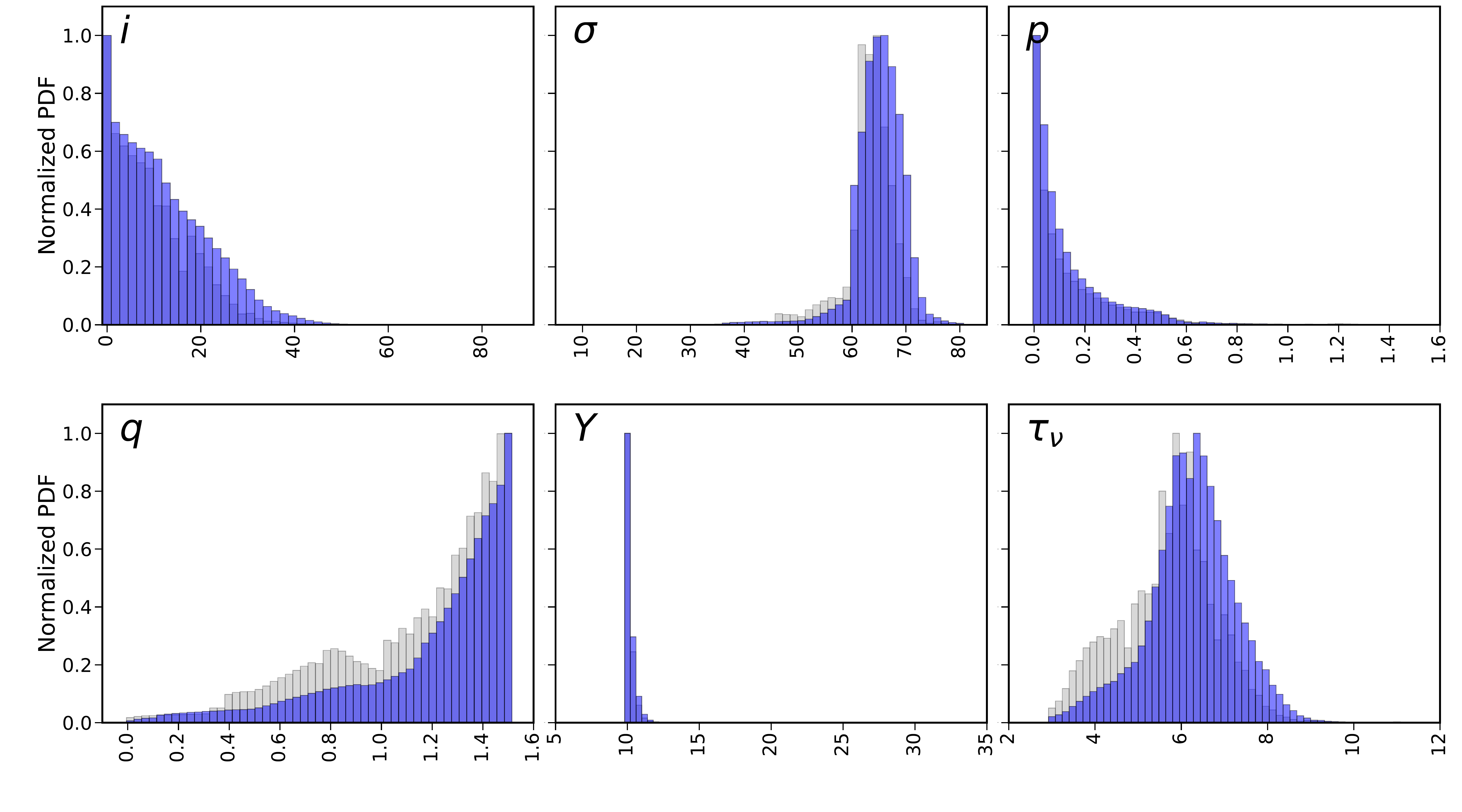}\\
\begin{tabular}{ccc}
\includegraphics[width=0.3\columnwidth]{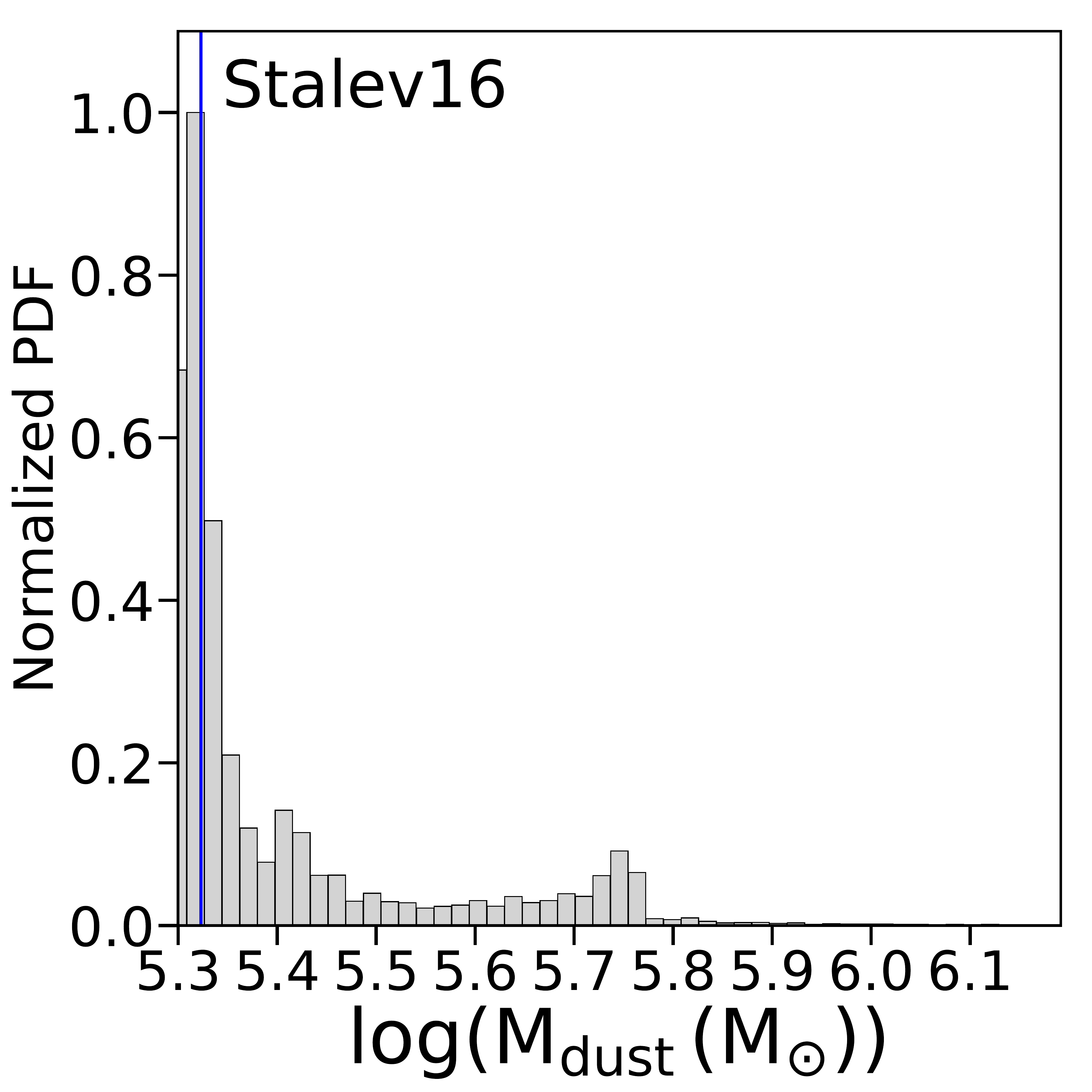}&\includegraphics[width=0.3\columnwidth]{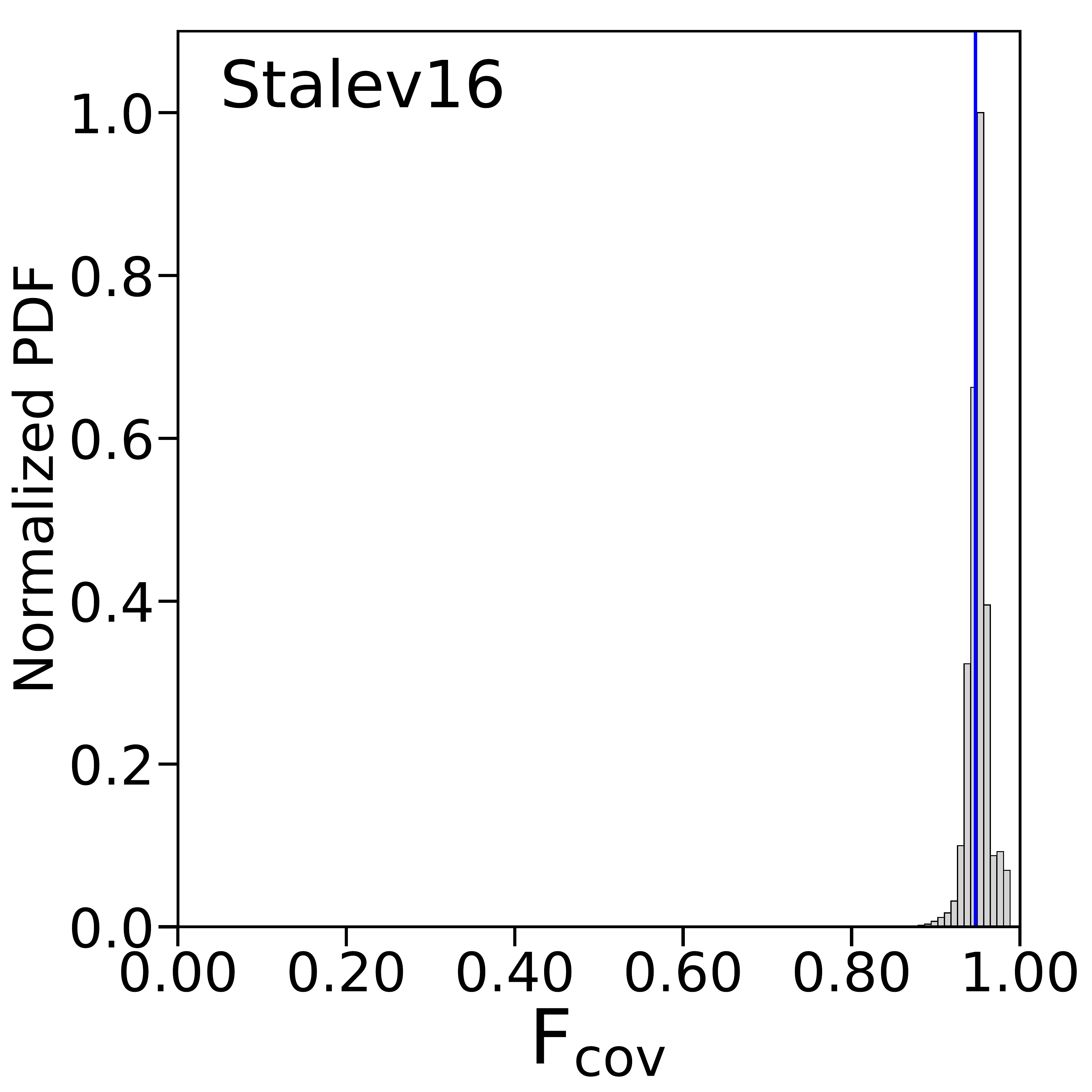}&
\includegraphics[width=0.3\columnwidth]{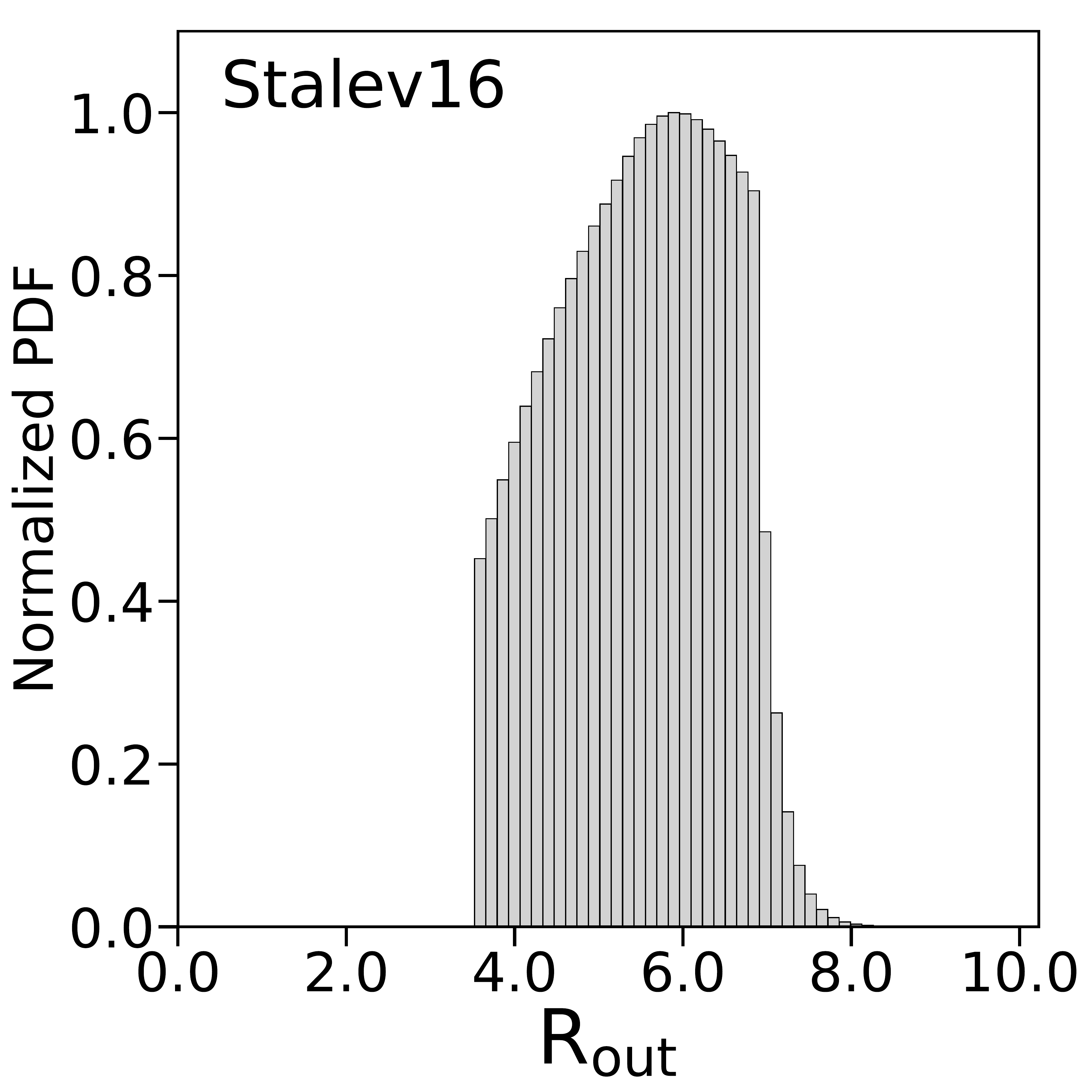} \\
\end{tabular}
\caption{{\bf The SED of PG~0050+124 best modeled by the Two-phase Media S16 model}. {\bf Upper panel}: the high angular resolution photometric points are plotted as blue points with their error bars, the black arrows are upper limits. The high angular resolution spectrum is plotted with a black solid line, the grey shaded region represent the errors. The {\it Spitzer}/IRS spectrum is plotted with a dark cyan solid line and its error with a cyan shaded region. The red solid line is the best model resulting from 
fit the high angular resolution data. The green and yellow dotted lines are the starburst and stellar components, respectively. The blue solid line represents the sum of the stellar, starburst and torus components that best fit the {\it Spitzer}/IRS spectrum.  {\bf Middle panel}. {\bf The model parameters derived}: normal probability distribution function of the free parameters. In grey we plot the parameters derived from fit the LSR spectrum, while in dark blue the distribution of the parameters derived from fit the FSR spectrum. {\bf Bottom panel}: normal PDF of the derived parameters. The blue vertical line indicate the mean.\label{fit13}}
\end{figure*}

We find that for all objects, several models can fit the same 
set of LSR and FSR data with a reduced $\chi^{2}<2$. 
However, according to the AIC criterion, the model that better reproduces the FSR data for 
Mrk~509, PG~2130+099, PG~1411+442, PG~1211+143, PG~1501+106, MR~2251-178, PG~1229+204, PG~0844+349, and PG~0804+761 is the Disk+Wind H17 model, while for PG~1440+356 is the Smooth F06 model, 
for PG~0050+124 is the Two-phase media S16 model, and for PG~1426+015 is the Clumpy N08 model. In the 
case of PG~0003+199 we find that none of the models can reproduce the 
FSR data probably because the shape of its ground-based MIR spectrum differs from the {\it Spitzer}/IRS spectrum between $\sim10.5-12.3\mu$m. These differences might be caused for the 
lost of signal in the border of the ground-based MIR high angular resolution spectrum \citep[see Figure C1 in][]{Martinez_Paredes17}. In 
Table~\ref{tab:stat} we highlight in bold letters the good fits according to the AIC criterion. For PG~2130+099, the data do not permit to distinguish between the Disk+Wind H17 and Smooth F06 models. For MR 2251-178, both the Disk+Wind H17 and Clumpy H10 models are equally valid.
However, based on a qualitative analysis, it is possible to note that the residuals obtained assuming the Disk+Wind H17 model are the flattest in both cases 
(see Figures~\ref{fit1} and \ref{fit2} for PG~2130+099 and Figure~\ref{fit3} and \ref{fit4} for MR~2251-178). For the rest of objects and models, the best-fitted SEDs and probability distribution functions (PDFs) of the constrained parameters are shown from  
Figure~\ref{fit5} to Figure~\ref{fit13}.

Considering only the models that resulted selected according to the AIC criterion, we find that the Disk+Wind H17 is the model that better reproduces the data ($64\%$ of the cases, 9/13) followed by 
the Smooth F06 ($14\%$, 2/13), the Clumpy H10 ($7\%$, 1/13), the Clumpy N08 ($1\%$, 1/13), and the Two-phase media S16 ($7\%$, 1/13) models. As an example, in Figure~\ref{models_Mrk509} we show the fits of the NIR-MIR SED for Mrk~509 assuming the Disk+Wind H17 model and in Figure~\ref{parameters_Mrk509} the  PDF of the parameters that best reproduce the data, 
and the PDF of 
the covering factor and the dust mass of the torus derived. On the other hand, if we assume the bluer accretion disk, none of the models fit the data for most cases. The exceptions are PG~0050+124 and PG~1501+106, for which the Disk+Wind H17 model was  the best model in fit the data. This result is consistent with those obtained when assuming the classic accretion disk. Additionally, in all cases we 
note that although the {\it Spitzer}/IRS spectra of these QSOs is mostly dominated 
by the contribution of the AGN ($>80\%$) and 
we remove the contribution from the accretion disk 
it is necessary to add both the stellar and/or starburst components for most objects (see Table~\ref{tab:stat}).

When using the classic accretion disk, the H and Ks bands are well fitted in most cases by the Wind+Disk H17 model. Only in two cases, MR~2251-174 and PG~1411+442, the flux in the H band, from GTC/EMIR and HST/NICMOS, respectively, is above the best fit. This excess could be due to some contamination by the host galaxy. Since in the case of MR~2251-174, the radial profile in the H band (see Section 4.1) is slightly above the radial profile of the standard star, and because  in some cases, a marginal emission from the host could be contaminating the PSF flux in the H band from {\it HST}/NICMOS \citep[see][]{Veilleux06, Veilleux09b}. We find that the Clumpy N08 model best fitted to PG~1426+015 and reproducers the H photometric point obtained from NICMOS/HST (the Ks flux is an upper limit). We also note that for PG~0050+124, the photometric points in the H band from HST/NICMOS and Ks band from GTC/CIRCE, are above the best fitted model (Two-phase Media S16). Additionally, for PG~1440+356, the photometric point in the H band from HST/NICMOS is slightly above the best fitted model (Smooth F06), the Ks flux is an upper limit. In these two last cases, this is probably due to the limitation of the model in well fit simultaneously the NIR and MIR SEDs \citep[see e.g.,][]{Martinez_Paredes20} and/or some marginal contamination from the host.

In Tables~\ref{tab:par_H17}, \ref{tab:par_H10}, \ref{tab:par_N08}, \ref{tab:par_F06}, and \ref{tab:par_S16} we list the mean and $1\sigma$ errors of the parameters derived for these models, while in  Tables~\ref{tab:Dpar_H17}, \ref{tab:Dpar_H10}, \ref{tab:Dpar_N08}, \ref{tab:Dpar_F06}, and \ref{tab:Dpar_S16} we list the derived (covering factor, dust mass, and outer radius) parameters. Additionally, in
Figure~\ref{ind_par} we plot the angular width (half angular width measured from the equator for the Disk+Wind H17 model) against the viewing angle obtained from the best fits for all models, as well as the dust mass and covering factor in size- and color-code. We note that in all cases, the parameters are consistent with the optical classification as type 1 AGN and that the 
dust mass is constrained between $\sim10^{4}-10^{5}\,\text{M}_{\odot}$ for 
most cases, while the covering factor ranges from $\sim0.14$ to 0.40 for the Disk+Wind H17 
model until 0.95 for the Two-phase media S16 model. These covering factors are 
consistent with the ones obtained in previous works for type 1 AGNs \citep[see e.g.,][]{Martinez_Paredes17, Gonzalez-Martin19b, Martinez_Paredes20}. However,  it is important to note that the covering factor is model dependent, as showed by \citet{Gonzalez-Martin19b}. For example, they found that for the Disk+Wind H17 model the covering factor only ranges from $\sim0.1-0.6$, while for the rest of models considered here, the parameters allow the covering factor to range from $\sim0.1-1$, showing a maximum towards the larger values and decreasing until they reach a value of $\sim0.1$.

\begin{table*}
	\begin{minipage}{1.\textwidth}
		\caption{{\bf Statistic of the acceptable fits}. Column 1 lists the name of the object. Columns 2 and 3 list the model assumed and the additional component included. Columns 4, 5, and 6 list the red-$\chi^{2}$, degree of freedom, and AIC values for the FSR fits. Column 7 lists the reduced-$\chi^{2}$ 
		for the LSR fits.\label{tab:stat}}
\centering
\begin{tabular}{lcccccc|lccccccc}
				\hline
 &  &Add.& \multicolumn{2}{c}{FSR} &  &LSR& & &Add. & \multicolumn{2}{c}{FSR} & & LSR\\

 Name & Model&Comp. & $\chi_{red}^{2}$& dof& AIC &  $\chi_{red}^{2}$& Name&Model & Comp. & $\chi_{red}^{2}$ & dof &AIC &$\chi_{red}^{2}$\\

\hline
\hline
Mrk~509	&	F06	& +Ste+SB&	1.2	&	772	&	584.88&	1.05&		PG~1211+143	&	F06	& +Ste+SB&	1.53	&	535	&	834.29&	1.96\\
	&	H10&+Ste+SB	&	0.85	&	773	&	314.63&0.59 	&		&N08& +Ste+SB	&1.33	&535	&727.77	&1.58	\\
	&	{\bf H17}&+Ste+SB	&	0.71	&	770	&	206.99&	0.47	&	&H10&+Ste+SB	&	1.24	&	536	&680.9	&1.27	\\
 	&	 	&	 	&	 	&	 & 	&	& &{\bf H17}&+Ste+SB	&	1.15	&533& 633.15	&1.20	\\
	&		&		&		&		&	&		&		&		&	&	&	& &	\\
PG~0050+124	&	F06&+Ste+SB	&	1.32	&	774	&	1038.01&	1.77&	PG~1501+106	&	F06&+Ste	&	1.28	&	521	&	638.39&	1.26\\
	&	{\bf S16}&+Ste+SB	&	1.25	&	774	&	980.92&1.65	&		&	S16&+Ste	&	1.12	&	521	&	599.03&	1.02\\
	&H17&+Ste+SB &	1.44	&772	&1131.31	 & 	1.71&	&	N08&+Ste	&	0.89	&	521	&	480.88&	0.68\\
	&	 &	&	 	&	 	&	 & 	&	&	H10&+Ste	&	0.86	&	522	&	465.3&	0.81\\
	&	&	&		&		&		&	& &	{\bf H17}&+Ste	&	0.72	&	519	&	394.67&	0.52\\
		&		&		&		&		&	&		&		&		&	&	&	& &	\\
PG~2130+099	&	{\bf F06}&+SB	&	0.42	&	601&	271.49&	0.39&MR~2251-178&{\bf H10}&+Ste	&0.6		&	1586	&973.41	&1.30\\
	&	S16&+Ste+SB	&	0.73	&	601	&	456.48&	0.70& &{\bf H17}&+Ste	&0.6		&1583	&964.12	&1.09	\\
	&	H10&+Ste+SB	&	0.64	&	602	&	402.31&0.58	&		& &	&		&	&	&	\\
	& 	{\bf H17}&+Ste+SB	&	0.42	&	599	&	271.03&0.32		&	& & & & & & \\
	&	&	&		&		&	&	&	& & 	& 		& 	& 	& 	\\
 PG~1411+442	&	F06&+Ste+SB	&	0.93 &	543	& 521.12&0.89& PG~1229+204	&	F06&+Ste	&	0.6	&	541	&	341.19 &0.74	\\
	&	 	    H10& +Ste+SB&	0.78&	544	& 437.5&0.78&	&	S16&+Ste	&	0.5	&	541	&	284.18 &0.56	\\
	&			{\bf H17}& +Ste+SB&	0.57&	541	& 328.27&0.27&	&	N08&+Ste+SB	&	0.56	&	541	&	317.12 &0.55	\\
	&			& &	&	& & &	&	H10&+Ste+SB	&	0.53	&	542	&	302.39&	0.51\\
	&	&	&		&		&		&	& &			{\bf H17}&+Ste+SB	&	0.41	&	539	&	241.34 &	0.42\\	
 	&	 & 	&	 	&	 	&	 & 	&	&	&	&		&		&		&	&	\\
PG~1440+356	&	{\bf F06}&+SB	&	0.58	&	547	&	332.65&0.42	&PG~0844+349	&	F06&+Ste	&	0.61	&	532	&	341.56 &0.48	\\
	&	S16&+Ste	&	1.12	&	547	&	628.59&0.91	&	&	S16&+Ste	&	0.65	&	532	&	362.13 &	0.49\\
	&	H10&+SB	&	1.52	&	548	&	844.65&	1.75&	&	N08&+Ste	&	0.6	&	532	&	335.28 &0.49	\\
	&	H17&+SB	&	0.65	&	545	&	374.16&	0.62	&		&	H10&+Ste	&	0.55	&	533	&	305.15 &0.43	\\
	&		&		&		&		&	& &		&{\bf H17}&+Ste	&	0.37	&	530	&	215.15&	0.20\\
 	&	 & 	&	 	&	 	&	 &	 &	&	&	&		&		&		&		& \\
PG~1426+015	&	{\bf N08}&+Ste+SB	&	0.91	&	536	&	504.47&	0.29	 	&PG~0804+761	&	F06&+Ste	&	0.8	&	554	&	457.7 &0.65	\\
 	&	 	&	 	&	 	&	 &	 &	&		&	S16&+Ste	&	1.95	&	554	&	1095.3 &1.37	\\
	&	 & 	&	 	&		&	 &	 &	&N08&+Ste	&	1.94	&	554	&	1089.87 &1.60	\\
	&	 	&	 	&	 	&	 &	 &	&		&	H10&+Ste	&	1.58	&	555	&	893.54 &1.72	\\
	&		&	&	&		&		&	&	&{\bf H17}&---	&	0.47	&	552	&	280.2 &	0.47\\

		\hline
			\end{tabular}\\
Note.-The best models according to the AIC criterion are highlighted in boldface.
	\end{minipage}
	\end{table*}

\begin{figure*}
\centering

\includegraphics[width=1\columnwidth]{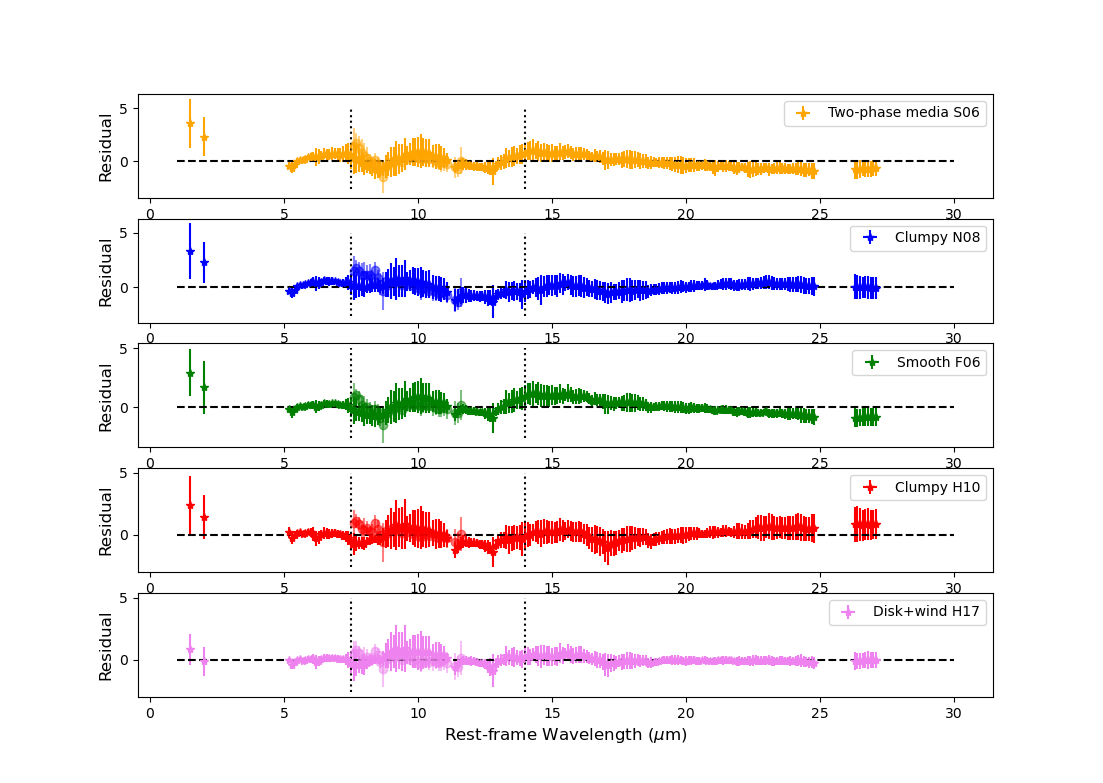}

\caption{{\bf Average residuals for each model}. The vertical lines separate the residuals in three parts for a better qualitative analysis (see the text). \label{residuals}}
\end{figure*}

To investigate how well the different models simultaneously 
reproduce the 
NIR and MIR SEDs of QSOs, we calculate the average residual from all fits and objects (see Figure~\ref{residuals}). 
From a qualitative analysis, we find that the Two-phase media S16 and the Clumpy N08 models are unable to fit the NIR data, while the Smooth F06 and Clumpy H10 models improve the fits at NIR, especially fitting the photometric point at Ks-band. However, the Disk+Wind H17 model is the best reproducing both photometric points at NIR, within the uncertainties. For the Two-phase media S16 and Clumpy N08 models it is very 
difficult to reproduce the bluer spectral range between 
$\sim5-7.5\,\mu$m, while for the other three models, the residuals look flatter. We observe that all models 
have difficulties reproducing the range between $\sim7.5$ and 14 $\mu$m where the 
$10\,\mu$m silicate feature lies, and none of the models can reproduce the spectral range around $12\,\mu$m. For wavelengths 
longer than $14\,\mu$m both the Clumpy N08 and H10 models, as well as the 
Disk+Wind H17 
model produces flatter residuals, while the Two-phase 
media S16 and Smooth F06 models show a steeper residual from the bluer to the redder range. 

\section{Discussion}
\subsection{The NIR and MIR spectral indexes}
 
Using the photometric IR data from the Infrared Space Observatory (ISO) 
\citet{Haas03} analyzed the IR SEDs ($5-200\,\mu$m) of 64 PG QSOs. Particularly, they found that 
the spectral slopes $\alpha_{1,10\,\mu\text{m}}$ range from -0.9 to -2.2, and do not correlate with the inclination-dependent extinction effects in the picture of a dusty torus. They suggested that the diversity of the SEDs that they observed could be 
explained in terms of an evolutionary scenario, in which those QSOs with a redder MIR SED 
are in early evolutionary stages among QSOs preceded by a dusty ultraluminous infrared galaxy (ULIRG) phase 
\citep[e.g.,][]{Sanders88}.

We 
measured four spectral indexes from ground-based high angular resolution SEDs. The first spectral index is calculated between the H (1.6 $\mu$m) and Si2 (8.7 $\mu$m) bands ($\alpha_{NIR-MIR(1.6,8.7)\mu\text{m}}$), the second between the Ks (2.2 $\mu$m) and Si2 (8.7 $\mu$m) bands ($\alpha_{NIR-MIR(2.2,8.7)\mu\text{m}}$),
the third one 
between 7.8 and 9.8 $\mu$m ($\alpha_{MIR(7.8,9.8\,\mu\text{m})}$), and the last one is 
calculated between 9.8 and 11.7 $\mu$m ($\alpha_{MIR(9.8,11.7\,\mu\text{m})}$). We measured the two later 
MIR spectral indexes instead of 
the $10\,\mu$m silicate strength due to the low S/N of the high angular resolution spectra. The spectral index is estimated 
according to the following definition $\alpha_{2,1}=-\frac{log(f_{\nu}(\lambda_{2})/f_{\nu}(\lambda_{1}))}{log(\lambda_{2}/\lambda_{1})}$, where $\lambda_{2}>\lambda_{1}$ \citep[see, e.g.,][]{Buchanan06}. In the case of the two MIR spectral indexes $\alpha_{MIR(7.8,9.8\,\mu\text{m})}$ and $\alpha_{MIR(9.8,11.7\,\mu\text{m})}$, the $\lambda_{2}$ and $\lambda_{1}$ 
wavelengths are chosen to fix the points 
on the continuum, avoiding
the spectral range of low atmospheric transmission around $9\,\mu\text{m}$, and the PAH feature at 11.3 $\mu$m that is present in some objects. 

In Table~\ref{tab:Alpha_Mdust},  we list the $\alpha_{NIR-MIR(1.6,8.7\,\mu\text{m})}$ and $\alpha_{NIR-MIR(2.2,8.7\,\mu\text{m})}$  spectral indexes measured after subtract the classic accretion disk from the data. On average, we find  $\alpha_{NIR-MIR(1.6,8.7\,\mu\text{m})}=-1.1\pm0.6$ and $\alpha_{NIR-MIR(2.2,8.7\,\mu\text{m})}=-1.2\pm0.4$. Additionally, we list the values obtained assuming the bluer accretion disk. Note that there are fewer objects in this case because, in most cases, the accretion disk overestimates the emission in the NIR bands (see Section 4.2). For these objects assuming a bluer accretion disk does not affect the estimation of the NIR to MIR spectral indexes within the uncertainties. Our measurements of both NIR-MIR spectral indexes are similar to the ones reported by \citet{Haas03}. 
%Note that we do not include the upper limits in these estimations.

\begin{figure*}
\begin{tabular}{cc}
\centering	
\includegraphics[width=0.48\columnwidth]{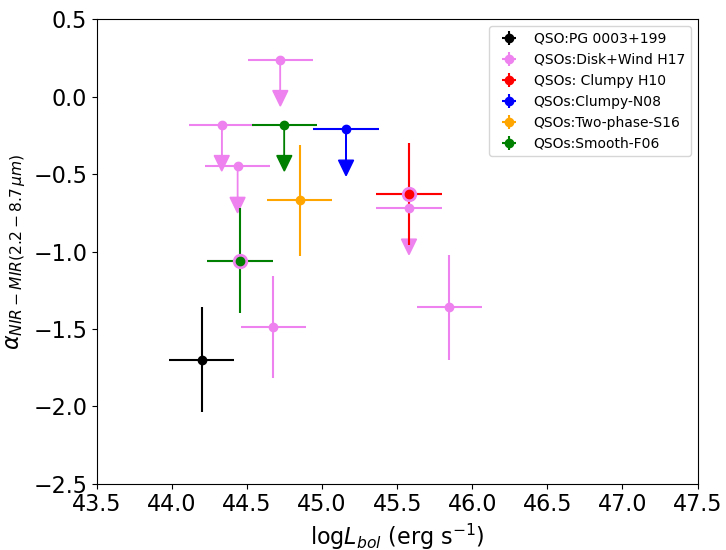} & 	\includegraphics[width=0.45\columnwidth]{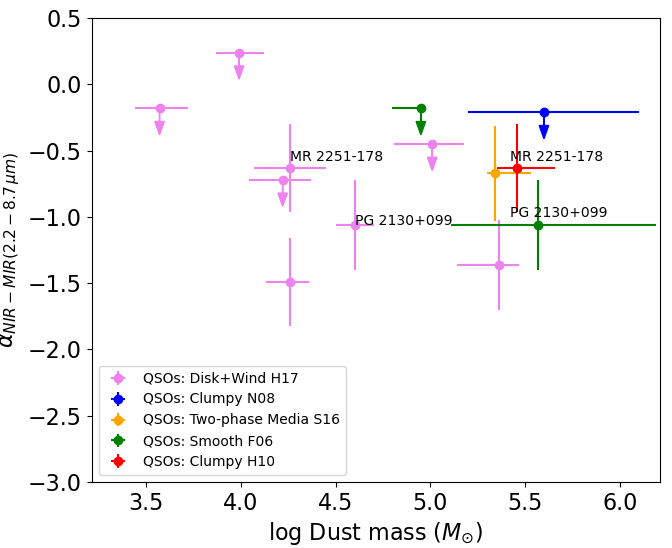}\\
\end{tabular}
\caption{The NIR to MIR spectral index $\alpha_{NIR-MIR(2.2,8.7\,\mu\text{m})}$ versus the bolometric luminosity estimated from the X-ray luminosity (left), and versus the dust mass inferred from the corresponding model (right).\label{Spec_index_Lbol}}
\end{figure*}

\begin{figure*}
	\centering
	\includegraphics[width=1\columnwidth]{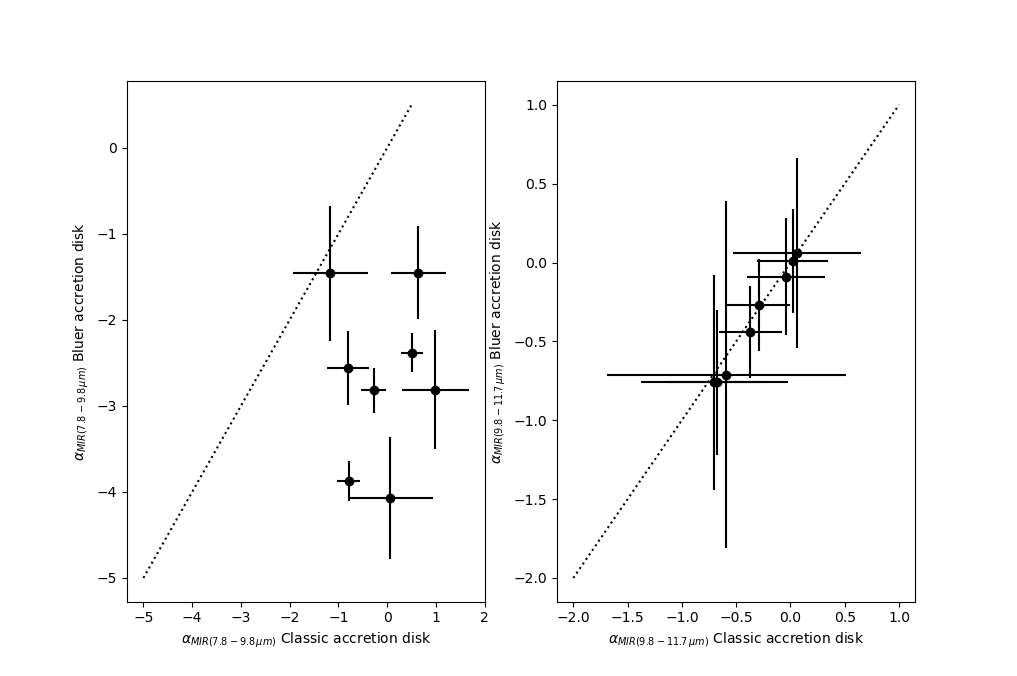}
	\caption{{\bf MIR spectral indexes obtained after subtract the bluer and classic accretion disk from the high angular resolution spectrum}.  \label{MIR_slopes1}}
\end{figure*}

\begin{figure*}
	\centering
	\includegraphics[width=1\columnwidth]{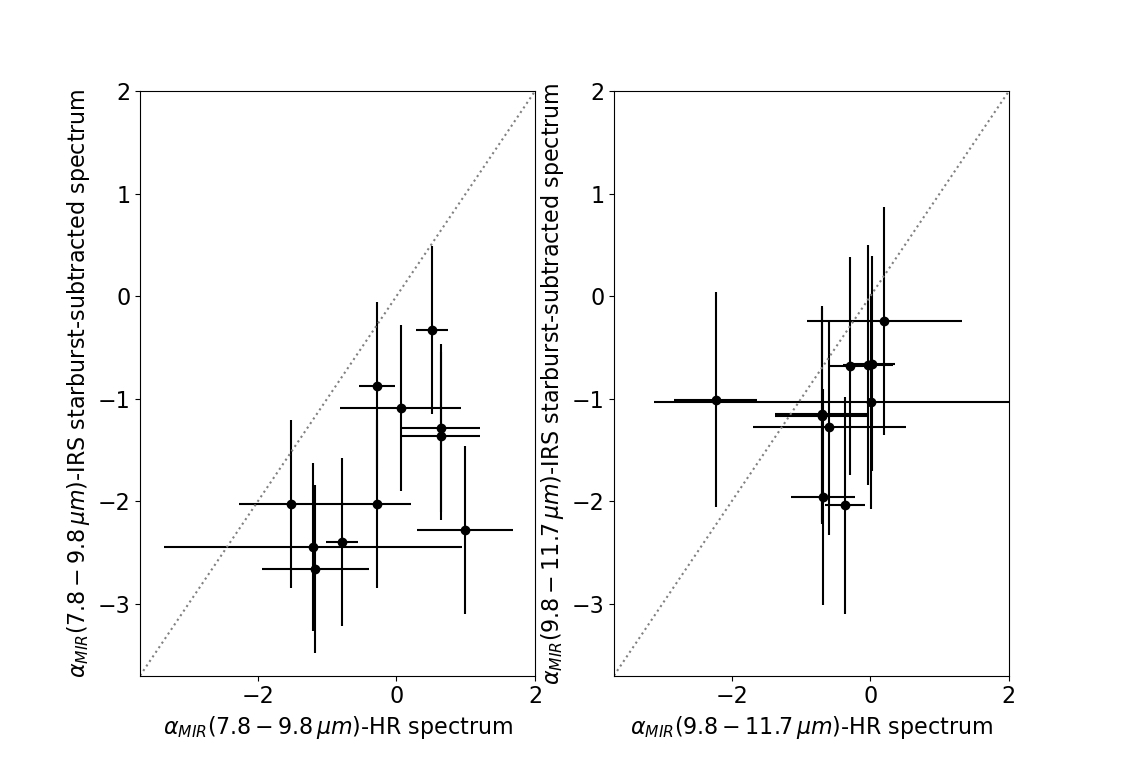}
	\caption{MIR spectral indexes were obtained after subtracting the classic accretion disk and the stellar and starburst components from the  {\it Spitzer}/IRS spectrum against the MIR spectral indexes measured on the high angular resolution spectrum.  \label{MIR_starburst}}
\end{figure*}

\begin{figure*}
\centering
\includegraphics[width=0.9\columnwidth]{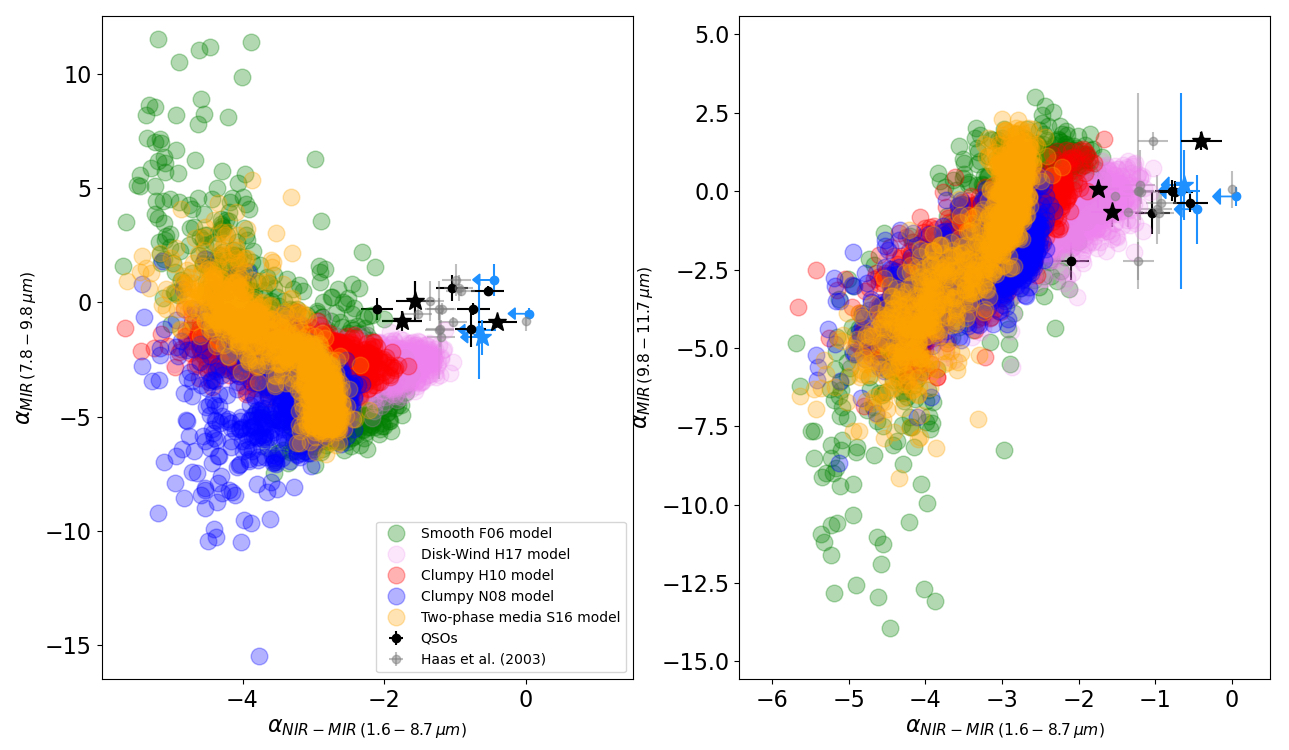}
\includegraphics[width=0.9\columnwidth]{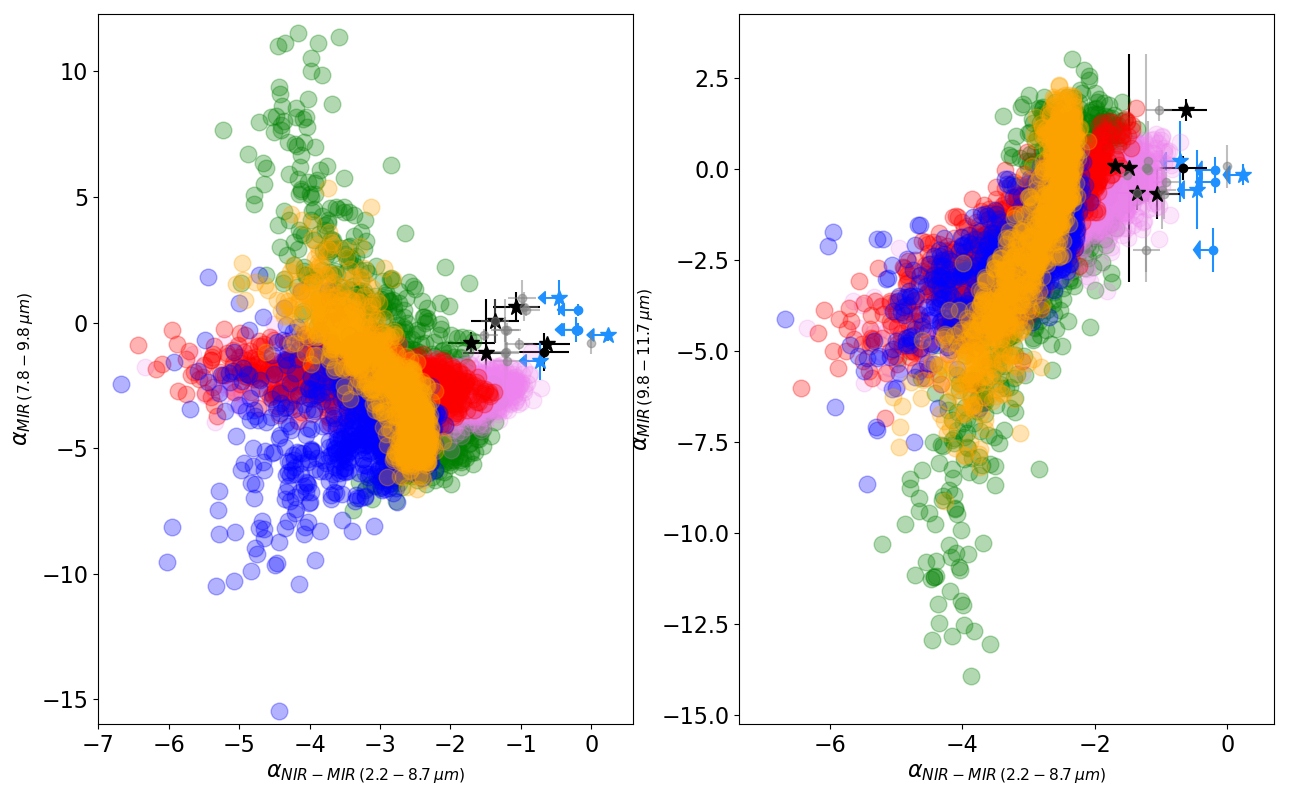}
\caption{The synthetic and observed spectral indexes $\alpha_{MIR(7.8,9.8\,\mu\text{m})}$ and $\alpha_{MIR(9.8,11.7\,\mu\text{m})}$ versus the NIR to MIR spectral indexes $\alpha_{NIR-MIR(1.6,8.7)}$ and $\alpha_{NIR-MIR(2.2,8.7)}$ for all objects and models. The stars indicate that the spectral index is derived using the new NIR high angular resolution data.\label{Alphas1}}
\end{figure*}

In the left panel of Figure~\ref{Spec_index_Lbol} we compare our measurements of
the NIR to MIR $\alpha_{NIR-MIR(2.2,8.7)}$ spectral index listed in Table~\ref{tab:Alpha_Mdust}, with the AGN bolometric luminosity $L_{bol}$ and the dust mass. We estimated the bolometric luminosity using the hard X-ray (2-10 keV) flux and the relation derived by \citet{Marconi04}, see \citet{Martinez_Paredes19}. In the plots, we identify each object with the model that better fits their data, except for PG~0003+199 for which none of the models can fit with a red-$\chi^{2}<2$. 
We note that most objects have a NIR to MIR spectral index between $\sim0$ and $-1.5$, which does not correlate with the bolometric luminosity nor with the dust mass. \citet{Haas03} estimated the dust mass of some 
objects in our sample using equation 6 in \citet{Stickel00} and assuming standard grain properties and a dust emissivity $\beta=2$. The mass that they derived varies from $\sim10^{6}-10^{8}$ M$_{\odot}$, with an uncertainty that can vary up to an order of magnitude. However, due to the large uncertainties in the far-IR fluxes, the dust masses might be overestimated, as pointed out by the authors.

We estimate the MIR spectral indexes for two cases, one in which we subtracted the classic accretion disk from the high angular resolution spectrum and another in which we subtracted the bluer accretion disk. We find that for the MIR spectral index $\alpha_{MIR(9.8-11.7\mu\text{m})}$ both measurements give similar results ($\sim-0.2$). However, for the $\alpha_{MIR(7.8-9.8\mu\text{m})}$ the values obtained by assuming the bluer  accretion disk are redder ($\sim-3$) than those obtained by assuming the classic  accretion disk ($\sim-0.4$), see Figure~\ref{MIR_slopes1}. 
Additionally, we investigate if the inner star formation could be affecting our estimations of the MIR spectral indexes. We measure the slopes $\alpha_{MIR(7.8-9.8\mu\text{m})}$ and $\alpha_{MIR(9.8-11.7\mu\text{m})}$  on the starburst subtracted {\it Spitzer}/IRS spectrum, and comparing with the slopes measured  on the  high angular resolution spectra. We find that the starburst subtracted {\it Spitzer}/IRS MIR slopes are redder ($\sim-2$ for the slope between $7.8-9.8\mu\text{m}$). Figure~\ref{MIR_starburst}  shows both spectral indexes, where the measurements are assuming the classic accretion disk.

These results suggest that although the high angular resolution spectrum is mostly dominated by the emission of the AGN, it is possible that the bluer spectral range of the spectrum is being affected by some dust components not directly related with the dust heated by the AGN.

\begin{table*}
\begin{minipage}{1.\textwidth}
\caption{{\bf NIR-MIR spectral indexes obtained after subtracting the classic accretion disk}. Column 1 list the names of the objects. Columns 2, 3 and 4 list the spectral index $\alpha_{NIR-MIR(1.6,8.7\,\mu\text{m})}$ and $\alpha_{NIR-MIR(2.2,8.7\,\mu\text{m})}$ calculated by us, and the spectral index $\alpha_{1,10\,\mu\text{m}}$ reported by \citet{Haas03}, respectively. \label{tab:Alpha_Mdust}}
\centering
%\resizebox{18cm}{!}{
\begin{tabular}{l|ccc}
				\hline
Name  &  This work & This work& \citet[][]{Haas03}\\
      &    $\alpha_{NIR-MIR(1.6,8.7\,\mu\text{m})}$ &  $\alpha_{NIR-MIR(2.2,8.7\,\mu\text{m})}$ & $\alpha_{1,10\,\mu\text{m}}$ \\
\hline
Mrk~509     & $-1.57\pm0.27$ & $-1.36\pm0.34$&... \\
PG~0050+124 & $-0.78\pm0.22$  &  $-0.67\pm0.36$&-1.21   \\ 
                       & $-1.15\pm0.25^{*}$  &  $-0.79\pm0.36^{*}$   \\ 
PG~2130+099 & $-1.04\pm0.23$  & $-1.06\pm0.34$&-1.52   \\
PG~1229+204$^{a}$ & $<-0.45$   &$<-0.45$ &-0.98  \\
PG~0844+349$^{a}$ & $<0.05$  & $<0.24$ & ... \\
PG~0003+199 & $-1.75\pm0.28$  & $-1.70\pm0.34$ & ... \\
PG~0804+761$^{a}$ & $<-0.62$  &$<-0.72$ &-1.20   \\ 
PG~1440+356 & $-0.54\pm0.23$  &$<-0.18$ &-0.92   \\
                       & ...  &$<-0.92^{*}$    \\
PG~1426+015 & $-2.10\pm0.23$  & $<-0.21$& -1.22   \\ 
PG~1411+442 & $-0.74\pm0.23$ &  $<-0.18$&... \\ 
                      & ... &  $<-0.71^{*}$ \\ 
PG~1211+143 & $<-0.66$  &$-1.49\pm0.33$  &-1.22 \\
PG~1501+106 & ...  & ...&-1.03  \\
MR~2251-178 & $-0.40\pm0.27$  &$-0.63\pm0.33$ &...  \\ 
\hline
\end{tabular}
%}
\end{minipage}
Note.- $^{a}$The accretion disk is not subtracted. $^{*}$This  values correspond to the case assuming the bluer accretion disk.
\end{table*}

We measure the same spectral indexes for the five models to investigate how well the different models sampled the observed NIR and MIR spectral slopes. In Figure~\ref{Alphas1} we compare the synthetic and observed values (measured on the high angular resolution spectrum after subtracting the classic accretion disk). In general, we note that none of the models matches the region of the observed spectral indexes. Although the closer match occurs for the Disk+Wind H17 model, especially when comparing the $\alpha_{MIR(9.8,11.9\,\mu\text{m})}$ and the $\alpha_{NIR-MIR(2.2,8.7\,\mu\text{m})}$. However, it is likely that the limited estimation of the $\alpha_{MIR(7.8,9.8\,\mu\text{m})}$  spectral index is affecting the comparison with the models since a spectral index around -2, as obtained from the starburst subtracted {\it Spitzer}/IRS MIR slopes, compares better with the models.

Particularly, we note that the Clumpy N08 and Two-phase media S16 models predict NIR to MIR spectral indexes  ($\alpha_{1.6,8.7}$ and $\alpha_{2.2,8.7}$ ) that are $\lesssim-2$, followed by the Clumpy H10 and Smooth F06 models that predict values $\lesssim-1.5$, and the Disk+Wind H17 model that predict values $\lesssim-0.8$. 
This result is consistent with the one obtained by \citet{Hoenig17} when compared the NIR to MIR  (3, 6 $\mu$m) versus  MIR (8, 14 $\mu$m) spectral indexes for the Clumpy H10 and the Disk+Wind H17 models. They showed that the Disk+Wind H17 models predict bluer NIR to MIR (3, 6 $\mu$m) spectral indexes,  and claimed that they come from SEDs with a $3-5\,\mu$m bump. Moreover, they say that these SEDs 
correspond to models where the disk has relatively steep dust
cloud distribution ($-2>a>-3$) and the wind shallower 
clouds distribution ($-0.5>a_{w}>-1.5$). We find that the average index of the cloud distribution in the disk and the wind is $a\sim-2.6$ and $a_{w}\sim-0.7$, respectively, suggesting that the most compact distribution of the clouds in the disk dominates the emission in the NIR, and therefore improves the fitting of this part of the SED.

From our analysis we found that the Disk+Wind SEDs predict the most bluer NIR to MIR spectral indexes. However, it is necessary for all models to better sample the region with NIR to MIR spectral index between $\sim-0.5$ and $\sim-2$, and MIR spectral index around zero, as it can be seen from Figure \ref{Alphas1}. This highlights the need for different dust distributions/compositions in the current phenomenological models available to the community. 

\citet{Martinez_Paredes20} found that the Disk+Wind H17 model is the best reproducing the AGN-dominated {\it Spitzer}/IRS spectrum for a sample of local type 1 AGNs, in which the $10\,\mu$m silicate emission feature is prominent. They obtained, on average, an 
index for the dust distribution in the disk $a\sim-1.9$ and an index for the dust distribution in the wind $a_{w}\sim-0.7$, which are similar to the values obtained for our sample of QSOs. Therefore, these results suggest that this model has a better treatment on the properties of the hot and warm dust, which dominates the emission between $\sim1$ and $\sim14\mu$m. 

\subsection{Torus/Disk geometry}

\citet{Mor09} used the {\it Spitzer}/IRS
($\sim2-35\,\mu$m) spectra of 26 QSOs ($z\sim0.06-0.33$) to investigate 
their main emission component. They found that the SED of these objects can be well reproduced by assuming three 
model components: 1) the Clumpy N08 model or 
dusty torus component, 2) the dusty narrow-line region cloud component, and 3) the blackbody-like dust component. 
They found that a substantial amount of the emission between $\sim2-7\,\mu$m originates from a hot dust component, likely situated 
in the innermost part of the torus.
Additionally, they found that the dust mass correlates with the bolometric luminosity, while the covering factor seems anti-correlated.

Later, \citet{Martinez_Paredes17} found that the starburst-subtracted 
{\it Spitzer}/IRS spectrum ($\sim8-15\,\mu$m) plus the NIR high angular resolution photometric point at H band from the NICMOS/HST of a sample of 20 
QSOs ($z<0.1$) could be well modelled assuming only the  
Clumpy N08 model. However, they noted that the inclusion of the spectral range between $\sim5-8\mu$m results in a poor fit of the silicate features at $9.7\,\mu$m, and therefore in a set of parameters inconsistent with an optically classified type 1 AGN.

Comparing the average residuals obtained from fitting the AGN-dominated {\it Spitzer}/IRS spectra of local type 1 AGNs, \citet{Martinez_Paredes20} found that in general, none of the models can well reproduce the spectrum from $\sim5-30\mu$m nor  the 10 and 18 $\mu$m, silicate emission features. However, they noted that the Wind+Disk H17 model produces the flattest residuals, especially for objects with higher bolometric luminosities ($\sim44.8-46.1$ erg s$^{-1}$ in log scale). Indeed, using a large sample of AGNs, \citet{Gonzalez-Martin19b} found that the Disk+Wind H17 models perform best for high luminosity AGNs ($44<log(L_{X (2-10\,\text{keV})})<44.5$ erg s$^{-1}$), while the Clumpy N08 model is better modelling low-luminosity AGNs, and the two-phase media S16 model {\it seems to be} better suited for intermediate luminosities.

In this work, we find that the NIR-MIR SED of QSOs is also better 
modelled by the Disk+Wind H17 model. Curiously, we note that for PG~1440+356
that fitted best with the Smooth F06 model, the next model that best fits the data is the Disk+Wind H17 one, while for PG~0050+124 that is best fitted with the Two-phase Media S16 model, the next model that best fit the data is the Smooth F06 model, suggesting that these QSOs always 
prefer the wind or smooth models rather than the clumpy models, probably 
since the former tend to produce 
bluer NIR SEDs. On the other hand, recently \citet{Woo20} claimed that AGNs with stronger outflow are hosted by galaxies  with  higher  star formation rates. Curiously, \citet{Martinez_Paredes19} found that the star formation rate of these QSOs is more centrally concentrated on scales of hundreds of pc as predicted by the evolutionary models. Therefore, it is possible that the outflow, if present, is playing an important role in the geometrical distribution of the central dust of these objects.
Recently, \citet{Garcia_Bernete21} found that the IR high angular resolution SED of local Seyfert 1 also prefer the Wind+Disk H17 model instead of the torus models. These results suggest the presence of a hot dust component in nearby type 1 AGNs. 
Note that, unlike Seyferts, in QSOs, the unresolved emission comes from a spatial region of $\lesssim1$ kpc. Considering these objects could be tracing different evolutionary stages \citep[see e.g, ][]{Haas03, Martinez_Paredes19}, in which the outflow could be responsible for the presence of polar dust on scales of hundreds of pc, could also explain the results that QSOs being best fitted by the Wind+Disk H17 model.

\section{Summary and conclusions}
\label{conclusion}
Understanding the nearby QSOs inner molecular dust geometry and dust composition has been partly limited for the lack of a well-sampled SED at IR wavelengths. In this work, we
use the cameras CIRCE and EMIR on the 10.4m Gran Telescopio CANARIAS to obtain new high angular resolution ($\sim0.3$arcsec) photometry in the H (1.6 $\mu$m) and/or Ks (2.1 $\mu$m) bands, for a sample of 13 QSOs. We analyse these NIR high angular resolution images and find that the H- and Ks-band emission is mostly unresolved in most QSOs. We find that only MR~2251-178 shows an extended ($\sim20\%$) emission in the H and Ks bands.
Additionally, we use the high angular resolution and unresolved photometry and spectroscopy at N band ($\sim7.5-13.5\,\mu$m) available for these objects. We complement these data 
with the low angular resolution ($\sim3.5$arcssec) AGN-dominated ($>80\%$) {\it Spitzer}/IRS spectrum 
($\sim5-30\,\mu$m) to take advantage of the high S/N ratio 
offered 
by these spectra. In this way, we build for the first time well-sampled NIR to MIR SEDs for local QSOs. 

To build a set of IR SEDs mainly dominated by the reprocessed emission of the dust, we decontaminate the unresolved high angular resolution data and the {\it Spitzer}/IRS spectrum from the emission of the accretion disk. We explore two power laws parametrizations for the accretion disk. The first is a classic accretion disk, for which we find that the contribution decreases from NIR to MIR wavelengths (H band $\sim35\%$, Ks band $\sim30\%$, and between $\sim5-7.5\,\mu$m $\sim2.4\%$) and becomes negligible at wavelengths $>7.5\mu$m. The second is a bluer accretion disk, for which we find the contribution in the NIR and MIR is larger when compared with the classic accretion disk. In the H and Ks bands, the contribution is $\sim41\%$ (two objects) and $\sim54\%$ (four objects), respectively. For the rest of the objects, the extrapolation of the bluer accretion disk overestimates the emission in the NIR. The contribution in the MIR  is $\sim17.5\%$ and $\sim5.5\%$ for the high angular resolution spectrum and in the {\it Spitzer}/IRS spectrum, respectively. This is nearly six to seven times larger than the contribution of the classic accretion disk.

Assuming the Smooth dusty torus model \citep[][ F06]{Fritz06}, the Clumpy dusty torus models \citep[][N08 and H10]{Nenkova08b, Hoenig10b}, the Two-phase media torus model 
\citep[][S16]{Stalev16}, and the Disk+Wind model \citep[][ H17]{Hoenig17}, we 
investigate which model better reproduces simultaneously the NIR and 
MIR data. To fit the models 
we use the interactive spectral fitting program XSPEC from the 
HEASOFT package and select as acceptable fits those with a red-$\chi^{2}<2$. Then, we use the Akaike Information Criterion (AIC) to investigate the quality of the model.
According to this criterion, we find that the Disk+Wind H17 model 
is the one better fits 
the objects ($64\%$, 9/13), 
followed by the Smooth torus model F06 ($14\%$, 2/13), the Clumpy 
torus H10 model ($7\%$, 1/13), the Clumpy torus N08 model ($7\%$, 1/13), and the 
Two-phase media S16 model ($7\%$, 1/13). 
We note that in general these objects prefer the wind and/or smooth models against the clumpy models, likely due to their flatter NIR SEDs. Additionally, we find that there is an object 
(PG~0003+199) that is not fitted by any model. Curiously this is the object 
with the lowest hard (2-10 keV) X-ray luminosity and with the poorest IR data set, since 
the {\it Spitzer}/IRS spectrum only covers $\sim10-35\mu$m, and the flux in the redder border of the ground-based high angular resolution spectrum drops due to the lack of signal. 

We make the same analysis using the observed SEDs that resulted from subtracting the bluer accretion disk. However, in this case, we find that the models fitted only 2/13 objects (PG~0050+124 and PG~1501+106). We find that the model that best reproduces the SED of these objects is the Disk+Wind H17 model, which is the same model obtained when assuming the classic accretion disk.

We calculate the average residual for each model. We find that 
the flattest residuals are obtained from the fits with the Disk+Wind H17 model. Comparing the angular width, viewing angle, covering factor, and dust mass derived from the best fits of each object, we note that the parameters are similar to the values previously obtained for a larger sample of type 1 AGNs using only the AGN-dominated {\it Spitzer}/IRS spectra.

Additionally, we use the unresolved fluxes at H ($1.6\,\mu$m), Ks ($2.2\,\mu$m), and Si2 (8.7 $\mu$m)
bands after subtracting the contribution from the accretion disk, 
and calculate two NIR to MIR spectral indexes, one between 1.6 and 8.7 $\mu$m, 
and another one between 2.2 and 8.7 $\mu$m, and two MIR spectral indexes, one between 7.8 and 9.8 $\mu$m, and another one between 9.8 and 11.7 $\mu$m. 

We compare the NIR to MIR spectral index  $\alpha_{NIR-MIR\,(2.2,8.7\,\mu\text{m})}$, with the bolometric luminosity and the dust mass derived from the best fit and find no correlation. Additionally, we measure the both NIR to MIR spectral indexes, $\alpha_{NIR-MIR\,(1.6,8.7\,\mu\text{m})}$ and $\alpha_{NIR-MIR\,(2.2,8.7\,\mu\text{m})}$, for all five models. Comparing the synthetic and observed 
values we find that none of the models simultaneously match the measured NIR to MIR and 7.8-9.8 slopes. However, we note that measuring the MIR slope between  7.8 and 9.8 $\mu$m on the starburst-subtracted {\it Spitzer}/IRS spectrum gives values that are more similar to the synthetic ones. Therefore, it is likely that some components not directly related to the dust heated for the AGN could affect the bluer spectral range of the high angular resolution spectrum.  Additionally, we note that the 
Disk+Wind H17 model has the closest values. This result is consistent with the fact that this model successfully reproduces most of the observed SEDs. We conclude that the differences among the synthetic and observed values highlight the need for different dust distributions/compositions in the phenomenological models.

Finally, we point out that a better understanding of the geometry 
and dust composition of the IR unresolved emission of QSOs will be possible 
with the forthcoming 
spectra from the Mid-Infrared Instrument on the James Webb 
Space Telescope. These spectra offer a similar angular resolution but better sensitivity, allowing to obtain a high angular resolution spectrum with a longer spectral range ($\sim5-30\,\mu$m) for the faint and bright nearby QSOs. While the future ground-based big-sized optical telescopes, 
like the European Extremely Large Telescope (40m), the 
Thirty Meter Telescope (30m), and the Giant Magellan Telescope (25m), will allow observing the inner nuclear dust 
in more distant objects.

\section*{Acknowledgements}
MM-P acknowledges support by the KASI postdoctoral fellowships. 
OG-M acknowledges supported from the UNAM PAPIIT [IN105720]. IGB  acknowledges support from STFC through grant ST/S000488/1. CRA acknowledges financial support from the Spanish Ministry of Science, Innovation
and Universities (MCIU) under grant with reference RYC-2014-15779, from
the European Union's Horizon 2020 research and innovation programme under Marie Sk\l odowska-Curie
grant agreement No 860744 (BiD4BESt), from the State Research Agency (AEI-MCINN) of the Spanish MCIU under grants
"Feeding and feedback in active galaxies" with reference PID2019-106027GB-C42 and "Quantifying the impact of quasar feedback on galaxy evolution (QSOFEED)" with reference EUR2020-112266. CRA also acknowledges support from the Consejería de Econom\' ia, Conocimiento y Empleo del Gobierno de 
Canarias and the European Regional Development Fund (ERDF) under grant with reference ProID2020010105 and from IAC project
P/301404, financed by the Ministry of Science and Innovation, through
the State Budget and by the Canary Islands Department of Economy, Knowledge and Employment, through the Regional Budget of the Autonomous Community. AA-H acknowledges support from PGC2018-094671-B-I00 (MCIU/AEI/FEDER,UE). AA-H work was done under project  No. MDM-2017-0737 Unidad de Excelencia “María de Maeztu”- Centro de Astrobiología (INTA-CSIC). IA acknowledges support from the project CB2016-281948.
JM acknowledges financial support from the State Agency for Research of the Spanish MCIU through the
"Center of Excellence Severo Ochoa" award to the Instituto de Astrofísica de Andalucía (SEV-2017-0709) and the  research projects AYA2016-76682-C3-1-P(AEI/FEDER, UE) and PID2019-106027GB-C41(AEI/FEDER, UE).
This work is
based on observations made with the 10.4m GTC located in the Spanish
Observatorio del Roque de Los Muchachos of the Instituto de
Astrof\'isica de Canarias, in the island La Palma. It is also based
partly on observations obtained with the \emph{Spitzer Space
Observatory}, which is operated by JPL, Caltech, under NASA contract
1407. This research has made use of the NASA/IPAC Extragalactic
Database (NED) which is operated by JPL, Caltech, under contract with
the National Aeronautics and Space Administration. CASSIS is a
product of the Infrared Science Center at Cornell University,
supported by NASA and JPL. This research made use of Photutils, an Astropy package for
detection and photometry of astronomical sources (Bradley et al.
2019).

\software{XSPEC (Arnaud 1996), HEAsoft (HEASARC 2014), astropy (The Astropy Collaboration 2013, 2018), Photutils (Bradley et al. 2019)}

%%%%%%%%%%%%%%%%%%%%%%%%%%%%%%%%%%%%%%%%%%%%%%%%%%

%%%%%%%%%%%%%%%%%%%% REFERENCES %%%%%%%%%%%%%%%%%%

% Alternatively you could enter them by hand, like this:
% This method is tedious and prone to error if you have lots of references

%%%%%%%%%%%%%%%%%%%%%%%%%%%%%%%%%%%%%%%%%%%%%%%%%%

%%%%%%%%%%%%%%%%% APPENDICES %%%%%%%%%%%%%%%%%%%%%
\section{Tables of parameters}

\begin{table*}
	\begin{minipage}{1.\textwidth}
		\caption{{\bf Parameters H17}. Column 1 lists the names of QSOs. Column 2 lists the viewing angle along the line of sight. Columns 3-5 list the parameters of the disk: the number of clouds along an equatorial line of sight, the power-law index of the radial distribution of dust clouds, and the dimensionless scale height. Columns 6-8 list the parameters of the wind: the angular width , half-opening angle of the wind, and power-law index of the radial distribution of dust clouds. Column 9 list the ratio between number of dust clouds along the cone and $N_{0}$. See equations in \citet{Hoenig17}.\label{tab:par_H17}}
%\centering
\resizebox{19cm}{!}{
\begin{tabular}{l|c|ccc|ccc|c}
				\hline
     & &\multicolumn{3}{c}{Disk} &\multicolumn{3}{|c}{Wind}&\\
     \hline
Name & $i$ & $N_{0}$ & $a$ & $h$&$\sigma_{\theta}$ &$\Theta_{w}$& $a_{w}$  & $f_{wd}$ \\
 &[0,90] deg& [5,10]&[-3.0,-0.5]& [0.1,0.5]&[7,15] deg&[30,45] deg &[-2.5,-0.5]  & [0.15,0.75]\\
  &mean (min, max)&mean (min, max)&mean (min, max)& mean (min, max)&mean (min, max)&mean (min, max) &mean (min, max) & mean (min, max)\\
\hline
\multicolumn{9}{c}{Good fits according to a $red-\chi^{2}<2$ and the AIC criterion}\\
\hline
{\bf Mrk509}	  & 70.5 (69.8, 90.0)& 10.0 (9.9, 10.0)& -3.0 (-3.0, -3.0)& 0.5 (0.1, 0.5)&12.9 (7.0, 13.3)& 45.0 (44.5, 45.0)& -0.5 (-0.5, -0.5) & 0.49 (0.50, 0.50)	\\		
{\bf PG1411+442}	& 0.01 (0.01, 0.04)& 10.0 (9.1, 10.0)& -3.0 (-3.0, -0.5)& 0.4 (0.3, 0.5)&7.5 (7.0, 8.8)& 36.7 (35.3, 45.0)& -1.2 (-2.5, -1.1)& 0.75 (0.68, 0.75)\\
{\bf PG1211+143}	& 58.1 (0.0, 90.0)& 10.0 (5.0, 10.0)& -2.5 (-3.0, -2.5)& 0.2 (0.1, 0.3)&15.0 (7.0, 15.0)& 45.0 (30.0, 45.0)& -2.2 (-2.5, -2.1)& 0.75 (0.15, 0.75)	\\
{\bf PG1501+106$^{*}$}&90.0 (89.3, 90.0)& 5.0 (5.0, 5.2)& -3.0 (-3.0, -0.5)& 0.2 (0.1, 0.5)&15.0 (14.8, 150)&. 45.0 (44.4, 45.0)& -0.5 (-2.5, -0.5)& 0.45 (0.44, 0.47) \\
{\bf MR2251-178}	&45.06 (44.67, 45.56) & 10.0 (9.9, 10.0) & -2.50 (-2.51 -2.50)& 0.22 (0.19, 0.25)&15.0 (14.8, 15.0) &45.0 (44.9. 45.0)& -2.0 (-2.5, -2.0)& 0.39 (0.38, 0.41)	\\
{\bf PG1229+204}	&21.20 (0.01, 21.72)& 10.0 (9.8, 10.0)& -2.5 (-2.6, -2.5)& 0.50 (0.46, 0.50)&15.0 (14.6, 15.0)& 45.0 (44.6, 45.0)& -1.50 (-1.51, -1.46)& 0.75 (0.73, 0.75)	\\
{\bf PG0844+349}	&17.11 (0.01, 20.37) & 10.0 (5.0, 10.0)& -2.87 (-2.94, -2.67)& 0.2 (0.1, 0.5)&15.0 (7.0, 15.0)& 35.5 (33.7, 45.0)& -0.5 (-0.8, -0.5)& 0.45 (0.33, 0.75)	\\
{\bf PG0804+761}	&0.01 (0.01, 0.01)& 9.5 (5.0, 10.0)& -3.0 (-3.0, -3.0)& 0.29 (0.10, 0.30)&7.6 (7.0, 8.4)& 34.9 (30.0, 35.9)& -0.5 (-0.5, -0.5)& 0.75 (0.15, 0.75)	\\
{\bf PG~2130+099} & 0.01 (0.01,0.01)& 10.0 (10.0,10.0) & -2.05 (-2.14,-2.00) & 0.5 (0.5,0.5)& 10.0 (9.7,10.2)& 45.0 (44.6,45.0) & -0.5 (-0.5,-0.5)  & 0.75 (0.72,0.75)\\
		\hline
\multicolumn{9}{c}{Accepted fits according to a $red-\chi^{2}<2$}\\
\hline
PG~0050+124 & 0.01 (0.01, 0.02) & 10.0 (9.9,10.0) & -1.6 (-1.7,-0.5) &  0.3 (0.3,0.5)&7.0 (7.0,7.7) & 45.0 (44.4,45.0) & -0.50 (-0.53,-0.50) &  0.70 (0.62,0.75) \\
PG~1440+356 & 0.01 (0.01,0.01) & 10.0 (9.9,10.0) & -3.0 (-3.0,-3.0) & 0.31 (0.30,0.5)&14.2 (13.3,15.0) & 42.0 (40.8,45.0) & -0.5 (-0.5,-0.5)  & 0.75 (0.74,0.75) \\
		\hline
			\end{tabular}}\\
	\end{minipage}
	\end{table*}

\begin{table*}
	\begin{minipage}{1.\textwidth}
		\caption{{\bf Derived parameters H17 and foreground extinction}. Column 1 lists the names. Column 2 and 3 list the dust mass in log scale and the covering factor (see the text for references on the equations). Column 4 lists the color excess for the foreground extinction $E(B-V)$.\label{tab:Dpar_H17}}
\centering
\begin{tabular}{l|cc|c}
				\hline
Name & $log_{10}Dust_{mass}$ (M$_{\odot}$)& $f_{cov}$&$E(B-V)$\\
  & mean (min, max) & mean (min, max)&\\
\hline
\hline
{\bf Mrk~509} & 5.36 (5.14;5.47)& 0.40 (0.22;0.44)& $<0.08$ \\
{\bf PG~1411+442} & 3.57 (3.44;3.72)& 0.19 (0.16;0.23)&  0.19 (0.17; 0.22)\\
{\bf PG~1211+143} & 4.26 (4.13;4.36)& 0.38 (0.28;0.45)& $<0.02$ \\
{\bf PG~1501+106} & 4.26 (4.07;4.45)& 0.23 (0.18;0.38)& 0.18 (0.15; 0.22) \\
{\bf MR~2251-178} & 5.01 (4.81;5.18)& 0.28 (0.19;0.39)&  0.22 (0.20;  0.24)\\
{\bf PG~1229+204} & 3.99 (3.87;4.12)& 0.40 (0.32;0.46)&  $<0.45$\\
{\bf PG~0844+349} & 4.22 (4.04;4.37)& 0.32 (0.23;0.41)&  $<0.06$\\
{\bf PG~0804+761} & 5.18 (5.03;5.29)& 0.38 (0.28;0.45)&  $<0.01$\\
{\bf PG~2130+099}      & 3.8 (3.7, 3.9)   &   0.19 (0.16, 0.23)&  0.39 (0.36;  0.43)\\
\hline
PG~0050+124      & 4.6 (4.5, 4.6)  &  0.41 (0.39, 0.44)&   $<0.01$\\
PG~1440+356      & 4.1 (3.9, 4.2)   &  0.23 (0.18, 0.28)&   0.33 (0.30; 0.37)\\
		\hline
			\end{tabular}\\
	\end{minipage}
	\end{table*}

\begin{table*}
	\begin{minipage}{1.\textwidth}
		\caption{{\bf Parameters H10}. Column 1 list the names. Columns 2-6 list the viewing angle, number of dust clouds along the equatorial line of the torus, the angular width od the torus, the power-law index of the radial distribution of the dust clouds, and the optical depth of the dust clouds. See equations in \citet{Hoenig10b}. \label{tab:par_H10}}
\centering
\begin{tabular}{l|ccccc}
				\hline
Name & $i$& $N_{0}$ & $\theta$ & $a$ & $\tau_{\nu}$ \\
  & [0,90] deg & [2.5,10.0] & [5,60] deg & [-2.0,0.0] & [30,80]\\
    & mean (min, max)&mean (min, max)&mean (min, max)&mean (min, max)&mean (min, max)\\
\hline
\multicolumn{6}{c}{Good fits according to a $red-\chi^{2}<2$ and the AIC criterion}\\
\hline
{\bf MR~2251-178} & 53.5 (51.5, 90.0)& 7.8 (2.5, 8.2)& 38.2 (34.9, 60.0)& -1.00 (-1.02, -0.98)& 30.0 (30.0, 30.4) \\
\hline
\multicolumn{6}{c}{Accepted fits according to a $red-\chi^{2}<2$}\\
\hline
Mrk~509 & 31.4 (30.3,32.3) & 10 (9.5,10.0) & 47.4 (46.3,60.0)& -0.91 (-0.98,-0.01) & 80 (79.4,80.0)  \\
PG1211+143 & 52.9 (50.5,54.9) & 10.0 (9.6,10.0) & 51.9 (5.0,54.9) & -0.9 (-2.0,-0.9) & 33.2 (30.1,36.9) \\
PG1440+356 & 51.3 (44.0, 90.0) & 3.0 (2.7,10.0) & 59.5 (5.0,60.0) & -1.7 (-1.7,-1.6) & 80.0 (78.5,80.0) \\
PG~1501+106 & 89.5 (77.6,90.0) & 10 (9.6,10.0) & 45.1 (40.2,54.0)  & -1.3 (-1.4,-1.3)  & 60.0 (30.0,65.4) \\
PG2130+099 & 45.00 (0.01,47.38) & 10.0 (9.8,10.0) & 54.5 (5.0,55.5)& -1.8 (-2.0,-1.8) & 80.0 (79.2,80.0) \\
PG~0804+761 &42.21 (39.38, 45.07)& 6.29 (6.14, 10.0)& 60.0 (59.2, 60.0)& -1.4 (-2.0, -1.4)& 80.0 (79.7 80.0)  \\ 
PG~0844+349 &29.9 (25.8, 31.3)& 8.8 (8.1, 9.5)& 57.0 (54.9, 59.1)& -1.1 (-1.1, -1.1)& 80.0 (79.2, 80.0) \\
PG~1229+204 &56.69 (0.01, 73.86)& 5.8 (4.9, 10.0)& 31.5 (5.0, 60.0)& -1.07 (-2.0, -1.03)& 66.2 (63.4, 80.0)\\
PG~1411+442 & 48.3 (44.7,90.0)& 10.0 (9.3,10.0)& 60.0 (5.0,60.0)&-1.91 (-1.93,-1.88) & 80.0 (79.1,80.0)\\
		\hline
			\end{tabular}\\
	\end{minipage}
	\end{table*}	
	
	\begin{table*}
	\begin{minipage}{1.\textwidth}
		\caption{{\bf Derived parameters H10 and foreground extinction}. Column 1 lists the names. Column 2 and 3 list the dust mass in log scale and the covering factor (see the text for references on the equations).  Column 4 lists the color excess for the foreground extinction $E(B-V)$.\label{tab:Dpar_H10}}
\centering
\begin{tabular}{l|cc|c}
				\hline
Name & $log_{10}Dust_{mass}$ & $f_{cov}$ & E(B-V)\\
  & mean (min, max) & mean (min, max) & \\
\hline
\hline
{\bf MR~2251-178} & 5.46 (5.35;5.66)&0.85 (0.78;0.92)  &  0.01  (0.01;  0.02) \\
\hline
Mrk~509 &  5.8 (5.7, 5.9)  &  0.92 (0.88, 0.93)  &  0.09 (0.07; 0.10) \\
PG~1211+143 &  4.4 (4.1, 4.6) &  0.74 (0.46, 0.88)  &   0.01 (0.01;   0.03)\\
PG~1440+356 &  5.0 (4.8, 5.1) &  0.93 (0.85, 0.95)  &  0.45 (0.41; 0.50) \\
PG~1501+106 &  4.9 (4.6, 5.1) & 0.74 (0.5, 0.92)  &   0.04 (0.01; 0.09) \\
PG~2130+099 & 4.3 (4.2, 4.4)  &  0.64 (0.53, 0.74)  & 0.55 (0.53; 0.57)  \\
PG~0804+761 &  5.3 (4.9, 5.5) &  0.64 (0.31, 0.88)  &  0.01  (0.01; 0.02) \\
PG~0844+349 & 4.5 (4.1, 4.8)  &  0.71 (0.32, 0.9)  &  0.01  (0.01; 0.02) \\
PG~1229+204 &  4.2 (3.9, 4.4) & 0.71 (0.44, 0.92)  &  0.28 (0.19; 0.36)  \\
PG~1411+442 & 4.61 (4.46;4.69) & 0.93 (0.88;0.95)  &   0.17 (0.15; 0.20) \\
		\hline
			\end{tabular}
	\end{minipage}
	\end{table*}

\begin{table*}
	\begin{minipage}{1.\textwidth}
		\caption{{\bf Parameters N08}. Column 1 lists the names. Column 2-7 list the viewing angle, the number of clouds along the equatorial line, the angular width, the outer to inner radius ratio of the torus, the index of the power-law distribution, and the optical depth of the dust clouds in the torus. See equations in \citet{Nenkova08b}. \label{tab:par_N08}}
\centering
\resizebox{18cm}{!}{\begin{tabular}{l|cccccc}
				\hline
Name & $i$& $N_{0}$ & $\sigma$ & $Y$ & $q$ & $\tau_{V}$ \\
     & [0,90] deg & [1,15] & [15,70] deg & [5,100] & [0,2.5] & [5,300]\\
     & mean (min,max)&mean (min,max)&mean (min,max)&mean (min,max)&mean (min,max)&mean (min,max)\\
\hline
\multicolumn{7}{c}{Good fits according to a $red-\chi^{2}<2$ and the AIC criterion}\\
\hline
{\bf PG~1426+015} &25.6 (13.4, 44.8)& 13.9 (9.4, 15.0)& 15.0 (15.0, 27.1)& 5.0 (5.0, 100)& 2.5 (2.4, 2.5)& 60.2 (55.4, 78.9)\\
\hline
\hline
{\bf PG~1426+015}$^{a}$ &$64^{+9}_{-8}$& $9^{+3}_{-3}$&$19^{+3}_{-2}$&$60^{+23}_{-27}$&$2^{+0.4}_{-0.4}$&$118^{+14}_{-23}$\\
{\bf PG~1426+015}$^{b}$ &30 &5&35&33&1&75\\
\hline
\multicolumn{7}{c}{Accepted fits according to a $red-\chi^{2}<2$}\\
\hline
PG1211+143 & 80.0 (74.9,90.0) & 9.0 (1.0,13.0) & 17.5 (15.0,70.0) & 5.0 (5.0,5.1) & 2.5 (2.4,2.5) & 44.01 (38.87,47.62) \\
PG1211+143$^{a}$& $80^{+6}_{-10}$& $2^{+13}_{-1}$&$45^{+12}_{-8}$&$40^{+4}_{-3}$& $1.43^{+0.05}_{0.05}$ & $38^{+3}_{-3}$ \\
%         \hline
PG1501+106$*$ & 90.0 (86.6,90.0) & 15.0 (14.0, 15.0) & 32.3 (15.0,42.2) & 9.99 (9.74,10.09) & 2.24 (2.15,2.25) & 18.49 (17.63,300.0) \\
PG1501+106$^{a}$  &$40^{+23}_{-21}$&$10^{+3}_{-3}$&$58^{+7}_{-8}$&$40^{+32}_{-18}$&$2.6^{+0.2}_{-0.2}$& $113^{+16}_{-14}$\\
%          \hline
PG~0804+761 & 90.0 (88.1, 90.0)& 1.0 (1.0, 1.0)& 45.1 (40.0, 54.5)& 5.5 (5.0, 5.5)& 2.5 (2.5, 2.5)& 55.0 (10.0, 55.4) \\
PG~0804+761$^{a}$ &$22^{+15}_{-12}$& $5^{+4}_{-2}$&$22^{+10}_{-4}$&$59^{+23}_{-29}$&$1.8^{+0.1}_{-0.2}$&$42^{+22}_{-17}$\\
%\hline
PG~0844+349 & 90.0 (87.3, 90.0)& 3.8 (3.4, 4.4)& 34.8 (16.0, 47.6)& 100.0 (74.8, 100.0)& 2.5 (2.5, 2.5)& 19.9 (18.5, 21.8) \\
PG~0844+349$^{a}$ & $88^{+1}_{-3}$&$2^{+13}_{-1}$ & $16^{+1}_{-1}$&$77^{+15}_{30}$&$1.3^{+0.2}_{-0.1}$&$77^{+25}_{-15}$\\
%\hline
PG~1229+204 &90.0 (78.1, 90.0)& 15.0 (14.4, 15.0)& 62.3 (47.0, 70.0)& 22.0 (19.1, 100.0)& 2.2 (2.1, 2.5)& 10.0 (10.0, 10.6) \\
PG~1229+204$^{a}$ & $73^{+2}_{-3}$ & $12^{+2}_{-3}$&$16^{+1}_{-1}$&$59^{+20}_{-18}$&$0.53^{+0.10}_{-0.07}$& $20^{+9}_{-5}$\\
PG~1229+204$^{b}$  & 28 & 5 & 34 & 33 & 2 & 91 \\
%\hline
MR2251-178 & 64.8 (62.2, 72.9)& 9.6 (1.0, 9.9)& 70.0 (66.6, 70.0)& 10.0 (9.8, 10.1)& 2.2 (0.0, 2.3)& 10.0 (10.0, 10.0)\\
		\hline
			\end{tabular}}\\
			Note.-$^{a}$Parameters obtained by \citet{Martinez_Paredes17}. $^{b}$Parameters obtained by \citet{Mor09}. $^{*}$NIR data not available.
	\end{minipage}
	\end{table*}
	
\begin{table*}
	\begin{minipage}{1.\textwidth}
		\caption{{\bf Derived parameters N08 and foreground extinction}. Column 1 list the name. Column 2, 3 and 4 list the dust mass in log scale and the covering factor, and outer radius of the torus (see the text for references on the equations).  Column 5 lists the color excess for the foreground extinction $E(B-V)$. Columns 6 and 7 list the covering factor previously derived by \citet{Martinez_Paredes17} and \citet{Mor09}, respectively.\label{tab:Dpar_N08}}
\centering
%\resizebox{18cm}{!}{
\begin{tabular}{l|ccc|c|cc}
				\hline
Name & $log_{10}Dust_{mass}$ (M$_{\odot}$) & $f_{cov}$&$R_{out}$ (pc)& $E(B-V)$ &$f_{cov}^{a}$&$f_{cov}^{b}$\\
 & mean (min, max) &mean (min, max) &mean (min, max) & & & \\
\hline
\hline
{\bf PG~1426+015} &5.6 (5.2; 6.1) &0.87 (0.77; 0.95) & 3.4 (3.4; 4.3)& 0.13 (0.07; 0.21)  &$0.14^{+0.02}_{-0.04}$& 0.33\\
\hline
PG~1211+143    &  3.1 (3.0, 3.5)  &  0.52 (0.36, 0.66) & 1.9 (1.9, 1.9)& 0.01 (0.01; 0.02) &$0.4^{+0.1}_{-0.1}$ & \\
PG~1501+106$^{*}$    &   5.8 (4.8, 7.3) & 0.80 (0.54, 0.93) &19.5 (7.9, 30.3) &0.01  (0.01; 0.02) &$0.9^{+0.1}_{-0.1}$ &  \\
PG~0804+761    &   4.0 (3.9, 4.2) & 0.84 (0.69, 0.92)  &  5.5 (5.5, 5.5)& 0.09 (0.06; 0.11) &$0.2^{+0.1}_{-0.1}$ &  \\
PG~0844+349    &  3.3 (3.0, 3.7)  & 0.64 (0.44, 0.87)  & 2.1 (2.1, 2.1)& 0.02 (0.01; 0.08) &$0.040^{+0.01}_{0.004}$ & \\
PG~1229+204    &   3.1 (2.9, 3.4) & 0.74 (0.52, 0.93)  & 1.5 (1.5, 1.5)& 0.05 (0.01; 0.13) &$0.1^{+0.01}_{-0.01}$  &0.31 \\
MR~2251-178 & 4.0 (3.9, 4.2)& 0.64 (0.48, 0.82)& 5.5 (5.5, 5.5)& 0.01  (0.01; 0.01) &$0.4^{+0.3}_{-0.2}$ &\\
		\hline
			\end{tabular}%}
			\\
			Note.-$^{a}$Parameters obtained by \citet{Martinez_Paredes17}. $^{b}$Parameters obtained by \citet{Mor09}. $^{*}$NIR data not available.
	\end{minipage}
	\end{table*}

\begin{table*}
	\begin{minipage}{1.\textwidth}
		\caption{{\bf Parameters F06}. Column 1 lists the names. Columns 2-7 list the viewing angle, the angular width of the torus, the power-law index of the polar density gradient, the power-law index of the radial density gradient, the outer to inner radius ratio of the torus, and the optical depth of the torus. See equations in \citet{Fritz06}. \label{tab:par_F06}}
\centering
\resizebox{19cm}{!}{
\begin{tabular}{l|cccccc}
				\hline
Name &$90-i^{*}$&$\sigma$ & $\gamma$ & $\beta$ & $Y$ & $\tau_{V}$\\
     & [0,90] deg & [20,60] deg & [0,6] & [-1.0,0.0] & [10,150] &[0.3,10]\\
\hline
\multicolumn{7}{c}{Good fits according to a $red-\chi^{2}<2$ and the AIC criterion}\\
\hline
{\bf PG~2130+099} &  49.0 (0.0, 51.1)& 38.2 (35.9, 39.9)& 0.01 (0.0, 6.0)& -0.50 (-0.58, -0.46)& 10.0 (10.0, 150.0)& 0.9 (0.1 1.0)\\
{\bf PG~1440+356} & 39.3 (0.0, 44.4)& 60.0 (57.9, 60.0)& 2.0 (1.3, 2.6)& -0.74 (-0.75, -0.74)& 11.7 (10.0, 11.9)& 0.5 (0.4 10.0)\\
\hline
\multicolumn{7}{c}{Good fits according to a $red-\chi^{2}<2$}\\
\hline
Mrk~509 & 69.3 (68.5,70.0) & 20.0 (20.0,20.4) & 0.07 (0.07,6.0) & -0.05 (-0.07,-0.01) & 10.0 (10.0,10.0) & 3.6 (3.4,3.7)\\
PG~0050+124 & 53.8 (48.4,90.0) & 22.1 (20.0,60.0) & 6.0 (6.0,6.0) & -0.57 (-0.59,-0.01) & 10.1 (10.0,10.2) & 6.7 (6.2,6.9) \\
PG~1211+143 & 14.48 (0.01,15.75) & 20.0 (20.0,20.2) & 6.0 (6.0,6.0) & -0.01 (-0.01,-0.01) & 10.0 (10.0,10.01) & 9.3 (9.0,9.8)\\
PG~1501+106 & 0.01 (0.01,0.02) & 20.0 (20.0,20.1) & 6.0 (5.9,6.0) & -0.15 (-0.16,-0.12) & 10.0 (10.0,10.01) & 10.0 (10.0,10.0) \\
PG~0804+761 &64.8 (61.4, 69.2)& 20.0 (20.0, 21.1)& 0.01 (0.01, 0.01)& -0.9 (-1.0, -0.8)& 10.0 (10.0, 10.0)& 0.60 (0.58, 0.61) \\
PG~0844+349 &0.01 (0.01, 9.62)& 20.0 (20.0, 20.9)& 3.37 (0.01, 3.62)& -0.25 (-1.0, -0.15)& 10.0 (10.0, 10.0)& 2.2 (0.1, 2.4) \\
PG~1229+204 &0.51 (0.01, 3.37)& 20.0 (20.0, 20.6)& 5.85 (0.01, 5.88)& -0.3 (-1.0, -0.2)& 10.0 (10.0, 10.0)& 10.0 (9.9, 10.0) \\
PG~1411+442 & 68.3 (0.0, 68.8)& 20.0 (20.0, 20.2)& 0.01 (0.0, 6.)& -1.0 (-1.0, 0.0)& 10.0 (10.0, 150.0)& 0.62 (0.61, 10.0) \\
		\hline
			\end{tabular}}\\
			Note.-$^{*}$In order to do a proper comparison between the orientation of the viewing angle 
			predicted by the several models we list the $90-i$ angle, since in F06 model the viewing angle increases from the polar to the equatorial axis.
	\end{minipage}
	\end{table*}
	
\begin{table*}
	\begin{minipage}{1.\textwidth}
		\caption{{\bf Derived parameters F06 and foreground extinction}. Column 1 lists the names. Column 2, 3 and 4 list the dust mass in log scale and the covering factor, and outer radius of the torus (see the text for references on the equations).  Column 5 lists the color excess for the foreground extinction $E(B-V)$.\label{tab:Dpar_F06}}
\centering
%\resizebox{18cm}{!}{
\begin{tabular}{l|ccc|c}
				\hline
Name & $log_{10}Dust_{mass}$ (M$_{\odot}$) & $f_{cov}$& $R_{out}$ (pc)& E(B-V))\\
  & mean (min, max) & mean (min, max) &mean (min, max) & \\
\hline
\hline
{\bf PG~2130+099} & 5.6 (5.1;6.2)& 0.86 (0.84;0.88)& 21.87 (21.87;22.47  &  0.48 (0.43; 0.52) \\
{\bf PG~1440+356} & 5.0 (4.8;5.0) & 0.14 (0.12;0.16) & 3.90 (3.90;4.74)  &  0.68 (0.66; 0.74)  \\
		\hline
Mrk~509   &  10.5 (10.2, 10.5)  & 0.15 (0.15, 0.17)   & 13.9 (13.9, 13.9)  &  0.01  (0.01; 0.02)   \\
PG~0050+124   &  5.2 (5.1, 5.4)  & 0.04 (0.02, 0.06)   & 27.2 (17.7, 39.5)  &  0.01 (0.01; 0.01)  \\
PG~1211+143   &  5.0 (5.0, 5.3)  & 0.20 (0.16, 0.22)   &  4.4 (3.6, 5.9)  &    0.01 (0.01; 0.01)\\
PG~1501+106   & 7.7 (7.1, 8.3)   &  0.22 (0.14, 0.35)  &  25.7 (10.7, 42.7)  &    0.01  (0.01; 0.03) \\
PG~0804+761   & 5.9 (5.8, 6.1)   & 0.04 (0.02, 0.06)   &  10.2 (10.2, 12.4)  &   0.01  (0.01; 0.02)  \\
PG~0844+349   &  5.1 (4.9, 5.2)  & 0.04 (0.02, 0.12)   &  7.1 (5.4, 14.4)  & 0.01 (0.01; 0.02)  \\
PG~1229+204   &  7.3 (5.7, 8.5)  &  0.90 (0.80, 0.96)  & 19.8 (11.0, 27.5)  &  0.01 (0.01; 0.03)   \\	
PG~1411+442 & 6.1 (5.9;6.5) & 0.92 (0.77;0.96)& 21.30 (15.54;26.02)  & 0.39 (0.36; 0.41) \\
\hline
			\end{tabular}%}\\
	\end{minipage}
	\end{table*}

\begin{table*}
	\begin{minipage}{1.\textwidth}
		\caption{{\bf Parameters S16}. Column 1 is for the name. Columns 2-7 show the viewing angle, the angular width, the power-law index of the radial density gradient, the power-law index of the polar density gradient, the outer to inner radius ratio of the torus, and the optical depth of the torus. See equations in \citet{Stalev16}.\label{tab:par_S16}}
\centering

\begin{tabular}{l|cccccc}
				\hline
Name & $i$& $\sigma$ & $p$ & $q$ & $Y$ & $\tau_{\nu}$ \\
  & [0,90] deg & [10,80] deg & [0,1.5] & [0,1.5] &[10,30] & [3,11]\\
\hline
\multicolumn{7}{c}{Good fits according to a $red-\chi^{2}<2$ and the AIC criterion}\\
\hline
{\bf PG~0050+124}  & 0.01 (0.01, 0.03)& 64.8 (63.6, 80.0)& 0.01 (0.01, 0.02)& 1.5 (0.0, 1.5)& 10.0 (10.0, 30.0)& 6.2 (6.0, 6.5) \\
\hline
\multicolumn{7}{c}{Good fits according to a $red-\chi^{2}<2$}\\
\hline
PG~1440+356 & 70.0 (69.7,70.3) & 10.0 (10.0,10.0) & 1.5 (1.5,1.5) & 0.01 (0.01,0.01) & 10.0 (10.0,10.1) & 11.0 (11.0,11.0) \\
PG~1501+106 & 11.7 (9.4,12.9) & 80.0 (79.8,80.0) & 1.45 (1.40,1.50) & 0.01 (0.01,0.01) & 10.0 (10.0,10.02) & 5.02 (4.9,5.2) \\
PG~2130+099 & 19.9 (15.4,22.4) & 60.0 (59.1,61.2) & 1.5 (1.5,1.5) & 1.5 (1.5,1.5) & 10.0 (10.0,10.0) & 8.7 (8.0,11.0) \\
MR~2251-178 &80.0 (79.2, 85.8)& 80.0 (79.8, 80.0)& 1.5 (1.5, 1.5)& 0.84 (0.82, 0.87)& 10.0 (10.0, 10.0)& 3.0 (3.0, 3.0) \\
PG0804+761 &42.3 (40.7, 43.7)& 70.0 (69.8, 70.2)& 1.5 (1.5, 1.5)& 1.5 (1.5, 1.5)& 10.0 (10.0, 10.0)& 3.0 (3.0, 3.0) \\
PG~0844+349 & 57.6 (44.1, 67.3)& 80.0 (69.8, 80.0)& 1.5 (1.5, 1.5)& 1.5 (1.4, 1.5)& 10.0 (10.0, 10.2)& 3.8 (3.6, 4.5) \\
PG~1229+204 & 63.6 (45.0, 68.0)& 79.2 (73.7, 80.0)& 1.5 (1.4, 1.5)& 0.90 (0.81, 1.50)& 10.0 (10.0, 10.2)& 6.9 (6.6, 7.2)  \\
		\hline
			\end{tabular}\\
	\end{minipage}
	\end{table*}

\begin{table*}
	\begin{minipage}{1.\textwidth}
		\caption{{\bf Derived parameters S16 and foreground extinction}. Column 1 is for the name. Column 2, 3 and 4 show the dust mass, the covering factor, and outer radius of the torus (see the text for references on the equations).  Column 5 lists the color excess for the foreground extinction $E(B-V)$.\label{tab:Dpar_S16}}
\centering
%\resizebox{18cm}{!}{
\begin{tabular}{l|ccc|c}
				\hline
Name & $log_{10}Dust_{mass}$ (M$_{\odot}$) & $f_{cov}$& $R_{out}$ (pc) & E(B-V)\\
   & mean (min, max) & mean (min, max) & mean (min, max) &  \\
\hline
\hline
{\bf PG~0050+124} &  5.3 (5.3;5.5)& 0.95 (0.94;0.96) & 5.49 (4.4;6.6)   &  0.01 (0.01; 0.01) \\
		\hline
PG~1440+356    &  5.4 (5.3, 5.4)  & 0.87 (0.84, 0.89)   &  3.3 (3.2, 3.5)   & 0.68 (0.66; 0.74)    \\
PG~1501+106    & 6.1 (5.9, 6.2)   &  0.97 (0.94, 0.98)  &  5.9 (4.4, 8.7)   &   0.01  (0.01; 0.03)  \\
PG~2130+099    & 5.2 (5.0, 5.3)   & 0.68 (0.66, 0.70)   &  6.3 (6.0, 6.4)   &   0.48 (0.43; 0.52)  \\
MR~2251-178    &  6.0 (6.0, 6.4)  & 0.95 (0.95, 0.95)   & 8.9 (8., 15.5)   &    0.01 (0.01; 0.01)  \\
PG~0804+761    &  6.7 (6.6, 6.9)  &0.98 (0.98, 0.98)    & 14.5 (11.1, 19.2)   &   0.01 (0.01; 0.02)   \\
PG~0844+349    &  5.9 (5.8, 6.0)  &0.98 (0.98, 0.98)    & 6.4 (5.0, 7.8)   &   0.01 (0.01; 0.02)   \\
PG~1229+204    &   5.3 (5.1, 5.4) & 0.96 (0.94, 0.98)   & 5.8 (4.9, 6.2)   &   0.01 (0.01; 0.03)   \\
\hline
			\end{tabular}%}\\
	\end{minipage}
	\end{table*}

\section{SED modeling}

%%%%%%%%%%%%%%%%%%%%%%%%%%%%%%%%%%%%%%%%%%%%%%%%%%%%

%%%%%%%%%%%%%%%%%%%%%%%%%%%%%%%%%%%%%%%%%%%%%%%%%%%%

%%%%%%%%%%%%%%%%%%%%%%%%%%%%%%%%%%%%%%%%%

%%%%%%%%%%%%%%%%%%%%%%%%%%%%%%%%%%%%%%%%%%%%%%%%%%%%%%%%%%%%%%%%%%%%%%%%%%%%%%%%%%%

%%%%%%%%%%%%%%%%%%%%%%%%%%%%%%%%%%%%%%%%%%%%%%%%%%%%%%%%%%%%%%%%%%%%%%%%%%%%%%%%%%%

%%%%%%%%%%%%%%%%%%%%%%%%%%%%%%%%%%%%%%%%%%%%%%%%%%%%%%%%%%%%%%%%%%%%%%%%%%%%%%%%%%%

%%%%%%%%%%%%%%%%%%%%%%%%%%%%%%%%%%%%%%%%%%%%%%%%%%%%%%%%%%%%%%%%%%%%%%%%%%%%%%%%%%%

%%%%%%%%%%%%%%%%%%%%%%%%%%%%%%%%%%%%%%%%%%%%%%%%%%%%%%%%%%%%%%%%%%%%%%%%%%%%%%%%%%%
%%%%%%%%%%%%%%%%%%%%%%%%%%%%%%%%%%%%%%%%%%%%%%%%%%%%%%%%%%%%%%%%%%%%%%%%%%%%%%%%%%%

%%%%%%%%%%%%%%%%%%%%%%%%%%%%%%%%%%%%%%%%%%%%%%%%%%%%%%%%%%%%%%%%%%%%%%%%%%%%%%%%%%

%%%%%%%%%%%%%%%%%%%%%%%%%%%%%%%%%%%%%%%%%%%%%%%%%%

\end{document}